\documentclass[%
 reprint,	
showpacs,
 amsmath,amssymb,
 aps,
 prc,
]{revtex4-1}

\usepackage{graphicx}
\usepackage{dcolumn}
\usepackage{bm}
\usepackage{url}
\usepackage{lipsum}
\usepackage{color}
\usepackage{hyperref}
\usepackage[mathlines]{lineno}
\usepackage{upgreek}
\usepackage{biolinum}

\begin{document}

\preprint{APS/123-QED}

\title{Centrality and transverse momentum dependence of $D^0$-meson production at mid-rapidity in Au+Au collisions at ${\sqrt{s_{\rm NN}} = \rm{200\,GeV}}$}


\author{
J.~Adam$^{12}$,
L.~Adamczyk$^{2}$,
J.~R.~Adams$^{34}$,
J.~K.~Adkins$^{25}$,
G.~Agakishiev$^{23}$,
M.~M.~Aggarwal$^{36}$,
Z.~Ahammed$^{56}$,
I.~Alekseev$^{3,30}$,
D.~M.~Anderson$^{50}$,
R.~Aoyama$^{53}$
% ...
}

% ...
\address{$^{2}$AGH University of Science and Technology, FPACS, Cracow 30-059, Poland}
\address{$^{3}$Alikhanov Institute for Theoretical and Experimental Physics, Moscow 117218, Russia}
% ...
\address{$^{12}$Creighton University, Omaha, Nebraska 68178}
% ...
\address{$^{23}$Joint Institute for Nuclear Research, Dubna 141 980, Russia}
% ...
\address{$^{25}$University of Kentucky, Lexington, Kentucky 40506-0055}
% ...
\address{$^{30}$National Research Nuclear University MEPhI, Moscow 115409, Russia}
% ...
\address{$^{34}$Ohio State University, Columbus, Ohio 43210}
% ...
\address{$^{36}$Panjab University, Chandigarh 160014, India}
% ...
\address{$^{50}$Texas A\&M University, College Station, Texas 77843}
% ...
\address{$^{53}$University of Tsukuba, Tsukuba, Ibaraki 305-8571, Japan}
% ...
\address{$^{56}$Variable Energy Cyclotron Centre, Kolkata 700064, India}
% ...

\collaboration{STAR Collaboration}
\noaffiliation

\date{\today}

\begin{abstract}
  We report a new measurement of $D^0$-meson production at mid-rapidity ($|y|$\,$<$\,1) in Au+Au collisions at ${\sqrt{s_{\rm NN}} = \rm{200\,GeV}}$ utilizing the Heavy Flavor Tracker, a high resolution silicon detector at the STAR experiment.
  Invariant yields of $D^0$-mesons with transverse momentum $p_{T}$ $\lesssim 9$\,GeV/$c$ are reported in various centrality bins (0--10\%, 10--20\%, 20--40\%, 40--60\% and 60--80\%). Blast-Wave thermal models are used to fit the $D^0$-meson $p_{T}$ spectra to study $D^0$ hadron kinetic freeze-out properties. The average radial flow velocity extracted from the fit is considerably smaller than that of light hadrons ($\pi,K$ and $p$), but comparable to that of hadrons containing multiple strange quarks ($\phi,\Xi^-$), indicating that $D^0$ mesons kinetically decouple from the system earlier than light hadrons. The calculated $D^0$ nuclear modification factors re-affirm that charm quarks suffer large amount of energy loss in the medium, similar to those of light quarks for $p_{T}$\,$>$\,4\,GeV/$c$ in central 0--10\% Au+Au collisions. At low $p_{T}$, the nuclear modification factors show a characteristic structure qualitatively consistent with the expectation from model predictions that charm quarks gain sizable collective motion during the medium evolution. 
The improved measurements are expected to offer new constraints to model calculations and help gain further insights into the hot and dense medium created in these collisions.

\end{abstract}

\maketitle
%
%
%

\section{Introduction}
\label{introduction}

The heavy ion program at the Relativistic Heavy Ion Collider (RHIC) and Large Hadron Collider (LHC) focuses on the study of strong interactions and Quantum Chromodynamics (QCD) at high temperature and density. Over the last few decades, experimental results from RHIC and LHC using light flavor probes have demonstrated that a strongly-coupled Quark-Gluon Plasma (sQGP) is created in these heavy-ion collisions. The most significant evidence comes from the strong collective flow and the large high transverse momentum ($p_{T}$) suppression in central collisions for various observed hadrons including multi-strange-quark hadrons $\phi$ and $\Omega$~\cite{StarWhitePaper,PhenixWhitePaper,LhcSummary,Adamczyk:2015ukd,Abelev:2014pua}.

Heavy quarks ($c$,$b$) are created predominantly through initial hard scatterings due to their large masses~\cite{Ziwei_Lin,Cacciari}. The modification to their production in transverse momentum due to energy loss and radial flow and in azimuth due to anisotropic flows is sensitive to heavy quark dynamics in the partonic sQGP phase~\cite{Moore}. Recent measurements of high-$p_{T}$ $D$-meson production at RHIC and LHC show a strong suppression in the central heavy-ion collisions~\cite{Alice_D_RAA_1,Alice_D_RAA_2,CMS_D_RAA_5TeV,Star_D_RAA}. The suppression is often characterized by the nuclear modification factor $R_{\rm AA}$, defined as
\begin{equation}
  R_{\rm AA}(p_T) = \frac{1}{\langle T_{\rm AA}\rangle} \frac{dN_{\rm AA}/dp_{T}}{d\sigma_{pp}/dp_{T}}.
\label{equ:equation0}
\end{equation}
where $dN_{\rm AA}/dp_T$ and $d\sigma_{pp}/dp_T$ are particle production yield and cross section in A+A and $p$+$p$ collisions, respectively. The nuclear thickness function $T_{\rm AA}=\langle N_{\rm bin}\rangle/\sigma_{pp}^{\rm inel}$ is often calculated using a Monte-Carlo Glauber model, where $\langle N_{\rm bin}\rangle$ is the average number of binary collisions and $\sigma_{pp}^{\rm inel}$ is the total inelastic $p$+$p$ cross section.
The $D$-meson $R_{\rm AA}$ is similar to that of light hadrons for $p_{T}$$\,>$\,4\,GeV/$c$, suggesting significant energy loss for charm quarks inside the sQGP medium. The measured $D$-meson anisotropic flow shows that $D$-mesons also exhibit significant elliptic and triangular flow at RHIC and LHC~\cite{Alice_D_v2_276TeV_PRL,Alice_D_v2_276TeV_PRC,CMS_D_vn_5TeV,Star_D_v2}. The flow magnitude when scaled with the transverse kinetic energy is similar to that of light and strange flavor hadrons. This indicates that charm quarks may have reached thermal equilibrium in these collisions at RHIC and LHC.

In this article, we report measurements of the centrality dependence of $D^0$-meson transverse momentum spectra at mid-rapidity ($|y|$\,$<$\,1) in Au+Au collisions at $\sqrt{s_{_{\rm NN}}} = 200$\,GeV. The measurements are conducted at the Solenoidal Tracker At RHIC (STAR) experiment utilizing the high resolution silicon detector (the Heavy Flavor Tracker, HFT)~\cite{Contin:2017mck}. The paper is organized in the following order: In Sec.~\ref{dataset}, we describe the detector setup and dataset used in this analysis. In Sec.~\ref{D0recon}, we present the topological reconstruction of $D^0$ mesons in the Au+Au collision data, followed by Sec.~\ref{correction} and Sec.~\ref{systematic} for details on efficiency corrections and systematic uncertainties. We present our measurement results and physics discussions in Sec.~\ref{result}. Finally, we end the paper with a summary in Sec.~\ref{summary} .

\section{Experimental setup and Dataset}
\label{dataset}

The dataset used in this analysis consists of Au+Au collision events at ${\sqrt{s_{\rm NN}} = \rm{200\,GeV}}$ collected in the 2014 year run. The main detectors used in this analysis are the Time Projection Chamber (TPC), the HFT, the Time of Flight (TOF) detector and the Vertex Position Detector (VPD). 

\subsection{Tracking and Particle Identification Subsystems}
\label{dataset:tpctof}

Precision tracking for this analysis is achieved with the TPC and HFT detectors. Particle identification for stable hadrons is performed with a combination of the ionization energy loss ($dE/dx$) measurement with the TPC and the time-of-flight ($tof$) measurement with the TOF detector. The event start time is provided by the VPD. Both the TPC and TOF detectors have full azimuthal coverage with a pseudo-rapidity range of $|\eta|$\,$<$\,1~\cite{TPC,TOF}. The TPC and TOF subsystems have been extensively used in many prior STAR analyses, including $D$-meson measurements ~\cite{Star_D_pp,Star_D_RAA,Adamczyk:2015ukd}. The HFT detector provides measured space points with high precision that are used to extend track trajectories and offer high-pointing resolution to the vicinity of the event vertex.

\subsection{Trigger and Dataset}
\label{dataset:trigger}

The minimum bias trigger used in this analysis is defined as a coincidence between the east and west VPD detectors located at 4.4\,$<$\,$|\eta|$\,$<$\,4.9~\cite{VPD}. Each VPD detector is an assembly of nineteen small detectors, each consisting of a Pb converter followed by a fast plastic scintillator read out by a photomultiplier tube. To efficiently sample the collision events in the center of the HFT acceptance, an online cut on the collision vertex position along the beam line (calculated via the time difference between the east and west VPD detectors) $|V_z^{\rm VPD}|$\,$<$\,6\,cm is applied. 
The decrease in the coincidence probability in the VPD degrades the online VPD vertex resolution in peripheral low multiplicity events. These inefficiencies are corrected in the offline analysis with a method discussed in the next section. 

Events used in this analysis are selected with the offline reconstructed collision vertex within 6 cm of the TPC and HFT centers along the beam direction to ensure uniform and large acceptance. The maximum total drift time of ionization electrons inside the TPC is about 40\,$\upmu$s while the hadronic Au+Au collision rate is typically around 40\,kHz for this dataset. There is a finite chance that more than one event is recorded in the TPC readout event frame. The VPD is a fast detector which can separate events from different bunch crossings (one bunch crossing interval at RHIC is 106\,ns). In order to suppress the chance of selecting a wrong vertex from collisions happening in bunch crossings different from that of the trigger, the difference between the event vertex $z$ coordinate $V_{z}^{\rm TPC}$ and the $V_{z}^{\rm VPD}$ is required to be less than 3\,cm. Approximately 9$\times 10^{8}$ minimum bias triggered events with 0--80\% centrality pass the selection criteria and are used in this analysis.

\subsection{Centrality Selection and Trigger Inefficiency}
\label{dataset:Centrality}

\begin{figure}[h]
\centering
\includegraphics[width=0.45\textwidth]{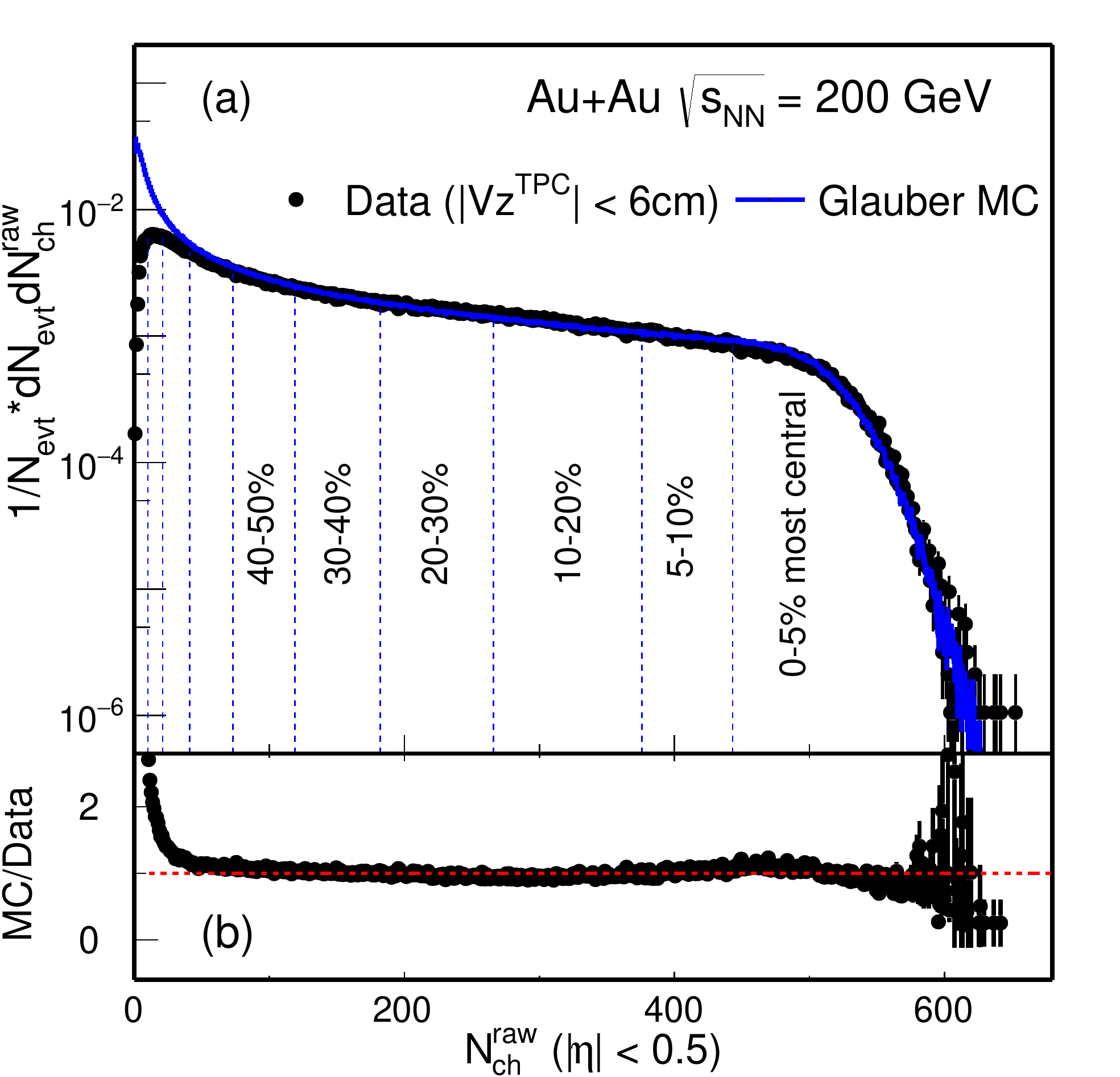}
  \caption{(a) Uncorrected charged particle multiplicity $N_{\rm ch}^{\rm raw}$ distribution measured within $|\eta|$\,$<$\,0.5 and $|V_z^{\rm{TPC}}|$\,$<$\,6\,cm. The solid curve depicts the multiplicity distribution from a MC Glauber simulation fit to the experimental data. (b) Ratio between MC simulation and real data.}
\label{fig:centrality} 
\end{figure}

The centrality is selected using the measured charged global track multiplicity $N_{\rm ch}^{\rm raw}$ at mid-rapidity within $|\eta|\ \rm{ < 0.5}$ and corrected for the online VPD triggering inefficiency using a Monte Carlo (MC) Glauber simulation. 0--X\% centrality is defined as the 0--X\% most central in terms of the total hadronic cross section determined by the impact parameter between two colliding nuclei. In this analysis, the dependence of $N_{\rm ch}^{\rm raw}$ on the collision vertex position and the beam luminosity has been taken into account. The measured track multiplicity distribution from Au+Au 200\,GeV from RHIC run 2014, corrected for the vertex and luminosity dependence, is shown in Fig.~\ref{fig:centrality}. The measured distribution is fit by the MC Glauber calculation in the high multiplicity region. One can observe that the fitted MC Glauber calculation matches the real data well for $N_{\rm ch}^{\rm raw}$\,$>$\,100, while the discrepancy in the low multiplicity region shows the VPD trigger inefficiency. Figure~\ref{fig:centrality} panel (b) shows the ratio between MC and data. Centrality is defined according to the MC Glauber model distribution shown in panel (a). Events in the low-multiplicity region are weighted with the ratio shown in panel (b) in all the following analysis as a correction for the inefficiency in trigger. 

\begin{figure*}
\centering
\includegraphics[width=0.78\textwidth]{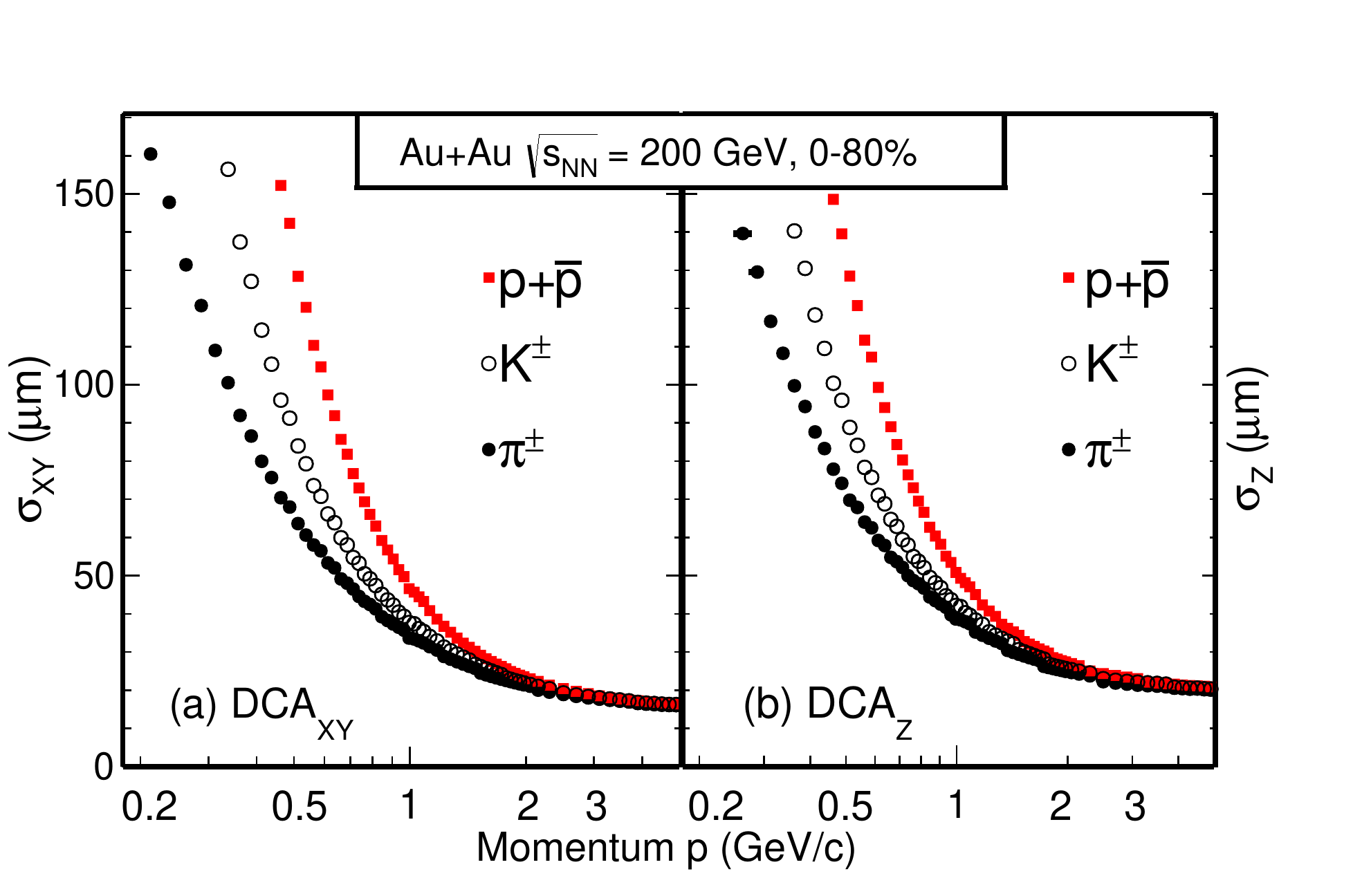}
\caption{Identified particle ($\pi^{\pm}$, $K^{\pm}$, and $p$+$\bar{p}$) pointing resolution in the transverse (a) and longitudinal (b) planes as a function of particle momentum in Au+Au 0--80\% collisions.}
\label{fig:DCAXy_Z} 
\end{figure*}

\begin{figure}[h]
\centering
\includegraphics[width=0.45\textwidth]{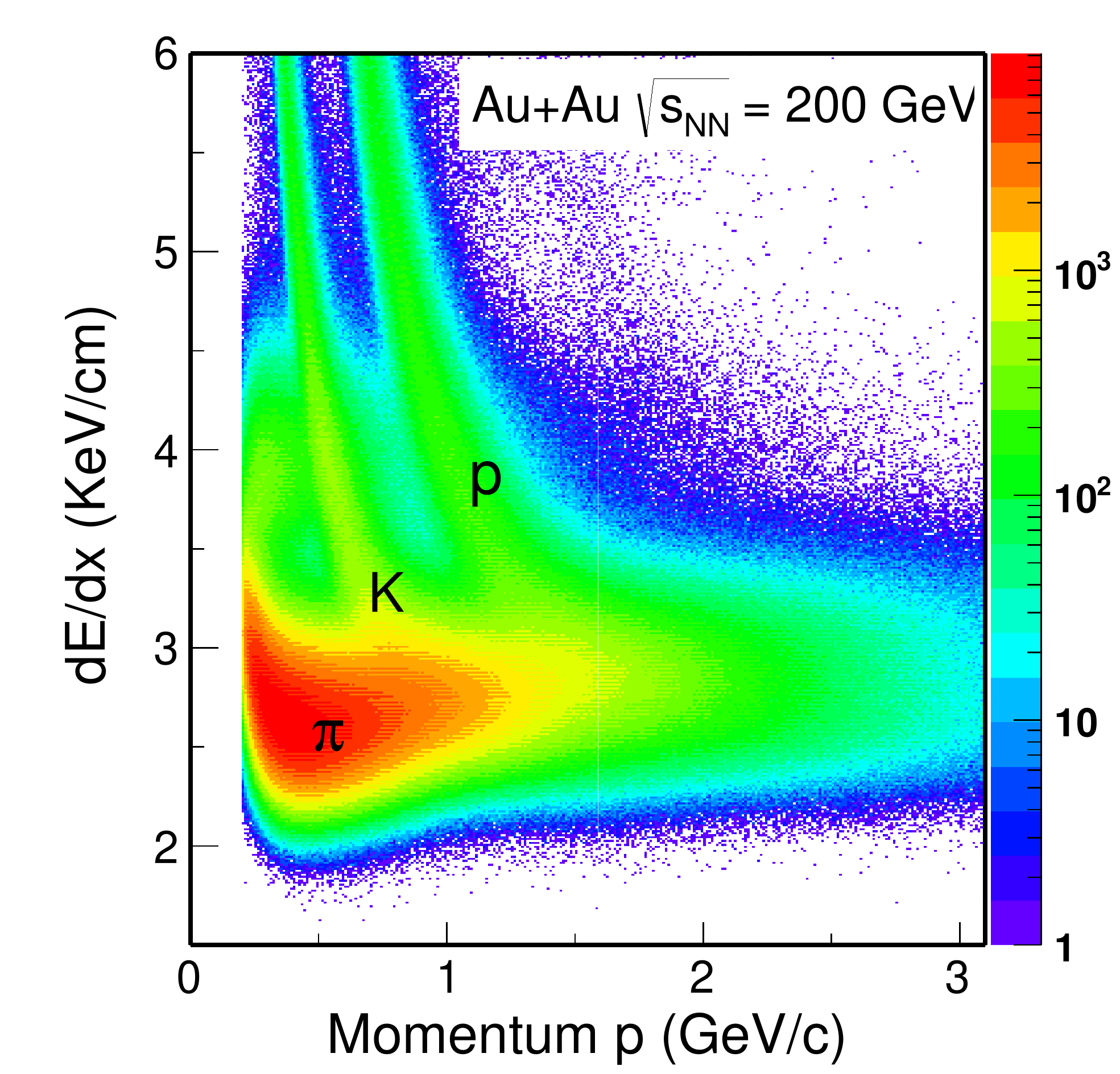}
\caption{TPC $dE/dx$ vs. particle momentum.}
\label{fig:PID_dEdx} 
\end{figure}

\begin{figure}[h]
\centering
\includegraphics[width=0.45\textwidth]{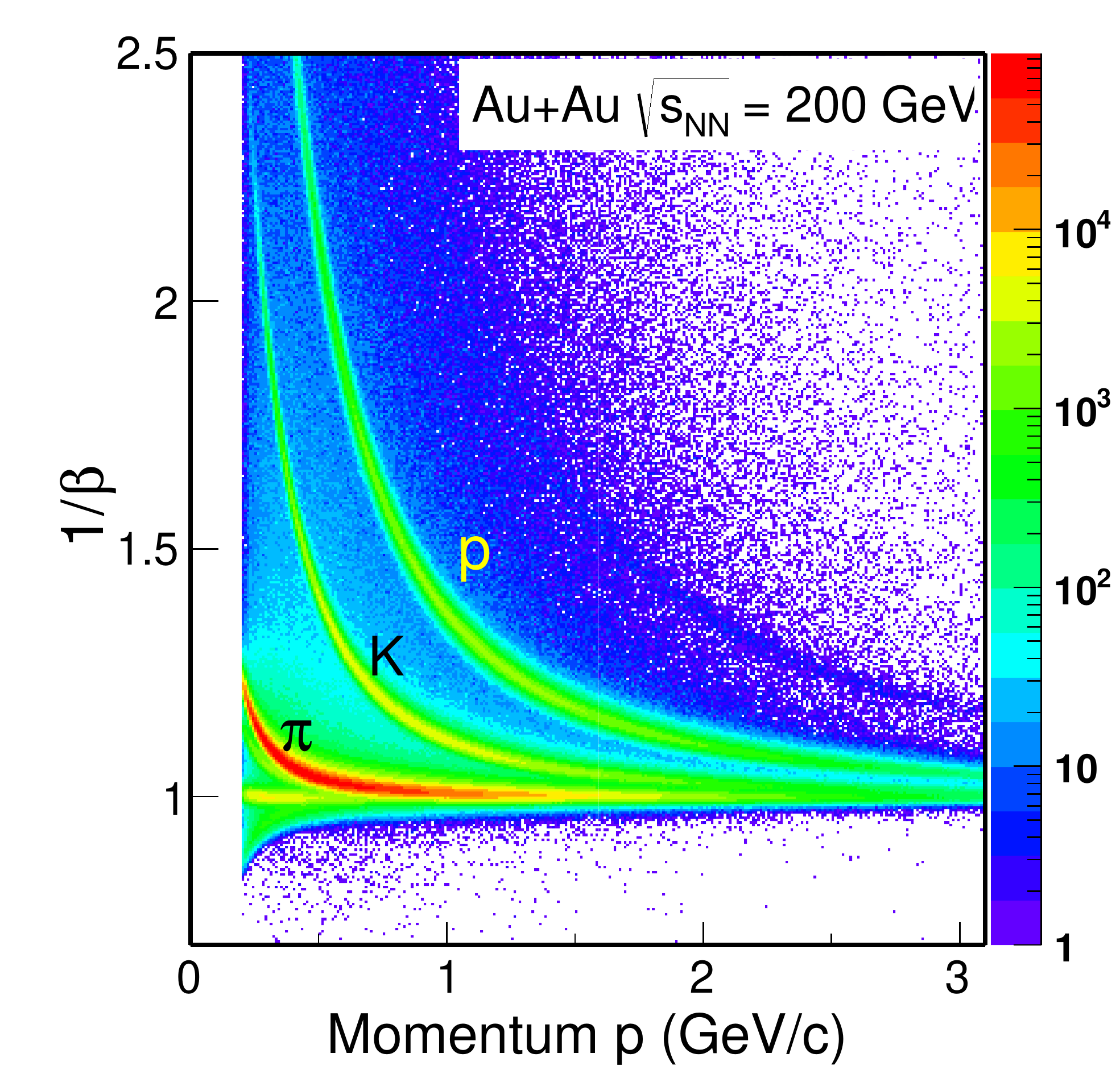}
\caption{TOF $1/\beta$ vs. particle momentum.}
\label{fig:PID_beta} 
\end{figure}

Table~\ref{table:ccentrality} lists the extracted values of average number of binary collisions ($N_{\rm bin}$), average number of participants ($N_{\rm part}$) and trigger inefficiency correction factors ($\varepsilon_{\rm trg}$) and their uncertainties in various centrality bins. 
The $\varepsilon_{\rm trg}$ correction factor is applied event-by-event in the analysis when combining centrality bins.

\begin{table*}[t]
\centering{
\caption{Estimated values of average number of binary collisions ($N_{\rm bin}$), average number of participants ($N_{\rm part}$) and trigger correction factors ($\varepsilon_{\rm trg}$, uncertainties negligible) for various centrality bins obtained from the MC Glauber model fit to the measured multiplicity distributions.}
\begin{tabular}{rcccccccc} \hline \hline
\hspace{1cm}Centrality\hspace{1cm} & \multicolumn{3}{c}{$\langle N_{\rm bin}\rangle$} & \hspace{1cm} & \multicolumn{3}{c}{$\langle N_{\rm part}\rangle$} & \hspace{1cm}$\varepsilon_{\rm trg}$\hspace{1cm} \\ \hline
0--10 \%\hspace{1cm}      & 938.8 & $\pm$ & 26.3 & & 319.4 & $\pm$ & 3.4  & 1.0 \\
10--20 \%\hspace{1cm}     & 579.9 & $\pm$ & 28.8 & & 227.6 & $\pm$ & 7.9  & 1.0 \\
20--40 \%\hspace{1cm}     & 288.3 & $\pm$ & 30.4 & & 137.6 & $\pm$ & 10.4 & 1.0 \\
40--60 \%\hspace{1cm}     & 91.3  & $\pm$ & 21.0 & & 60.5  & $\pm$ & 10.1 & 0.92 \\
60--80 \%\hspace{1cm}     & 21.3  & $\pm$ & 8.9  & & 20.4  & $\pm$ & 6.6  & 0.65 \\ \hline \hline
\end{tabular}
\label{table:ccentrality}
}
\end{table*}

\subsection{Heavy Flavor Tracker}
\label{dataset:hft}

The HFT~\cite{Contin:2017mck} is a high resolution silicon detector system that aims for topological reconstruction of secondary vertices from heavy flavor hadron decays. It consists of three silicon subsystems: the Silicon Strip Detector (SSD), the Intermediate Silicon Tracker (IST), and two layers of the PiXeL (PXL) detector. 
Table~\ref{table:HFT} lists the key characteristic parameters of each subsystem. The SSD detector was still under commissioning when the dataset was recorded, and therefore is not used in the offline data production and this analysis.
The PXL detector uses the new Monolithic Active Pixel Sensors (MAPS) technology~\cite{Contin:2017mck}. This is the first application of this technology in a collider experiment. It is specifically designed to measure heavy flavor hadron decays in the high multiplicity heavy-ion collision environment.

\begin{table*}[t]
\centering{
\caption{Several key characteristic parameters for each subsystem of the STAR HFT detector.}
\begin{tabular}{ccccc} \hline \hline
  \hspace{0.5cm}Subsystem\hspace{0.5cm} & \hspace{0.5cm}Radius (cm)\hspace{0.5cm} & \hspace{0.5cm}Length (cm)\hspace{0.5cm} & \hspace{0.5cm}Thickness at $\eta$\,$=$\,0 ($X_{0}$)\hspace{0.5cm} & \hspace{0.5cm}Pitch Size ($\upmu \textup{m}^2$)\hspace{0.5cm} \\ \hline
PXL inner layer & 2.8 & 20 & 0.52\% (0.39\%$^{\dagger}$) & 20.7$\times$20.7 \\
PXL outer layer & 8.0 & 20 & 0.52\% & 20.7$\times$20.7 \\
IST & 14.0 & 50 & 1.0\% & 600$\times$6000 \\
SSD$^{\dagger\dagger}$ & 22.0 & 106 & 1.0\% & 95$\times$40000 \\ \hline \hline
\end{tabular} \\
$^{\dagger}$ - PXL inner detector material is reduced to 0.39\%$X_0$ in 2015/2016 runs. \\
$^{\dagger\dagger}$ - SSD is not included in this analysis.
\label{table:HFT} 
}
\end{table*}

In the offline reconstruction, tracks are reconstructed in the TPC first and then extended to the HFT detector to find the best fit to the measured high resolution spatial points. A Kalman filter algorithm that considers various detector material effects is used in the track extension~\cite{Kalman}. Considering the level of background hits in the PXL detector due to pileup hadronic and electromagnetic collisions, tracks are required to have at least one hit in each layer of the PXL and IST sub-detectors. Figure~\ref{fig:DCAXy_Z} shows the track pointing resolution to the primary vertex in the transverse plane ($\sigma_{\rm XY}$) in panel (a) and along the longitudinal direction ($\sigma_{\rm Z}$) in panel (b) as a function of total momentum ($p$) for identified particles in 0--80\% centrality Au+Au collisions. The design goal for the HFT detector was to have a pointing resolution better than 55 $\upmu$m for 750\,MeV charged kaon particles. Figure~\ref{fig:DCAXy_Z} demonstrates that the HFT detector system meets the design requirements. This performance enables precision measurement of $D$-meson production in high multiplicity heavy-ion collisions.

\section{$D^0$-meson reconstruction}
\label{D0recon}

$D^0$ and $\overline{D}^{0}$ mesons are reconstructed via the hadronic decay channel $D^0\rightarrow K^-+\pi^+$ and its charge conjugate channel with a branching ratio ($B.R.$) of 3.89\%. In what follows, we imply $(D^0 +\overline{D}^{0})/2$ when using the term $D^0$ unless otherwise specified. $D^0$ mesons decay with a proper decay length of $c\tau\sim123\ \upmu$m after they are produced in Au+Au collisions. We utilize the high-pointing resolution capability enabled by the HFT detector to topologically reconstruct the $D^0$ decay vertices that are separated from the collision vertices, which drastically reduces the combinatorial background ($\sim$five orders of magnitude) and improves the measurement precision.

Charged pion and kaon tracks are reconstructed with the TPC and HFT. Tracks are required to have at least 20 measured TPC points out of maximum 45 to ensure a good momentum resolution. To enable high pointing precision, both daughter tracks are required to have at least one measured hit in each layer of the PXL and IST as described above. Particle identification is achieved via a combination of the ionization energy loss measurement in the TPC and the time-of-flight measurement in the TOF. The resolution-normalized $dE/dx$ deviation from the expected values is defined as:
\begin{equation}
  n\sigma_X = \frac{1}{R}\ln\frac{\langle{dE/dx}\rangle_{\rm {mea.}}}{\langle{dE/dx}\rangle_{X}},
\label{equ:equation1}
\end{equation}
where $\langle{dE/dx}\rangle_{\rm {mea.}}$ and $\langle{dE/dx}\rangle_{X}$ represent measured and expected values with a hypothesis of particle $X$, and $R$ is the $dE/dx$ resolution (typically $\sim$\,8\%~\cite{TPC}). The $n\sigma_X$ distribution should be close to a standard Gaussian for each corresponding particle species (mean $=$ 0, $\sigma = $ 1) with good $dE/dx$ calibration.
Pion (kaon) candidates are selected by a requirement of the measured $dE/dx$ to be within three (two) standard deviations ($|n\sigma_{X}|$) from the expected value. When tracks have matched hits in the TOF detector, an additional requirement on the measured inverse particle velocity ($1/\beta$) to be within three standard deviations from the expected value ($|\Delta 1/\beta|$) is applied for either daughter track. Figures~\ref{fig:PID_dEdx} and ~\ref{fig:PID_beta} show examples of the particle identification capability from the TPC and TOF. Tracks within the kinematic acceptance $p_{T}$\,$>$\,0.6\,GeV/$c$ and $|\eta|$\,$<$\,1 are used to combine and make pairs. The choice of the $p_T$\,$>$\,0.6\,GeV/$c$ cut is an optimized consideration to balance the loss of signal acceptance when tightening the cut, and the increase in background due to the HFT fake matches when loosening this cut (see Sec.~\ref{correction:hft}). The threshold has been varied for systematic uncertainty evaluation. See Sec.~\ref{systematic} for details. Table~\ref{table:singlecut} lists the TPC and TOF selection cuts for daughter kaon and pion tracks used for $D^0$ reconstruction.

\begin{table}
\centering{
\caption{TPC and TOF selection cuts for $K$ and $\pi$ tracks.}
\begin{tabular}{cccc} \hline \hline
Variable &\hspace{0.8cm} & $K^{\mp}$  & $\pi^{\pm}$ \\ \hline
$p_{T}$ (GeV/$c$)   & $>$ & 0.6 & 0.6 \\
$|\eta|$			    & $<$ & 1.0 & 1.0 \\
nHitsFit (TPC)		    & $>$ & 20  & 20 \\
$|n\sigma_{X}|$         & $<$ & 2.0 & 3.0 \\
$|\Delta 1/\beta|$ (if TOF matched)     & $<$ & 0.03 & 0.03 \\ \hline \hline
\end{tabular}
}
\label{table:singlecut} 
\end{table}

\begin{figure*}
\centering
\includegraphics[width=0.50\textwidth]{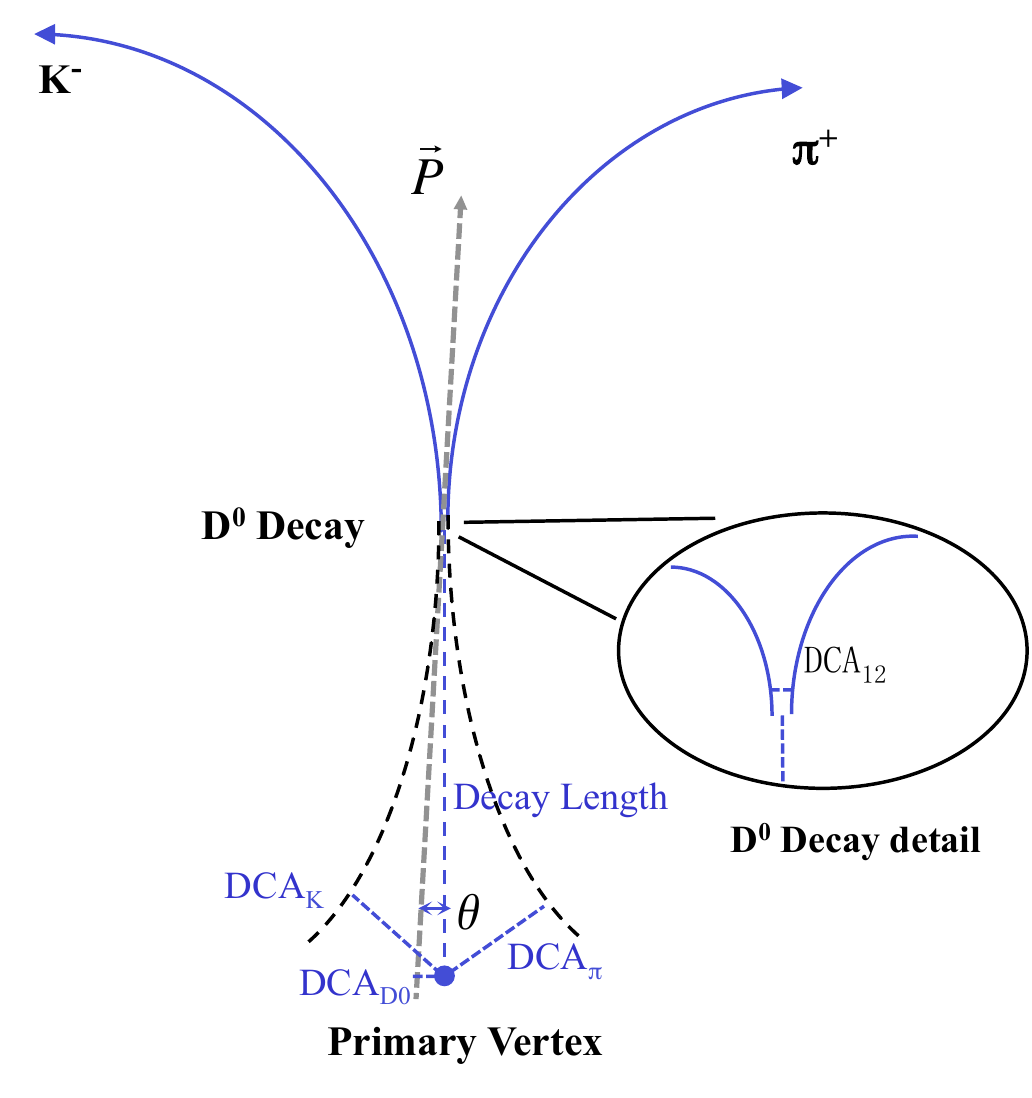}
\caption{A cartoon picture for $D^0\rightarrow K^-+\pi^+$ decay and definition of topological variables used in the reconstruction.}
\label{fig:D0carton} 
\end{figure*}

With a pair of two daughter tracks, pion and kaon, the $D^0$ decay vertex is reconstructed as the middle point on the distance of the closest approach between the two daughter trajectories. One of the dominant background sources is the random combination of $K\pi$ pairs directly from the collision point. With the selection of the following topological variables, the background level can be greatly reduced.

\begin{itemize}
  \item Decay Length: the distance between the reconstructed decay vertex and the Primary Vertex (PV).
  \item Distance of Closest Approach (DCA) between the two daughter tracks ($\rm{DCA_{12}}$).
  \item DCA between the reconstructed $D^0$ and the PV ($\rm{DCA_{D^{0}}}$).
  \item DCA between the pion and the PV ($\rm{DCA_{\pi}}$).
  \item DCA between the kaon and the PV ($\rm{DCA_{K}}$).
  \item Angle between the $D^0$ momentum and the direction of the decay vertex with respect to the PV ($\theta$).
\end{itemize}

The schematic in Fig.~\ref{fig:D0carton} shows the topological variables used in the analysis, where $\vec{P}$ represents the $D^0$ momentum. The Decay Length and angle $\theta$ follow the formula: $\rm DCA_{D^{0}}$ = Decay Length $\times$ sin($\theta$). The cuts on the topological variables for this analysis are optimized using a Toolkit for Multivariate Data Analysis (TMVA) package integrated in the ROOT framework in order to obtain the greatest signal significance~\cite{TMVA}. The Rectangular Cut optimization method from the TMVA package is chosen in this analysis, similar as in our previous publication~\cite{Star_D_v2}. The optimization is conducted for different $D^0$ $p_{T}$ bins and different centrality bins. Table~\ref{table:topocut} lists a set of topological cuts for 0--10\% central Au+Au collisions.

\begin{table*}[t]
\centering {
\caption{Topological cuts used for $D^0$ reconstruction in 0--10\% most central collisions for separate $p_T$ intervals.}
\begin{tabular}{cc|ccccccc} \hline \hline
$0-10\% \ |\ p_{T}$ (GeV/$c$) &    & \hspace{0.5cm}(0,0.5)\hspace{0.5cm} & \hspace{0.5cm}(0.5,1)\hspace{0.5cm} & \hspace{0.5cm}(1,2)\hspace{0.5cm} & \hspace{0.5cm}(2,3)\hspace{0.5cm} & \hspace{0.5cm}(3,5)\hspace{0.5cm} &  \hspace{0.5cm}(5,8)\hspace{0.5cm} &  \hspace{0.5cm}(8,10)\hspace{0.5cm} \\ \hline
  Decay Length ($\upmu$m) & $>$ & 100 & 199 & 227 & 232 & 236 & 255 & 255 \\
  $\rm DCA_{12}$        ($\upmu$m) & $<$ & 71  & 64 & 70 & 63 & 82 & 80 & 80 \\
  $\rm DCA_{D^{0}}$       ($\upmu$m) & $<$ & 62  & 55 & 40 & 40 & 40 & 44 & 44 \\
  $\rm DCA_{\pi}$  ($\upmu$m) & $>$ & 133 & 105 & 93 & 97 & 67 & 55 & 55 \\
  $\rm DCA_{K}$    ($\upmu$m) & $>$ & 138 & 109 & 82 & 94 & 76 & 54 & 54 \\ 
  $\cos(\theta)$          & $>$ & 0.95 & 0.95 & 0.95 & 0.95 & 0.95 & 0.95 & 0.95 \\ \hline \hline
\end{tabular}
\label{table:topocut}
}
\end{table*}

\begin{figure*}
\centering
\includegraphics[width=1.0\textwidth]{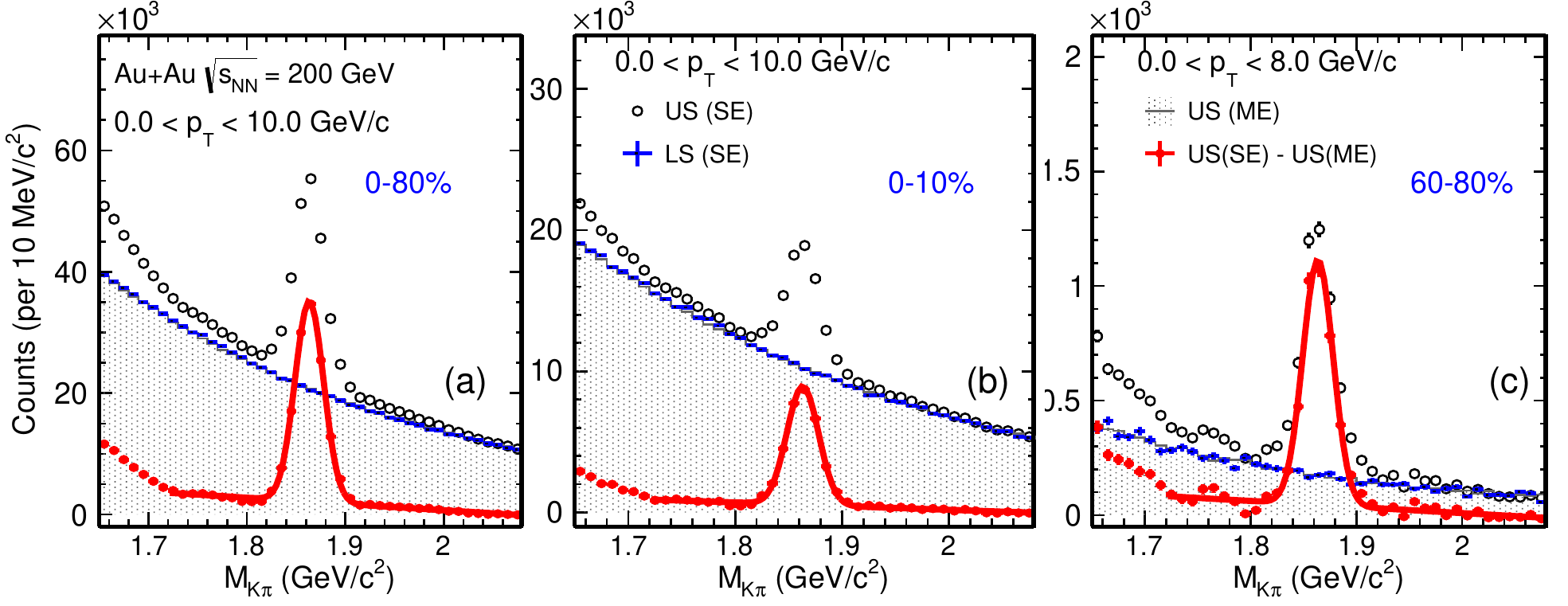}
\caption{Invariant mass $M_{K\pi}$ distributions in 0\,$<$\,$p_{T}$\,$<$\,10\,GeV/$c$ from centrality bins 0--80\% (a), 0--10\% (b), and in 0\,$<$\,$p_{T}$\,$<$\,8\,GeV/$c$ for 60--80\% (c), respectively. Black open circles represent the same-event (SE) unlike-sign (US) distributions. Blue and grey shaded histograms represent the SE like-sign (LS) and mixed-event (ME) US distributions that are used to estimate the combinatorial background. The red solid circles depict the US (SE) distributions with the combinatorial background subtracted using the US (ME) distributions.
}
\label{fig:signal_0} 
\end{figure*}

\begin{figure*}
\centering
\includegraphics[width=1.0\textwidth]{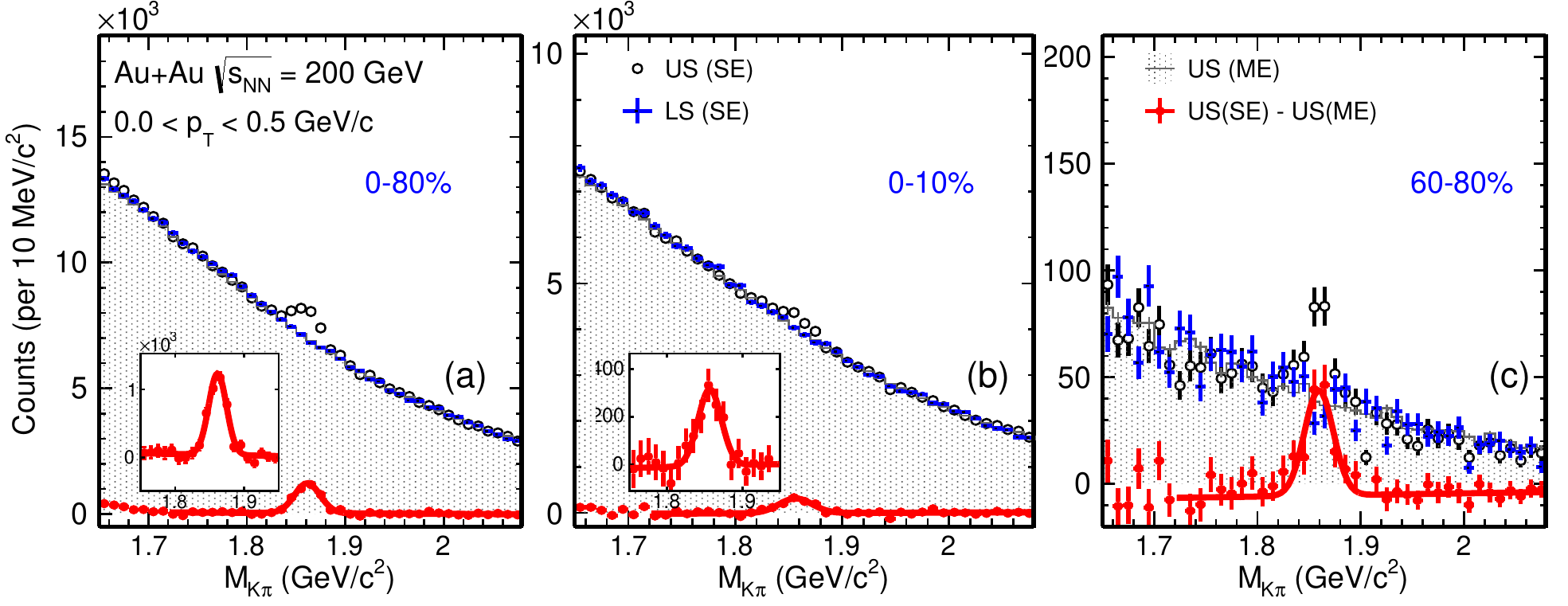}
\caption{Invariant mass $M_{K\pi}$ distributions in 0\,$<$\,$p_{T}$\,$<$\,0.5\,GeV/$c$ from centrality bins 0--80\% (a), 0--10\% (b) and 60--80\% (c), respectively. All histograms and markers use the same notation as in Fig.~\ref{fig:signal_0}.}
\label{fig:signal_1} 
\end{figure*}

\begin{figure*}
\centering
\includegraphics[width=1.0\textwidth]{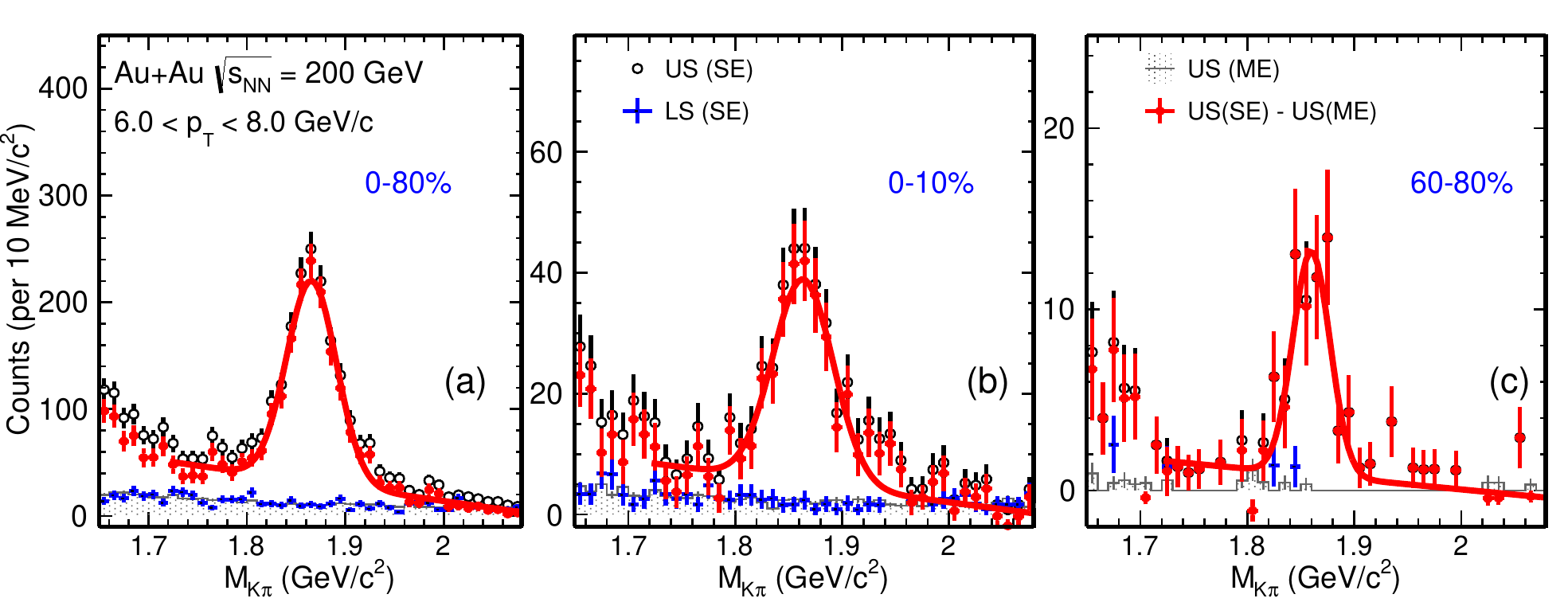}
\caption{Invariant mass $M_{K\pi}$ distributions in 6\,$<$\,$p_{T}$\,$<$\,8\,GeV/$c$ from centrality bins 0--80\% (a), 0--10\% (b) and 60--80\% (c), respectively. All histograms and markers use the same notation as in Fig.~\ref{fig:signal_0}.}
\label{fig:signal_2} 
\end{figure*}

Figure~\ref{fig:signal_0} shows the invariant mass distributions of $K\pi$ pairs in the $p_{T}$ region of 0--10 GeV/$c$ for 0--80\% minimum bias and the 0--10\% most central collisions, and 0--8 GeV/$c$ for 60--80\% peripheral collisions, respectively. The reason of choosing a different $p_T$ range for the 60--80\% centrality bin is because no signal is observed beyond the current statistics. The combinatorial background is estimated with the same-event (SE) like-sign (LS) pairs (blue histograms) and the mixed-event (ME) unlike-sign (US) (grey histograms) technique in which $K$ and $\pi$ from different events of similar characteristics ($V_{Z}$, centrality, event plane angle) are paired. The mixed-event spectra are normalized to the like-sign distributions in the mass range of 1.7--2.1\,GeV/$c^2$. After the subtraction of the mixed-event unlike-sign combinatorial background from the same-event unlike-sign pairs (black open circles), the remainder distributions are shown as red solid circles in each panel. Compared to the previous $D^0$ measurement~\cite{Star_D_RAA}, the $D^0$ signal significance is largely improved by a factor of $\sim$\,15 using the same amount of event statistics. 

Figures~\ref{fig:signal_1} and ~\ref{fig:signal_2} show the invariant mass distributions in the same centrality bins as Fig.~\ref{fig:signal_0} but for different $p_T$ ranges: 0\,$<$\,$p_{T}$\,$<$\,0.5\,GeV/$c$ in Fig.~\ref{fig:signal_1} and 6\,$<$\,$p_{T}$\,$<$\,8\,GeV/$c$ in Fig.~\ref{fig:signal_2}.

After the combinatorial background is subtracted, the residual $K\pi$ invariant mass distributions are then fit to a Gaussian plus linear function. The linear function is used to represent remaining correlated background from either partial reconstruction of charm mesons or jet fragments.
The $D^0$ raw yields are extracted from the Gaussian function fit results while different choices of fit ranges, background functional forms, histogram counting vs. fitting methods etc. have been used to estimate systematic uncertainties on the raw yield extraction. See Sec.~\ref{systematic} for details.

\begin{figure}
\centering
\includegraphics[width=0.43\textwidth]{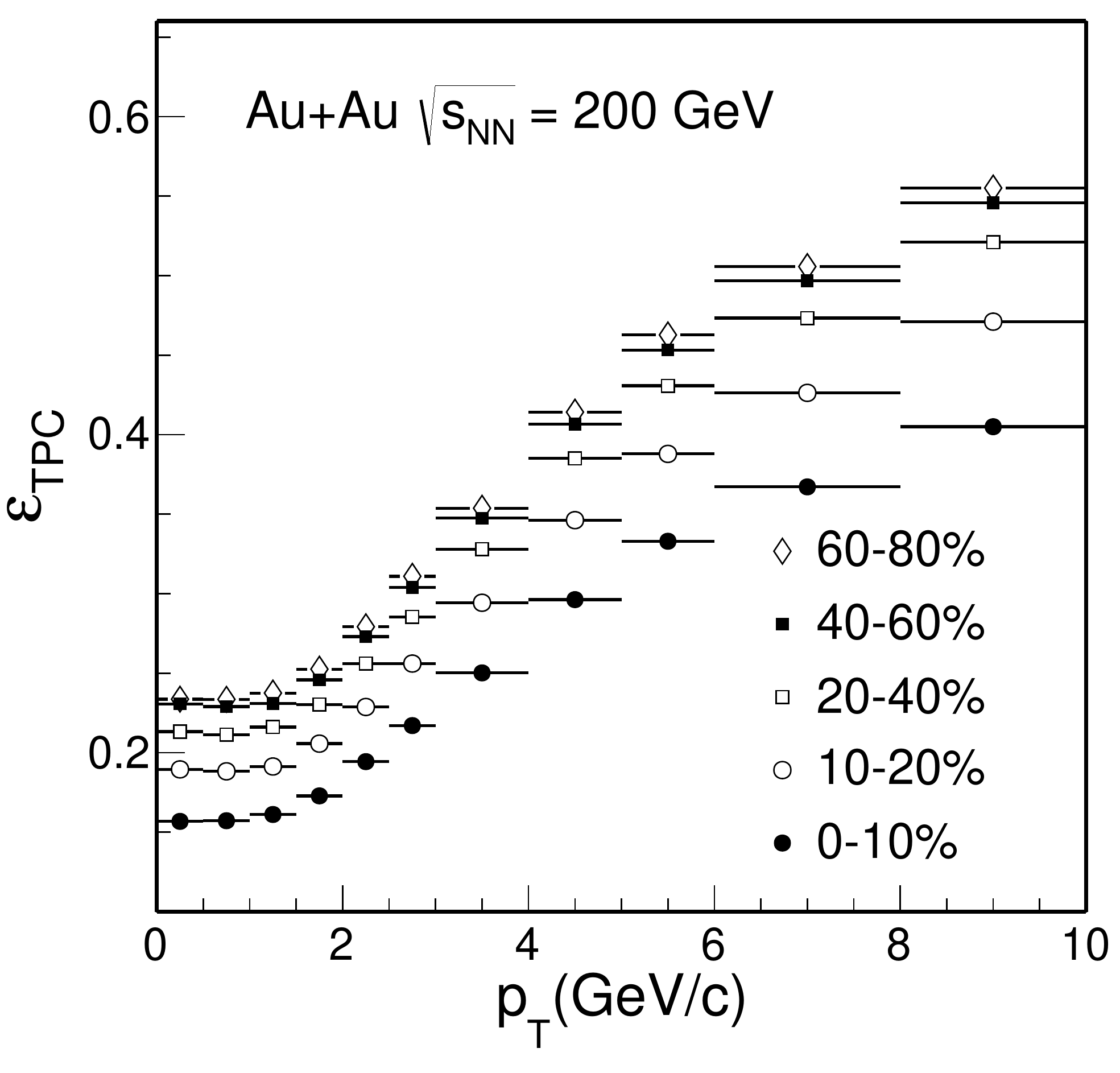}
  \caption{$D^{0}$ TPC acceptance and tracking efficiencies from different centrality classes in Au+Au collisions at $\sqrt{s_{_{\rm NN}}}$ = 200\,GeV.}
\label{fig:Datad0Eff_tpc} 
\end{figure}

\begin{figure*}
\centering
\includegraphics[width=0.84\textwidth]{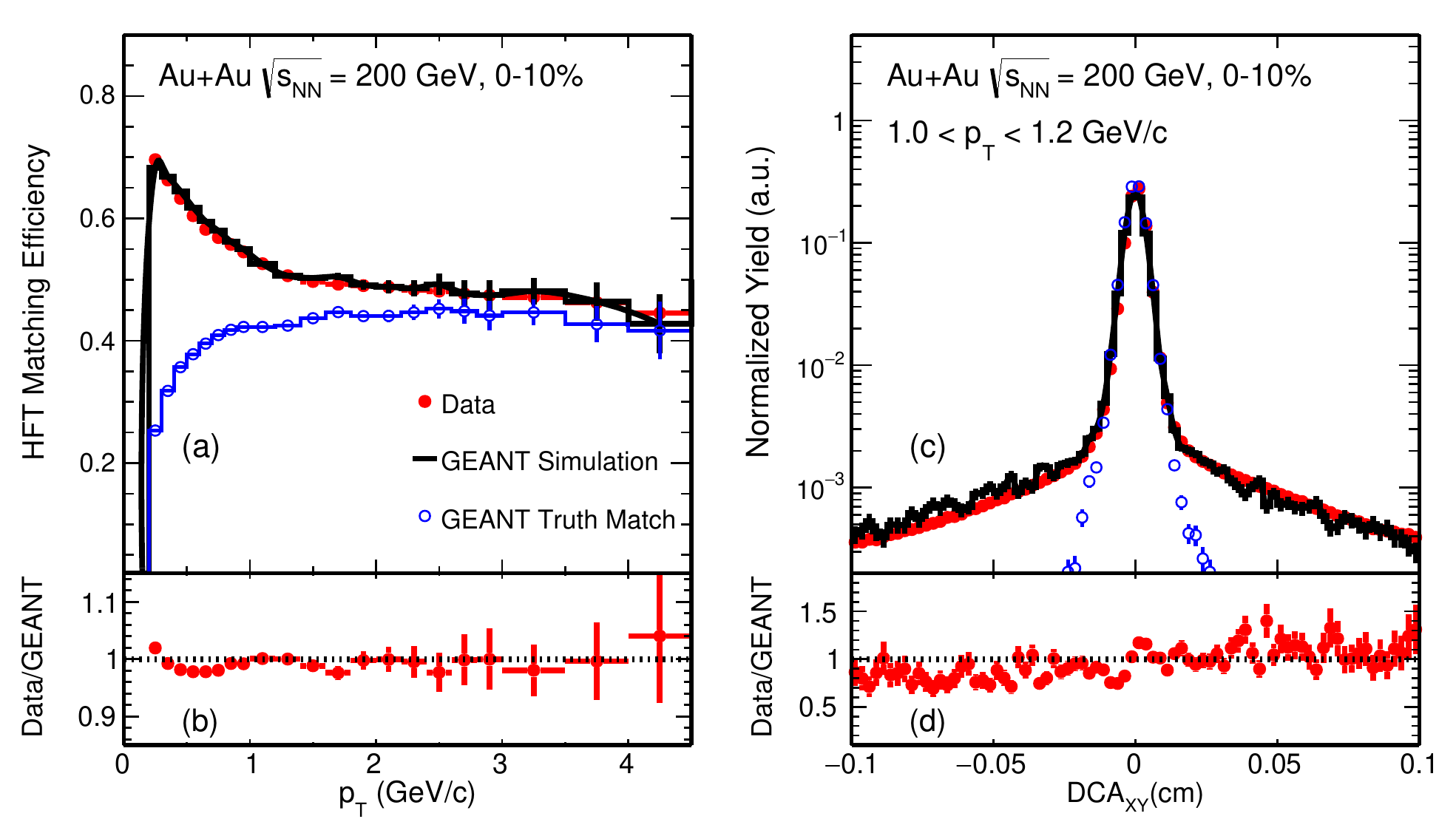}
\caption{HFT matching efficiency $\varepsilon_{\rm HFT}^{\rm match}$ (a) and DCA$_{\rm XY}$ (c) distributions of inclusive charged pions from real data and MC simulation in 0--10\% Au+Au collisions. The ratios between real data and GEANT simulation are shown in the bottom panels. The blue histogram depicts the true matches for which the reconstructed tracks pick up the correct MC hits in the HFT detector induced by the associated MC tracks in the GEANT simulation.}
\label{fig:HijingRatioDca} 
\end{figure*}

\section{Efficiencies and Corrections}
\label{correction}

The reconstructed $D^0$ raw yields are calculated in each centrality, $p_{T}$ bin, and within the rapidity window $|y|$\,$<$\,1. The fully corrected $D^0$ production invariant yields are calculated using the following formula:
\begin{equation}
  \begin{aligned}
& \frac{d^2N}{2\pi p_{T}dp_{T}dy} = \frac{1}{\rm B.R.} \times \frac{N^{\rm raw}}{N_{\rm evt} 2\pi p_{T}\Delta p_{T}\Delta y} \\
& \times \frac{1}{\varepsilon_{\rm trg}\times\varepsilon_{\rm TPC}\times\varepsilon_{\rm HFT}\times\varepsilon_{\rm PID}\times\varepsilon_{\rm vtx}},
  \end{aligned}
\label{equ:invariantyield}
\end{equation}
where B.R. is the $D^0\rightarrow K^-\pi^+$ decay branching ratio, (3.89$\pm$0.04)\%~\cite{pdg}, $N^{\rm raw}$ is the reconstructed $D^0$ raw counts, $N_{\rm evt}$ is the total numbers of events used in this analysis, $\varepsilon_{\rm trg}$ is the centrality bias correction factor described in Sec.~\ref{dataset:trigger}. The raw yields need to be corrected for the TPC acceptance and tracking efficiency - $\varepsilon_{\rm TPC}$, the HFT acceptance and tracking plus topological cut efficiency - $\varepsilon_{\rm HFT}$, the particle identification efficiency - $\varepsilon_{\rm PID}$, and the finite vertex resolution correction - $\varepsilon_{\rm vtx}$.

\subsection{TPC Acceptance and Tracking Efficiency - $\varepsilon_{\rm TPC}$}
\label{correction:tpc}

The TPC acceptance and tracking efficiency is obtained using the standard STAR TPC embedding technique, in which a small amount of MC tracks (typically 5\% of the total multiplicity of the real event) are processed through the full GEANT simulation~\cite{GEANT3}, then mixed with the raw Data Acquisition (DAQ) data in real events and reconstructed through the same reconstruction chain as the real data production. The TPC efficiency is then calculated as the ratio of the number of reconstructed MC tracks with the same offline analysis cuts for geometric acceptance and other TPC requirements to that of the input MC tracks.

Figure~\ref{fig:Datad0Eff_tpc} shows the TPC acceptance and tracking efficiency $\varepsilon_{\rm TPC}$ for $D^0$ mesons within $|y|$\,$<$\,1 in various centrality classes in this analysis. The efficiencies include the TPC and analysis acceptance cuts $p_{T}$\,$>$\,0.6\,GeV/$c$ and $|\eta|$\,$<$\,1 as well as the TPC tracking efficiency for both pion and kaon daughters. The lower efficiency observed in central collisions is due to the increased multiplicity resulting higher detector occupancy which leads to reduced tracking efficiency in these collisions.

\begin{figure*}
\centering
\includegraphics[width=0.90\textwidth]{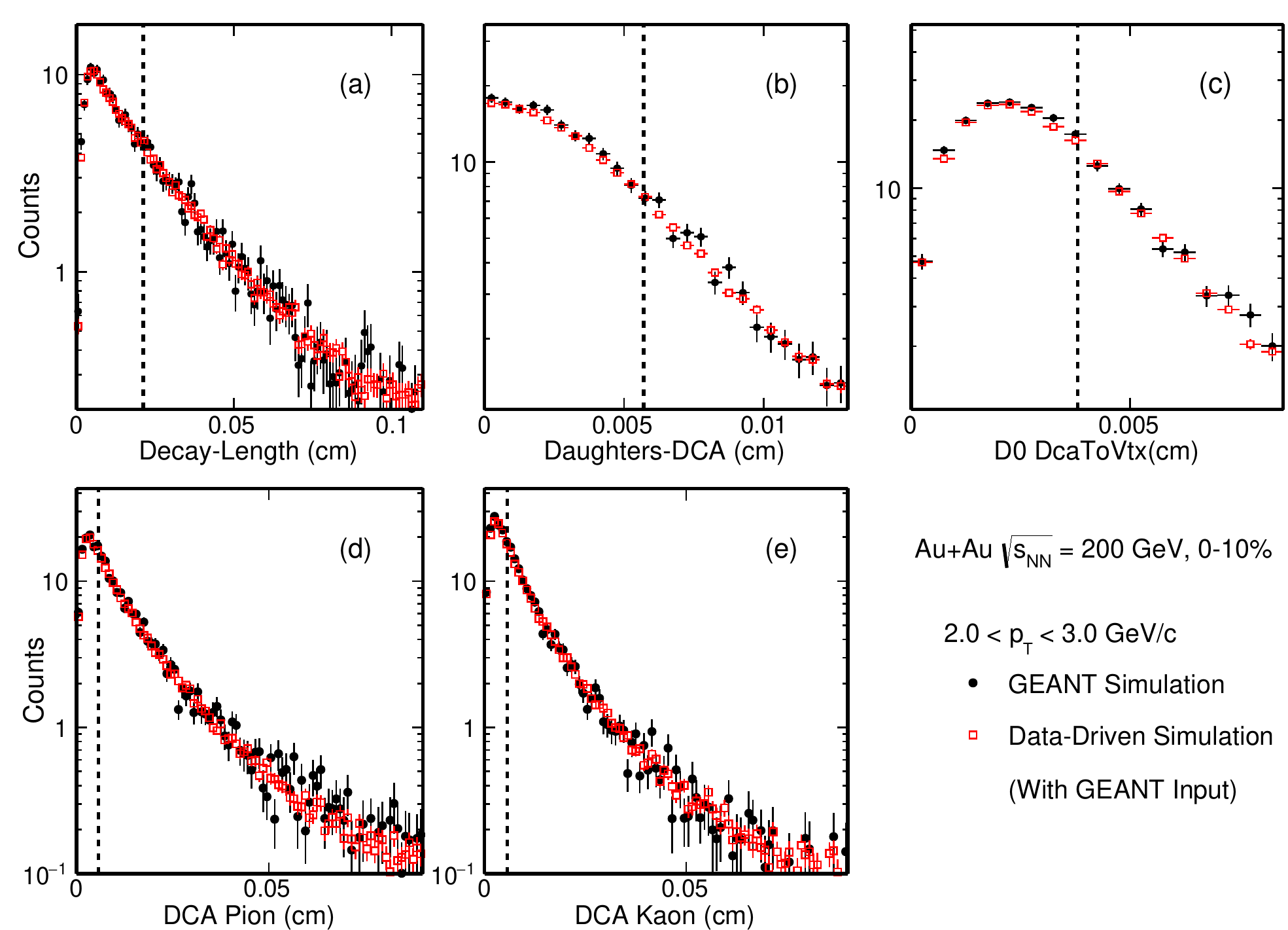}
\caption{Comparisons in topological variable distributions between MC GEANT simulation $(black)$ and data-driven fast simulation with reconstructed MC data as the input $(red)$ in 0--10\% Au+Au collisions for $D^0$ mesons at 2$<$\,$p_T$\,$<$\,3\,GeV/$c$.}
\label{fig:McTopo} 
\end{figure*}

\begin{figure}
\centering
\includegraphics[width=0.42\textwidth]{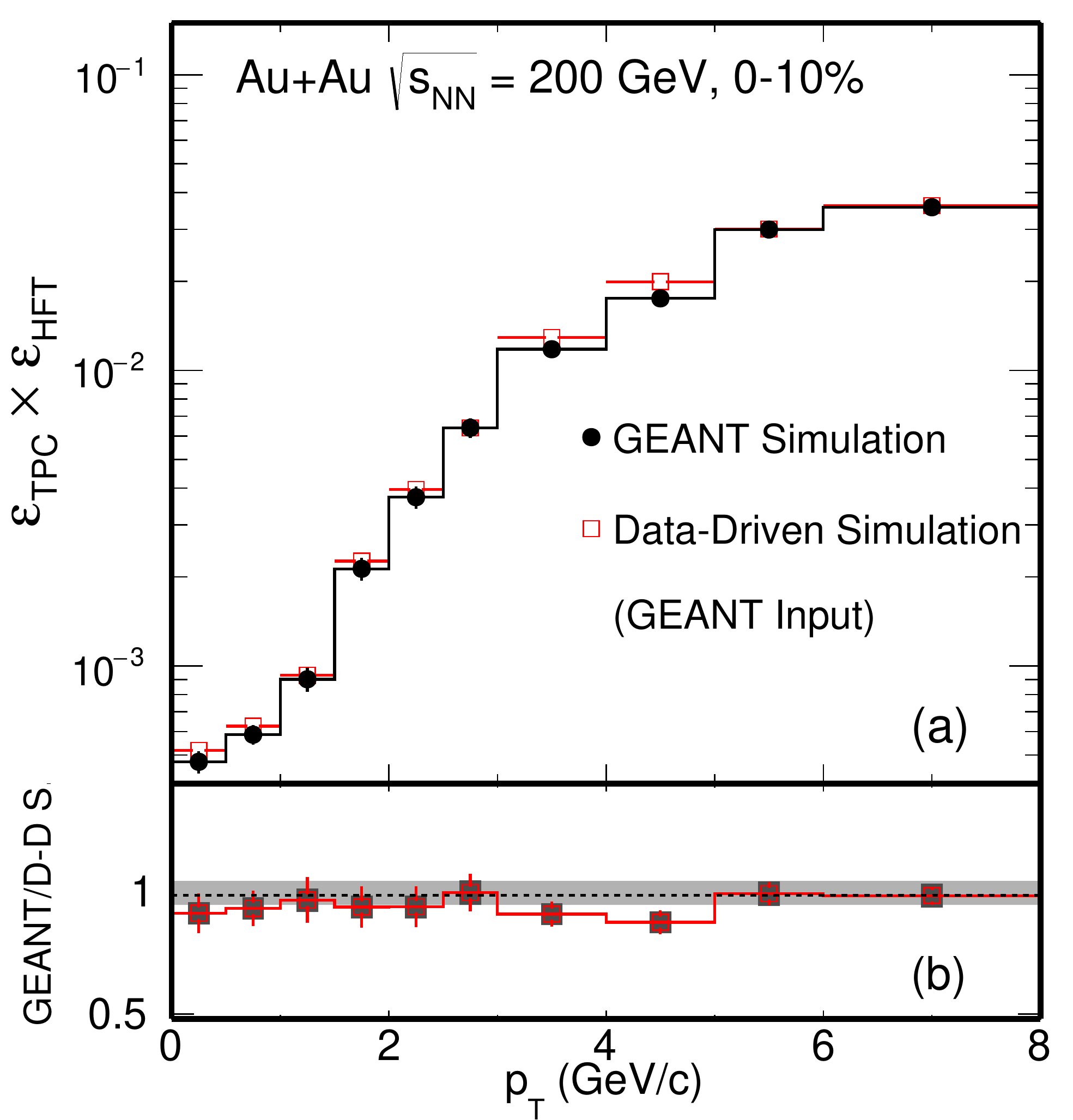}
  \caption{(a) $D^{0}$ reconstruction efficiency comparison between MC GEANT simulation $(black)$ and data-driven fast simulation with reconstructed MC data as the input $(red)$ in central 0--10\% Au+Au collisions. (b) The ratio between the two methods. The grey band around unity represents the 5\% systematic uncertainties.}
\label{fig:Mcd0Eff_0_10} 
\end{figure}

\subsection{HFT Acceptance, Tracking and Topological Cut Efficiency - $\varepsilon_{\rm HFT}$}
\label{correction:hft}

\subsubsection{Data-driven Simulation} 
\label{correction:hft:fastsim}

In order to fully capture the real-time detector performance, the HFT-related efficiency is obtained using a data-driven simulation method in this analysis. The performance of inclusive HFT tracks is characterized by a TPC-to-HFT matching efficiency ($\varepsilon_{\rm HFT}^{\rm match}$) and the DCA distributions with respect to the primary vertex. The HFT matching efficiency $\varepsilon_{\rm HFT}^{\rm match}$ is defined as the fraction of reconstructed TPC tracks that satisfy the requirement on the number of HFT hits. In this analysis, the requirement is to have at least one hit in each PXL and IST layer. The $\varepsilon_{\rm HFT}^{\rm match}$ includes the HFT geometric acceptance and the tracking efficiency that associate HFT hits to the extended TPC tracks. It contains the true matches for which the reconstructed tracks pick up real hits induced by these charged tracks when passing through the HFT, as well as some random fake matches. The latter has a decreasing trend as a function of $p_T$ as the track pointing resolution gets better at high $p_T$, resulting in a smaller search window when associating HFT hits in the tracking algorithm. The DCA distributions are obtained for those tracks that satisfy the HFT hit requirement. Figure~\ref{fig:HijingRatioDca} shows an example of the HFT matching efficiency and the 1-D projection of the $\textup{DCA}_{\textup{XY}}$ distribution for single pions at 1.0$<$\,$p_{T}$\,$<$\,1.2\,GeV/$c$ and 0--10\% central collisions. Such distributions obtained from real data are fed into a MC decay generator for $D^0\rightarrow K^-\pi^+$, followed by the same reconstruction of $D^0$ secondary vertex as in the real data analysis. The same topological cuts are then applied and the HFT related efficiency for the $D^0$ reconstruction is calculated.

To best represent the real detector performance, we obtain the following distributions from real data in this Monte Carlo approach:
\begin{itemize}
\item Centrality-dependent $\textup{V}_\textup{z}$ distributions.
\item HFT matching efficiency $\varepsilon_{\rm HFT}^{\rm match}$, including the dependence on particle species, centrality, $p_T$, $\eta$, $\phi$, and $\textup{V}_\textup{z}$.
\item $\textup{DCA}_{\textup{XY}}$--$\textup{DCA}_{\textup{Z}}$ 2-dimensional (2D) distributions including the dependence on particle species, centrality, $p_T$, $\eta$, and $\textup{V}_\textup{z}$.
\end{itemize}
The $\textup{DCA}_{\textup{XY}}$--$\textup{DCA}_{\textup{Z}}$ 2D distributions are the key to represent not only the true matches, but also the fake matches when connecting the TPC tracks with HFT hits. The distributions are obtained in 2D to consider the correlation between the two quantities and this is necessary and essential to reproduce the 3D DCA position distributions observed in real data. The $\phi$ dependence of these distributions are integrated over due to computing resource limits. We have checked the $\phi$ dependence (by reducing other dependencies for the same reason) and it gives a consistent result compared to the $\phi$-integrated one.

In total, there are 11 ($\phi$) $\times$ 10 ($\eta$) $\times$ 6 ($\textup{V}_\textup{z}$) $\times$ 9 (centrality) $\times$ 2 (particles) 1D histograms (36 $p_T$ bins) used for the HFT matching efficiency distributions and 5 ($\eta$) $\times$ 4 ($\textup{V}_\textup{z}$) $\times$ 9 (centrality) $\times$ 2 (particles) $\times$ 19 ($p_T$) 2D histograms (144 $\textup{DCA}_{\textup{XY}}$ $\times$ 144 $\textup{DCA}_{\textup{Z}}$ bins) for 2D DCA distributions. The number of bins chosen is optimized to balance the need of computing resources as well as the stability of the final efficiency. All dimensions have been checked so that further increase in the number of bins (in balance we need to reduce the number of bins in other dimensions)  will not change the final obtained efficiency.

The procedure for this data-driven simulation package for efficiency calculation is as follows:

\begin{itemize}
\item Sample $\textup{V}_\textup{z}$ distribution according to the distribution obtained from the real data.
\item Generate $D^0$ at the event vertex position with desired $p_T$ (Levy function shape fitted to $D^0$ spectra~\cite{Star_D_RAA}) and rapidity (flat) distributions.
\item Propagate $D^0$ and simulate its decay to $K^-\pi^+$ daughters following the decay probability.
\item Smear daughter track momentum according to the values obtained from embedding.
\item Smear daughter track starting position according to the $\textup{DCA}_{\textup{XY}}$--$\textup{DCA}_\textup{Z}$ 2D distributions from the reconstructed data.
\item Apply HFT matching efficiency according to that extracted from the reconstructed data.
\item Perform the topological reconstruction of $D^0$ decay vertices with the same cuts as applied in the data analysis and calculate the reconstruction efficiency.
\end{itemize}
The DCA and HFT matching efficiency distributions used as the input in this simulation tool can be obtained from the real data or the reconstructed data in MC simulation. The latter is used when we validate this approach using the MC GEANT simulation (see Sec.~\ref{correction:hft:validation}). 

This approach assumes these distributions obtained from real data are good representations for tracks produced at or close to the primary vertices. The impact of the secondary particle contribution will be discussed in Sec.~\ref{correction:hft:secondary}. The approach also neglects the finite event vertex resolution contribution which will be discussed in Sec.~\ref{correction:vtx}.

Lastly in this MC approach, we also fold in the TPC efficiency obtained from the MC embedding so the following presented efficiency will be the total efficiency of $\varepsilon_{\rm TPC}\times\varepsilon_{\rm HFT}$.

\subsubsection{Validation with GEANT Simulation}
\label{correction:hft:validation}

In this subsection, we will demonstrate that the data-driven MC approach has been validated with the GEANT simulation plus the offline tracking reconstruction with realistic HFT detector performance to reproduce the real $D^0$ reconstruction efficiency. We should point out that in this validation procedure, what we are after is the efficiency difference between two calculation methods: one from the MC simulation directly, and the other one from the data-driven simulation package using the reconstructed MC simulation data as the input.

The GEANT simulation uses the HIJING~\cite{HIJING} generator as its input with $D^0$ particles embedded to enrich the signal statistics. The full HFT detector materials (both active and inactive) have been included in the GEANT simulation as well as the offline track reconstruction. The pileup hits in the PXL detector due to finite electronic readout time have been added to realistically represent the HFT matching efficiency and DCA distributions. The overall agreement between the GEANT simulation and real data is fairly good, as can be seen in Fig.~\ref{fig:HijingRatioDca}. The small deviations between real data and MC simulation are not considered in the systematic uncertainty estimation since the latter is not used to calculate the absolute efficiency directly, but to validate the data-driven simulation procedure as described below.

The increase in the HFT matching efficiency at low $p_{T}$ range is due to the increased fake matches (in contrast to true HFT matches) and the efficiency stays flat in the high $p_{T}$ range. The matching efficiency includes the tracking efficiency when associating the HFT hits as well as the HFT geometric acceptance. Therefore, the ratio has a strong dependence on the event $\textup{V}_\textup{Z}$ and the track $\eta$. The DCA distributions used in the package are 2-dimentional distributions, as $\textup{DCA}_{\textup{XY}}$ and $\textup{DCA}_\textup{Z}$ are strongly correlated.

With the tuned simulation setup, we use this sample to validate our data-driven simulation approach for $D^0$ efficiency calculation. We follow the same procedure as described in Sec.~\ref{correction:hft:fastsim} to obtain the HFT matching efficiency as well as the 2D $\textup{DCA}_\textup{XY}$-$\textup{DCA}_\textup{Z}$ distributions for primary particles from the reconstructed data in this simulation sample. Then these distributions are fed into the data-driven simulation framework to calculate the $D^0$ reconstruction efficiency. The calculated $D^0$ efficiency from the data-driven simulation framework will be compared to the real $D^0$ reconstruction efficiency directly obtained from the GEANT simulation sample.

To validate the data-driven simulation tool, Fig.~\ref{fig:McTopo} shows the comparisons of several topological variables used in the $D^0$ reconstruction obtained from the GEANT simulation directly and from the data-driven simulation with reconstructed GEANT simulation data as the input in the most central (0--10\%) centrality and in 2\,$<$\,$p_{T}$\,$<$\,3\,GeV/$c$. The topological variables shown here are $D^0$ decay length, DCA between two $D^0$ decay daughters, $D^0$ DCA with respect to the collision vertex, pion DCA and kaon DCA with respect to the collision vertex. As seen in this figure, the data-driven simulation tool reproduces all of these topological distributions quite well. The agreements for the other $p_{T}$ ranges are also decent.

Figure~\ref{fig:Mcd0Eff_0_10} shows the $D^0$ reconstruction efficiency $\varepsilon_{\rm TPC}$ $\times$ $\varepsilon_{\rm HFT}$ calculated with the following two methods in this GEANT simulation. The first method is the standard calculation by applying the tracking and topological cuts for reconstructed $D^0$ mesons in the simulation sample. In the second method, we employ the data-driven simulation method and take the reconstructed distributions from the simulation sample as the input and then calculate the $D^0$ reconstruction efficiency in the data-driven simulation framework. In panel (a) of Fig.~\ref{fig:Mcd0Eff_0_10}, efficiencies from two calculation methods agree well in the whole $p_{T}$ region in central 0--10\% Au+Au collisions, and the ratio between the two is shown in panel (b). This demonstrates that the data-driven simulation framework can accurately reproduce the real $D^0$ reconstruction efficiency in central Au+Au collisions.

\subsubsection{Efficiency for real data}
\label{correction:hft:fordata}

\begin{figure*}
\centering
\includegraphics[width=0.9\textwidth]{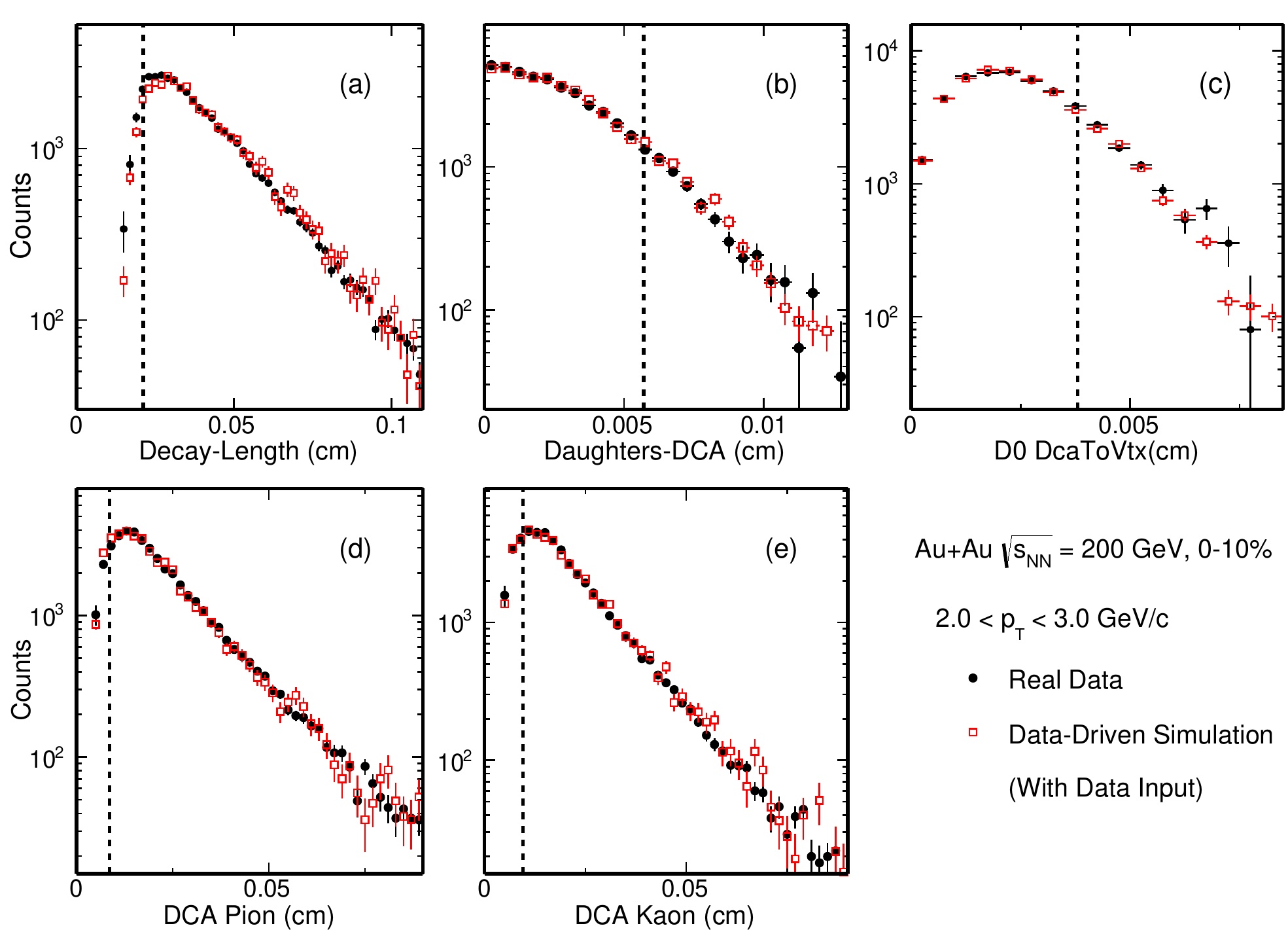}
\caption{Comparison of topological variable distributions between $D^0$ signals in real data $(black)$ and in data-driven Simulation with real data distributions as the input $(red)$ in 0--10\% Au+Au collisions for $D^0$ mesons at 2\,$<$\,$p_T$\,$<$\,3\,GeV/$c$. The dashed lines indicate the final topological cuts chosen for each individual topological variable.}
\label{fig:DataTopo} 
\end{figure*}

\begin{figure}[h]
\centering
\includegraphics[width=0.43\textwidth]{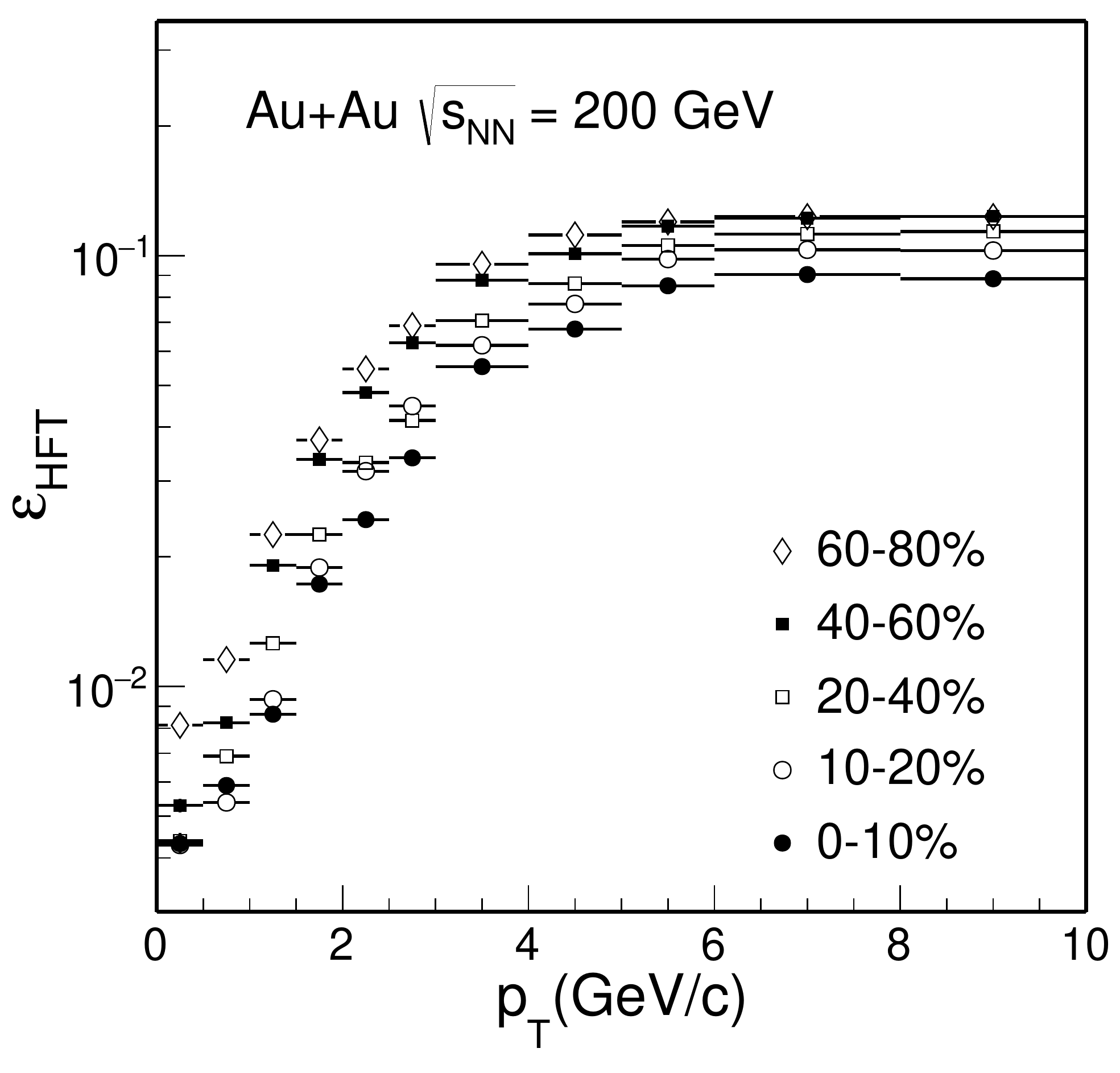}
\caption{$D^{0}$ HFT tracking and topological cut efficiencies $\varepsilon_{\rm HFT}$ from different centrality classes in Au+Au collisions at $\sqrt{s_{_{\rm NN}}}$ = 200\,GeV.}
\label{fig:Datad0Eff_hftTopo} 
\end{figure}

We employ the validated data-driven simulation method for the real data analysis. Figure~\ref{fig:DataTopo} shows comparisons of the same five topological variables between $D^0$ signals in real data and data-driven simulated distributions with real data as the input in central 0--10\% collisions for $D^0$ mesons at 2\,$<$\,$p_{T}$\,$<$\,3\,GeV/$c$. The real data distributions are extracted by reconstructing $D^0$ signals with the same reconstruction cuts as in Sec.~\ref{D0recon} except for the interested topological variable to be compared. The distributions for $D^0$ candidates are generated by statistically subtracting the background using the side-band method from the same-event unlike-sign distributions within the $D^0$ mass window. The cut on the interested topological variable is loosened, but one needs to place some pre-cuts to ensure reasonable $D^0$ signal reconstruction for the extraction of these topological variable distributions. These pre-cuts effectively reduce the low-end reach for several topological variables, e.g. the $D^0$ decay length. In the data-driven simulation method, charged pion and kaon HFT matching efficiencies and 2D DCA distributions are used as the input to calculate these topological variables for $D^0$ signals. Figure~\ref{fig:DataTopo} shows that in the selected ranges, the data-driven simulation method reproduces topological variables distributions of $D^0$ signals, which supports that this method can be reliably used to calculate the topological cut efficiency.

Figure~\ref{fig:Datad0Eff_hftTopo} shows the HFT tracking and topological cut efficiency $\varepsilon_{\rm HFT}$ as a function of $D^0$ $p_{T}$ for different centrality bins obtained using the data-driven simulation method described in this section with the input distributions taken from the real data. The smaller efficiency seen in central collisions is in part because the HFT tracking efficiency is lower in higher occupancy central collisions, and in addition because we choose tighter topological cuts in central collisions for background suppression.

\subsubsection{Secondary particle contribution}
\label{correction:hft:secondary}

\begin{figure}[h]
\centering
\includegraphics[width=0.38\textwidth, angle = 0]{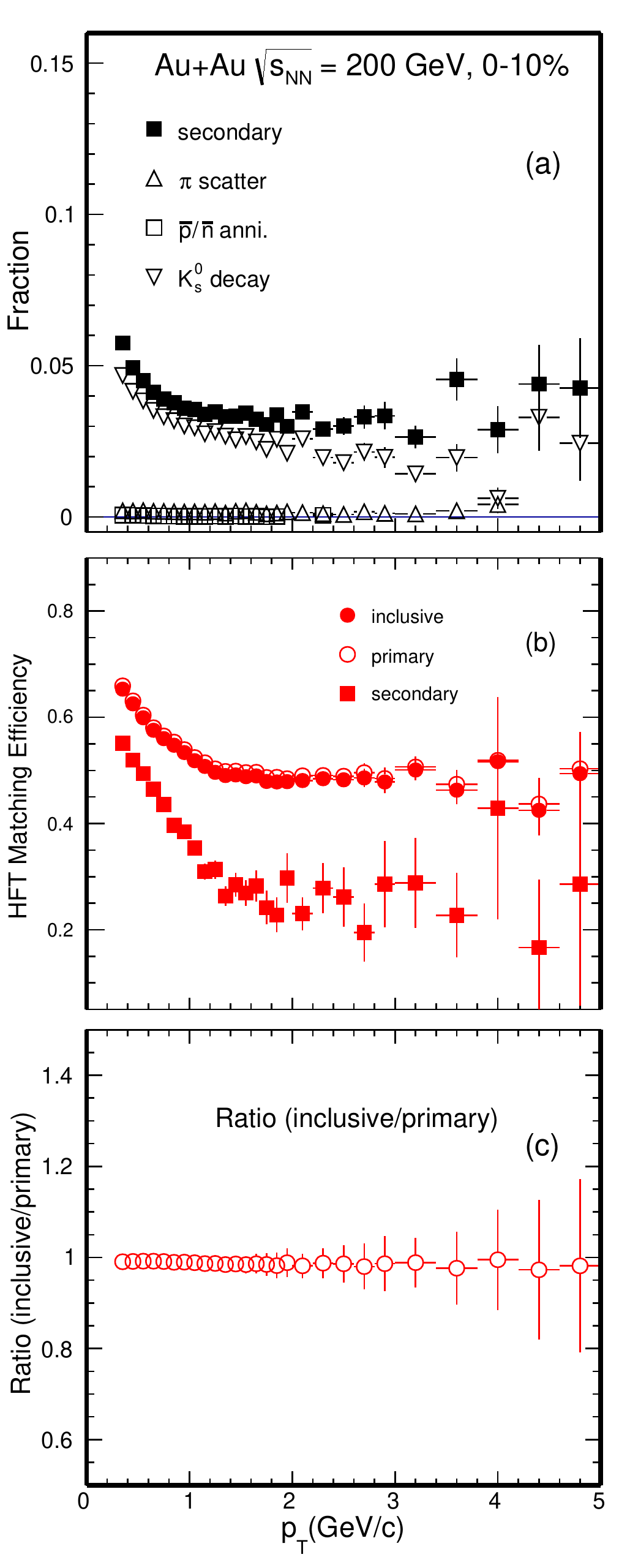}
  \caption{Secondary pion contribution estimated from Hijing+GEANT simulation with FLUKA hadronic package. Panel (a) shows the fraction of different sources for secondary pion tracks. Panel (b) shows the HFT matching efficiency $\varepsilon_{\rm HFT}^{\rm match}$ for inclusive, primary and secondary pions. Panel (c) shows the ratio of HFT matching efficiencies between inclusive and primary pions.}
\label{fig:Fraction_Pion} 
\end{figure}

\begin{figure}
\centering
\includegraphics[width=0.43\textwidth]{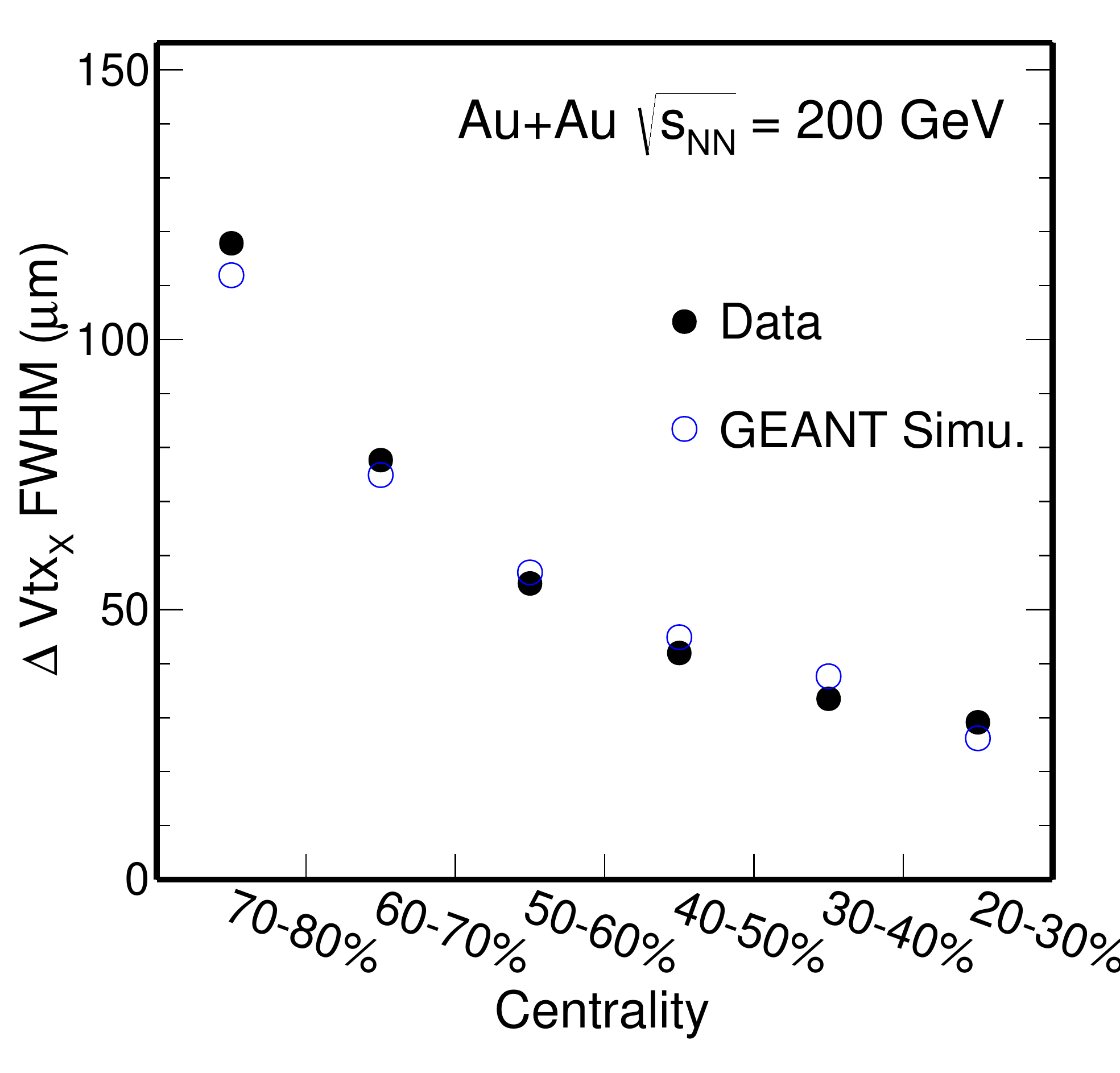}
\caption{Full-Width-at-Half-Maximum (FWHM) of vertex position difference in the X dimension between two randomly-divided sub-events in various centrality bins. Black solid circles present the FWHM values from real data while blue empty circles are from Hijing+GEANT simulation. Statistical uncertainties are smaller than the marker size.}
\label{fig:vtxX_vsCent} 
\end{figure}

\begin{figure*}
\centering
\includegraphics[width=1.05\textwidth]{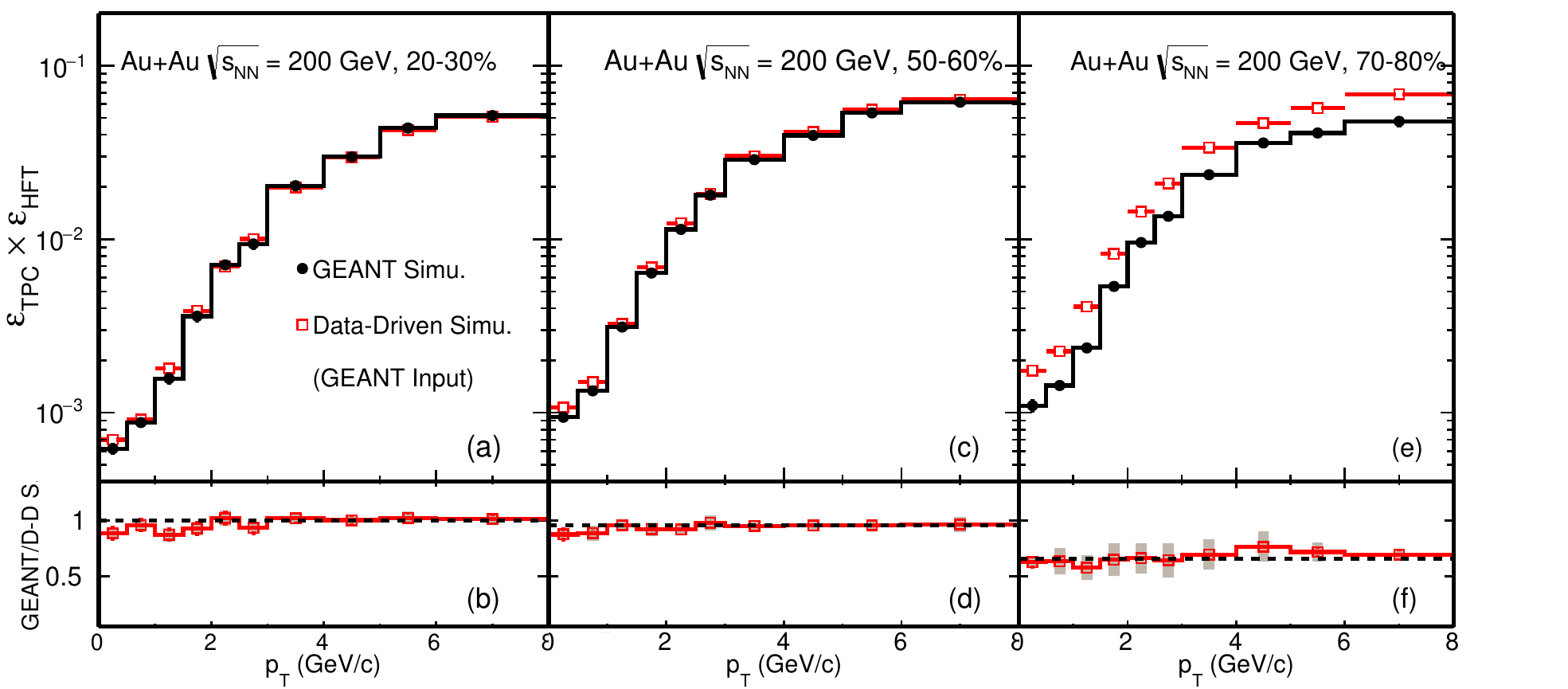}
  \caption{$D^{0}$ reconstruction efficiency comparison between MC GEANT simulation (GEANT, $black$) and  data-driven simulation with the reconstructed MC data as the input (D-D S.,$red$) for 20--30\% (a), 50--60\% (c) and 70--80\% (e) Au+Au collisions. Bottom panels (b,d,f) show the ratios between the two distributions above.}
\label{fig:Mcd0Eff_20_80} 
\end{figure*}

In the data-driven method for obtaining the efficiency correction, inclusive pion and kaon distributions are taken from real data as the input while the validation with GEANT simulation is performed with primary particles. There is a small amount of secondary particle contribution (e.g. weak decays from $K^0_S$ and $\Lambda$) to the measured inclusive charged pion tracks.

The impact of secondary particle contribution to the charged pions is studied using the HIJING events processed through the GEANT simulation and the same offline reconstruction. The fraction of secondary pions from weak decay of strange hadrons ($K^0_S$ and $\Lambda$) to the total inclusive charged pions within ${\rm DCA}<$ 1.5\,cm cut is estimated to be around 5\% at pion $p_{T}$ = 0.3\,GeV/$c$ and decreases to be $<2\%$ above 2\,GeV/$c$. This is consistent with what was observed before in measuring the prompt charged pion spectra~\cite{Adams:2003xp}. There is another finite contribution of low momentum anti-protons and anti-neutrons annihilated in the detector material and producing secondary pions. The transverse momenta of these pions are mostly around 2--3\,GeV/$c$ and the fraction of total inclusive pions is $\sim$ 10--12\% at $p_{T} =2\textup{-}3$\,GeV/$c$ based on this simulation and contribute $\sim$ 5--8\% to the HFT matching efficiency. This is obtained using the GEANT simulation with GHEISHA hadronic package. With a different hadronic package, FLUKA, the secondary pion fraction in 2--3\,GeV/$c$ region is significantly reduced to be negligible. The difference between the primary pions and the inclusive pions in the HFT matching efficiency has been considered as one additional correction factor in our data-driven simulation method when calculating the final efficiency. The maximum difference with respect to the result obtained using the GHEISHA hadronic package is used as the systematic uncertainty for this source. Figure~\ref{fig:Fraction_Pion} shows the secondary pion contribution in Au+Au collisions with FLUKA hadronic package. Panel (a) shows the fraction of different sources for secondary tracks including the weak decays, the scattering and the $\bar{p}/\bar{n}$ annihilation in the detector material. Panel (b) shows the HFT matching efficiencies for inclusive, prompt and secondary pions. Panel (c) is the ratio of the HFT matching efficiencies between the inclusive and the primary pions from panel (b). The effect of such secondary contribution to charged kaons is found to be negligible~\cite{Adams:2003xp}.

\subsection{Vertex Resolution Correction - $\varepsilon_{\rm vtx}$}
\label{correction:vtx}

In the data-driven approach, $D^0$ mesons are injected at the event vertex. In the real data, the reconstructed vertex has a finite resolution with respect to the real collision vertex. This may have some effect on the reconstructed $D^0$ signal counts after applying the topological cuts in small multiplicity events where the event vertex resolution decreases. We carry out similar simulation studies as described in Sec.~\ref{correction:hft:fastsim} for other centrality bins. Figure~\ref{fig:vtxX_vsCent} shows the Full-Width-at-Half-Maximum (FWHM) of the difference in the vertex x-position of two randomly-divided sub-events in various centrality bins between data and MC simulation. We choose the FWHM variable here as the distributions are not particularly Gaussian. The MC simulation reproduces the vertex difference distributions seen in the real data reasonably well. This gives us confidence for using this MC simulation setup to evaluate the vertex resolution correction $\varepsilon_{\rm vtx}$.

\begin{figure}
\centering
\includegraphics[width=0.43\textwidth]{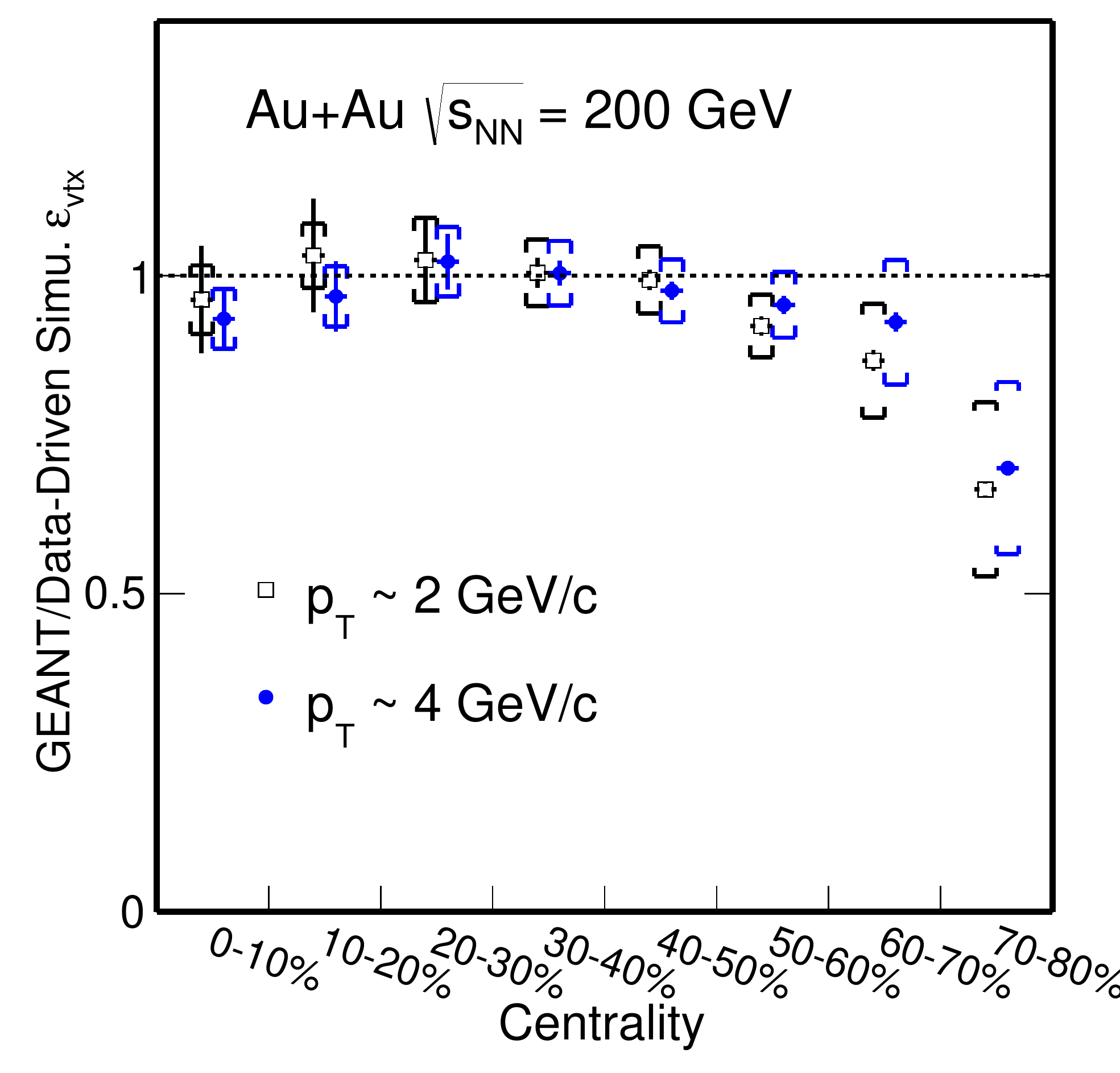}
\caption{$\varepsilon_{\rm vtx}$, $D^0$ reconstruction efficiency ratios between MC GEANT simulation and data-driven simulation with the reconstructed MC data as the input versus collision centrality for $p_{T}$ at 2 and 4\,GeV/$c$. The brackets depict the estimated systematic uncertainties.}
\label{fig:Mcd0Eff_20_80_vsCent} 
\end{figure}

To estimate the vertex resolution effect, we embed single PYTHIA $c\bar{c}$ event into a HIJING Au+Au event, and the whole event is passed through the STAR GEANT simulation followed by the same offline reconstruction as in the real data production. The PYTHIA $c\bar{c}$ events are pre-selected to have at least one $D^0\rightarrow K^-\pi^+$ decay or its charge conjugate to enhance the statistics. Figure~\ref{fig:Mcd0Eff_20_80} shows the comparison in the obtained $D^0$ reconstruction efficiency between MC simulation $(black)$ and data-driven simulation using reconstructed MC data as the input $(red)$ for 20--30\% (left), 50--60\% (middle) and 70--80\% (right) centrality bins, respectively. The bottom panels show the ratios of the efficiencies obtained from the two calculation methods. In central and mid-central collisions, the data-driven simulation method can properly reproduce the $D^0$ real reconstruction efficiency. This is expected since the vertex resolution is small enough so that it has negligible impact on the obtained efficiency using the data-driven simulation method. However, in more peripheral collisions, the data-driven simulation method overestimates the $D^0$ reconstruction efficiency as shown in the middle and right panels. The vertex resolution correction factor $\varepsilon_{\rm vtx}$, denoted in Eq.~\ref{equ:invariantyield}, has a mild $p_{T}$ dependence but strong centrality dependence as shown in Fig.~\ref{fig:Mcd0Eff_20_80_vsCent} for $p_T$ = 2 and 4 GeV/$c$. The brackets denote the systematic uncertainties in the obtained correction factor $\varepsilon_{\rm vtx}$. They are estimated by changing the multiplicity range in the HIJING + GEANT simulation so that the variation in the sub-event vertex difference distributions from the real data can be covered by distributions obtained from different simulation samples. The vertex resolution corrections are applied as a function of $p_{T}$ in each individual centrality class.

\subsection{PID Efficiency - $\varepsilon_{\rm PID}$ and Doubly-mis-PID Correction}
\label{correction:PID}

\begin{figure}
\centering
\includegraphics[width=0.43\textwidth]{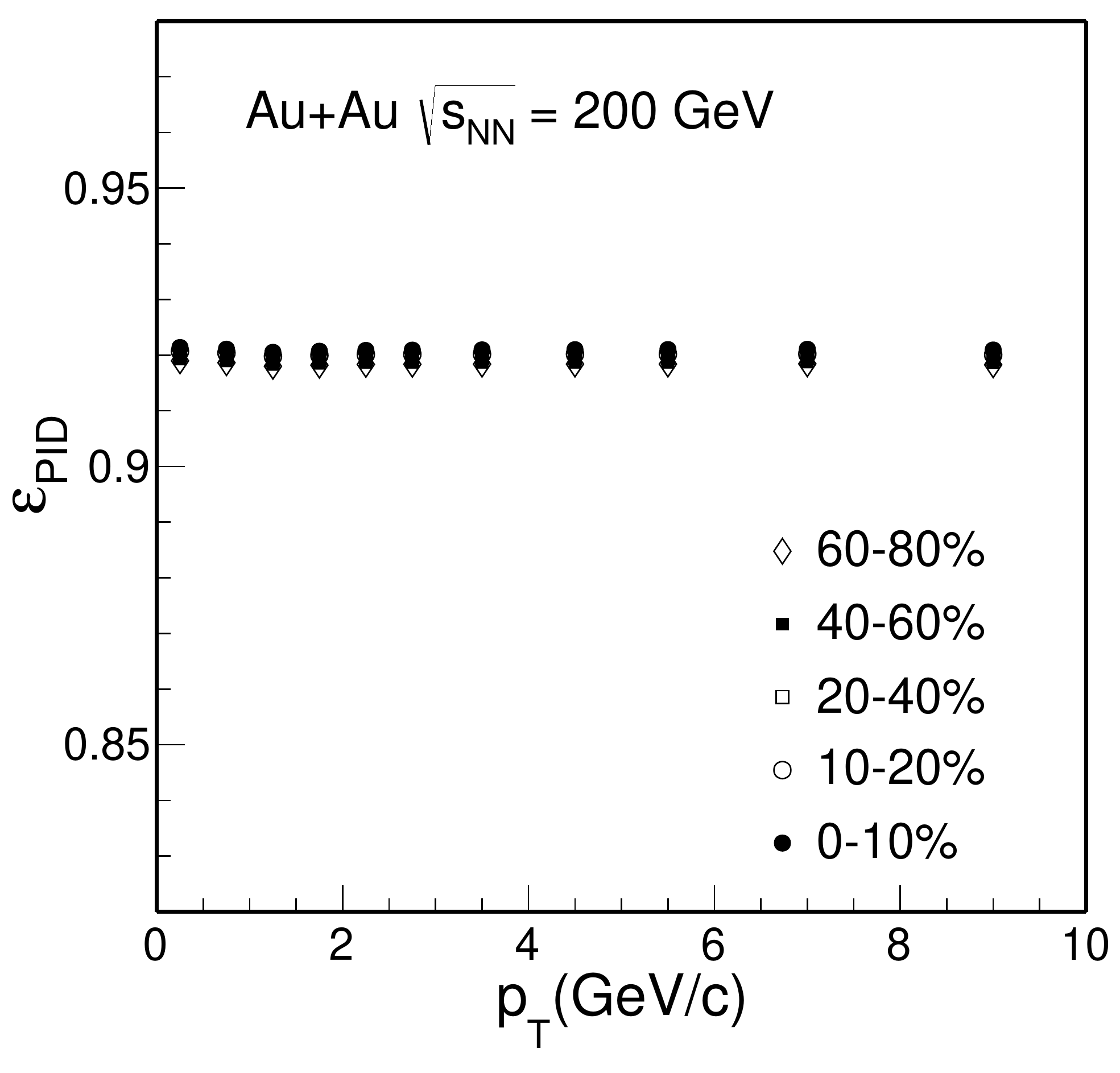}
\caption{Particle identification efficiency ($\varepsilon_{\rm PID}$) of $D^0$ mesons in different centrality classes.}
\label{fig:Datad0Eff_pid} 
\end{figure}

\begin{figure}
\centering
\includegraphics[width=0.43\textwidth]{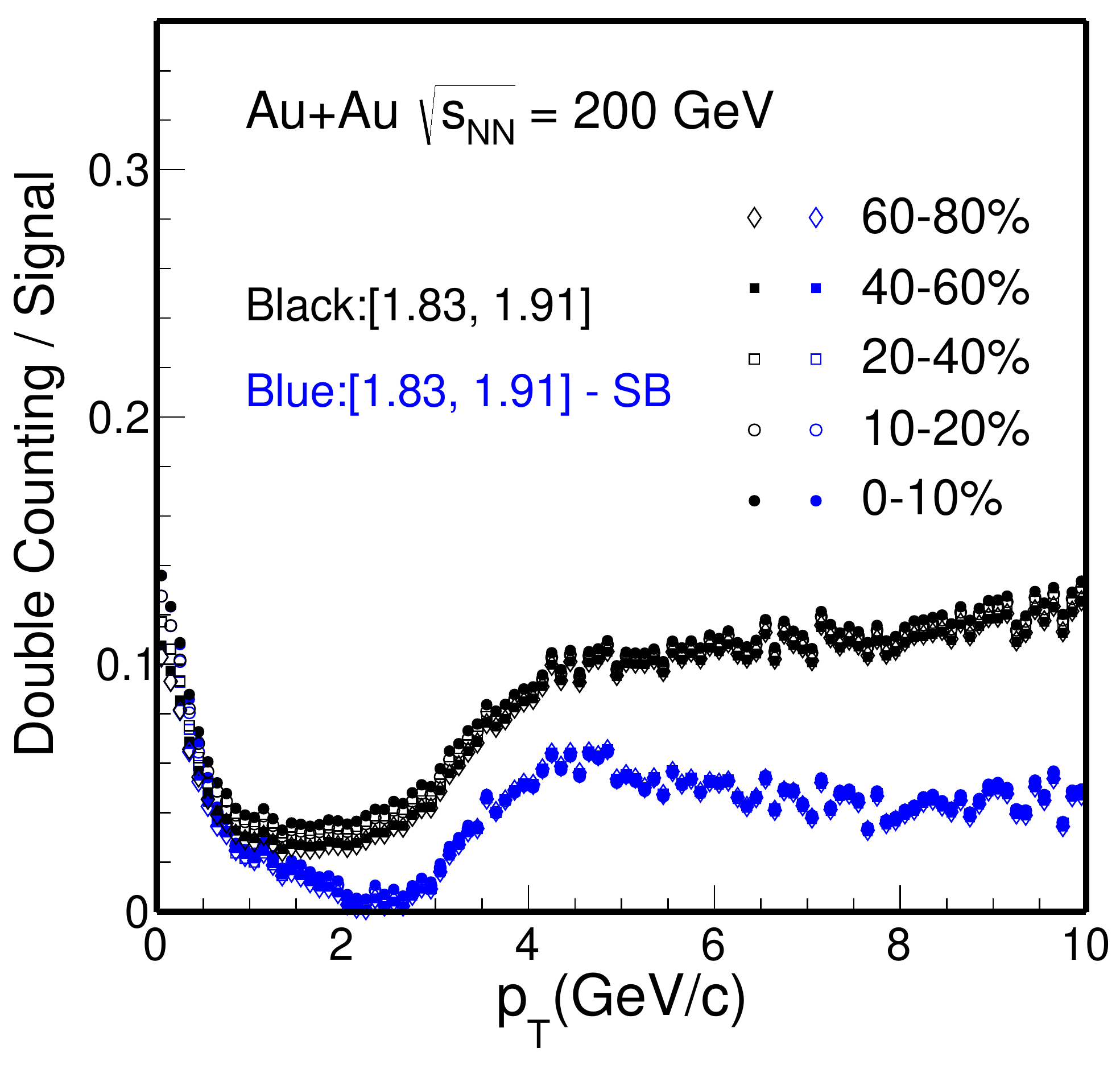}
\caption{$D^{0}$ yield double counting fraction due to doubly mis-PID in different centrality classes. The black markers depict an estimation taking the total double counting yield in the $D^0$ mass window while the blue markers depict an estimation with an additional side-band (SB) subtraction. Note that most data points from different centrality bins overlap with each other.}
\label{fig:Datad0Eff_doublecounting} 
\end{figure}

The $D^0$ daughter particle identification (PID) cut efficiency includes contributions from the $dE/dx$ selection cut efficiency as well as the TOF matching and $1/\beta$ cut efficiency. To best estimate the selection cut efficiency, we select the enriched kaon and pion samples from $\phi,K_{S}^{0}$ decays following the same procedure as in~\cite{Shao:2005iu,Xu:2008th} and obtain the mean and width in the $dE/dx$ $n\sigma_X$ distributions. The $dE/dx$ cut efficiencies for pion and kaon daughter tracks are calculated correspondingly. The TOF $1/\beta$ cut efficiency is determined by studying the $1/\beta$ distributions for kaons and pions in the clean separation region, namely $p_{T}$\,$<$\,1.5\,GeV/$c$. There is a mild dependence for the offset and width of $\Delta 1/\beta$ distributions vs. particle momentum and our selection cuts are generally wide enough to capture nearly all tracks once they have valid $\beta$ measurements. The total PID efficiency of $D^0$ mesons is calculated by folding the individual track TPC and TOF PID efficiencies following the same hybrid PID algorithm as implemented in the data analysis. Figure~\ref{fig:Datad0Eff_pid} shows the total PID efficiencies for $D^0$ reconstruction in various centrality bins. The total PID efficiency is generally high and has nearly no centrality or $p_T$ dependence.

When the $D^0$ daughter kaon track is mis-identified as a pion track and the other daughter pion track is mis-identified as a kaon track, the pair invariant mass distribution will have a bump structure around the real $D^0$ signal peak, but the distribution is much broader in a wide mass region due to the mis-assigned daughter particle masses. Based on the PID  performance study described above, we estimate the single kaon and pion candidate track purities. After folding the realistic particle momentum resolution, we calculate the reconstructed $D^0$ yield from doubly mis-identified pairs (double counting) underneath the real $D^0$ signal and the double counting fraction is shown in Fig.~\ref{fig:Datad0Eff_doublecounting}. The black markers show the fraction by taking all doubly mis-identified pairs in the $D^0$ mass window while the blue markers depict it with an additional side-band (SB) subtraction. The latter is used as a correction factor to the central values of reported $D^0$ yields while the difference between the black and blue symbols is considered as the systematic uncertainty in this source. The double counting fraction is below 10\% in all $p_{T}$ bins, and there is little centrality dependence.

\begin{figure}
\centering
\includegraphics[width=0.43\textwidth]{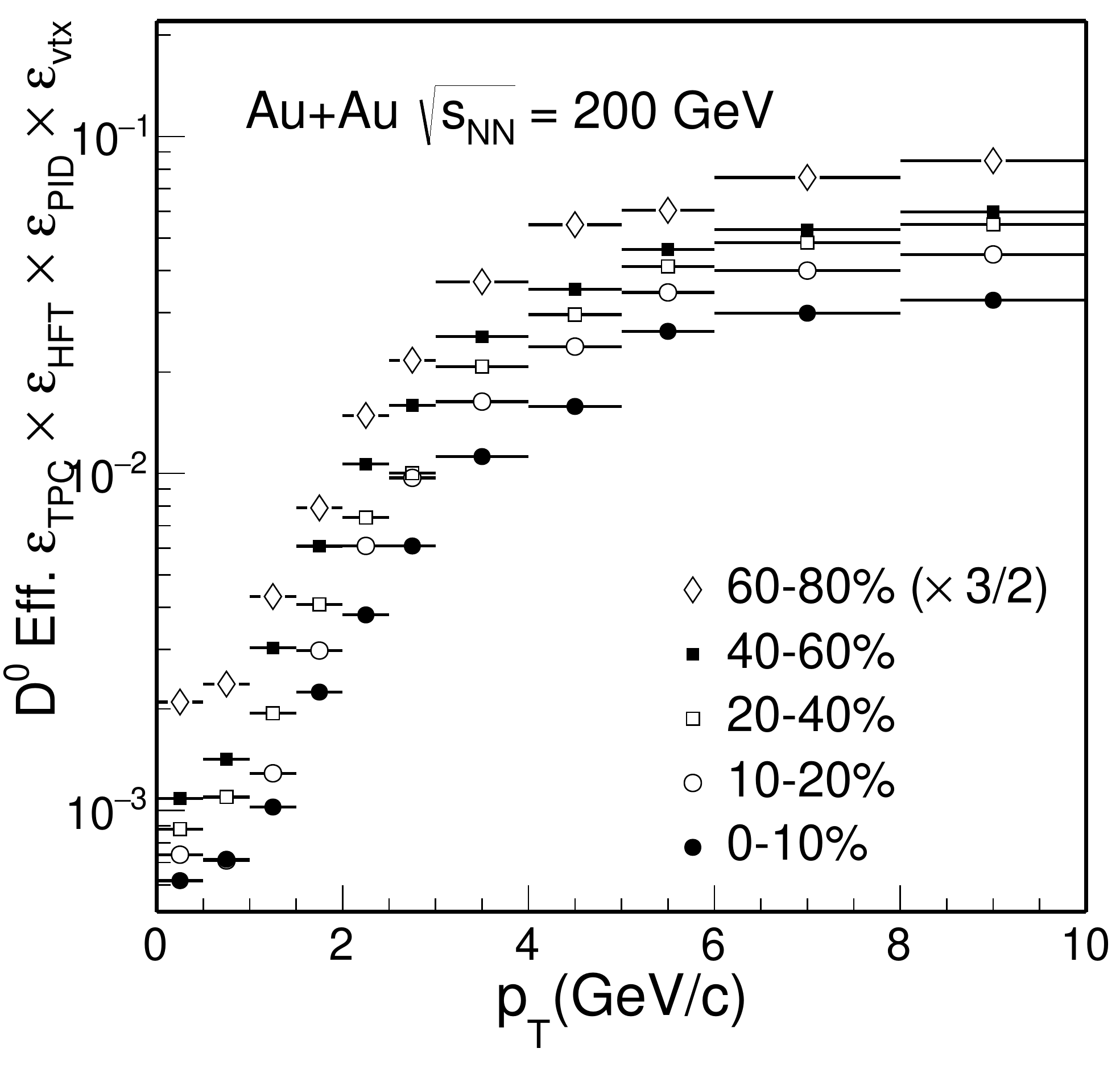}
\caption{The total $D^{0}$ reconstruction efficiency from different centrality classes.}
\label{fig:Datad0Eff} 
\end{figure}

Figure~\ref{fig:Datad0Eff} shows the total $D^{0}$ reconstruction efficiency from different centrality classes in Au+Au collisions including all of the individual components discussed above.

\section{Systematic Uncertainties}
\label{systematic}

The systematic uncertainty on the final measured $D^0$ $p_{T}$ spectra can be categorized as the uncertainty of the raw $D^0$ yield extraction and the uncertainty of efficiencies and corrections.

The uncertainty of the raw yield extraction is estimated by a) changing the $D^0$ raw yield counting method from the Gaussian fit to histogram bin counting. b) varying invariant mass ranges for fit and for side bands and c) varying background estimation from mixed-event and like-sign methods. The maximum difference between these scenarios is then converted to the standard deviation and added to the systematic uncertainties. It is the smallest in the mid-$p_{T}$ bins due to the best signal significance and grows at both low and high $p_{T}$. The double counting contribution in the $D^0$ raw yield  due to mis-PID is included as another contribution to the systematic uncertainty for the $D^0$ raw yield extraction as described in Sec.~\ref{correction:PID}.

The uncertainty of the TPC acceptance and efficiency correction $\varepsilon_{\rm TPC}$ is estimated via the standard procedure in STAR by comparing the TPC track distributions between real data and the embedding data. It is estimated to be $\sim$5--7\% for 0--10\% collisions and $\sim$5--8\% for 60--80\% collisions, and is correlated for different centralities and $p_{T}$ regions. 

The uncertainty of the PID efficiency correction is estimated by varying the PID selection cuts and then convoluting to the final corrected $D^0$ yield.

To estimate the uncertainty of the HFT tracking and topological cut efficiency correction $\varepsilon_{\rm HFT}$, we employ the following procedures: a) We vary the topological variable cuts such that the $D^0$ $\varepsilon_{\rm HFT}$ is changed to 50\% and 150\% from the nominal (default) efficiency and compare the efficiency--corrected final $D^0$ yields. The maximum difference between the two scenarios is then added to the systematic uncertainties. b) We also vary the lower threshold cut on the daughter $p_{T}$ between 0.3 to 0.6 GeV/$c$ and the maximum difference in the final corrected $D^0$ yield is also included in the systematic uncertainties. c) We add the systematic uncertainty due to limitation of the data-driven simulation approach, $\sim$5\%, and the impact of the secondary particles, $\sim$2\%, to the total $\varepsilon_{\rm HFT}$ systematic uncertainty.

With the corrected $D^0$ transverse momentum spectra, the nuclear modification factor $R_{\rm CP}$ is calculated as the ratio of $N_{\rm bin}$--normalized yields between central and peripheral collisions, as shown in the following formula:
\begin{equation}
  R_{\rm CP} = \frac{d^2N/dp_{T}dy}{N_{\rm bin}} |_{\rm cen\ } \times \frac{N_{\rm bin}} {d^2N/dp_{T}dy} |_{\rm peri }.
\label{equ:equation3}
\end{equation}

The systematic uncertainties in the raw signal extraction in central and peripheral collisions are propagated as they are uncorrelated, while the systematic uncertainties from the other sources are correlated or partially correlated in contributing to the measured $D^0$ yields. To best consider these correlations, we vary selection cuts simultaneously in central and peripheral collisions, and the difference in the final extracted $R_{\rm CP}$ value is then directly counted as systematic uncertainties in the measured $R_{\rm CP}$.

The nuclear modification factor $R_{\rm AA}$ is calculated as the ratio of $N_{\rm bin}$--normalized yields between Au+Au and $p$+$p$ collisions. The baseline for $p$+$p$ collisions is chosen the same as in Ref.~\cite{Star_D_RAA}. The uncertainties from the $p$+$p$ reference dominates the systematic uncertainty for $R_{\rm AA}$. They include the 1$\sigma$ uncertainty from the Levy function fit to the measured spectrum and the difference between Levy and power-law function fits for extrapolation to low and high $p_T$, expressed as one standard deviation.

With the corrected $D^0$ and $\overline{D}^{0}$ transverse momentum spectra, the $\overline{D}^{0}/D^0$ ratio is calculated as a function of the transverse momentum. The systematic uncertainties in the raw signal extraction for $\overline{D}^{0}$ and $D^0$ are propagated as they are uncorrelated, while the systematic uncertainties from the other sources are correlated or partially correlated in contributing to the measured $\overline{D}^{0}/D^0$ ratio. As in the $R_{\rm CP}$ systematic uncertainty estimation, we vary selection cuts simultaneously for $D^0$ and $\overline{D}^{0}$, and the difference in the final extracted $\overline{D}^{0}/D^0$ value is then directly counted as systematic uncertainties for the measured $\overline{D}^{0}/D^0$ ratio.

\begin{table*}
\centering{
\caption{Summary of systematic uncertainties, in percentage, on the $D^0$ invariant yield in 0--10\% and 60--80\% collisions and on $R_{\rm CP}$(0--10\%/60--80\%).}
\begin{tabular}{c|cc|c|c} \hline \hline
  Source & \multicolumn{3}{c|}{Systematic uncertainty [\%]} & Correlation in $p_{T}$\\ \cline{2-4}
       & \hspace{1cm}0--10\%\hspace{1cm} & \hspace{1cm}60--80\%\hspace{1cm} & \hspace{1cm}$R_{\rm CP}$(0--10\%/60--80\%)\hspace{1cm} &  \\ \hline \hline
Signal extra. & 1-6  & 1-12  & 2-13\ & uncorr. \\ \hline
Double mis-PID & 1-7  & 1-7  & negligible & uncorr. \\ \hline \hline
$\varepsilon_{\rm TPC}$ & 5-7 & 5-8 & 3-7 & largely corr. \\ \hline
$\varepsilon_{\rm HFT}$ & 3-15 & 3-20 & 3-20 & largely corr. \\ \hline
$\varepsilon_{\rm PID}$ & 3 & 3 & negligible & largely corr. \\ \hline
$\varepsilon_{\rm vtx}$ & 5 & 9-18 & 10-18 & largely corr. \\ \hline
BR. & \multicolumn{2}{c|}{0.5} & 0 & global \\ \hline \hline
$N_{\rm bin}$ & 2.8 & 42 & 42 & global \\ \hline \hline
\end{tabular}
}
\label{table:syserror}
\end{table*}

Table~\ref{table:syserror} summarizes the systematic uncertainties and their contributions, in percentage, on the $D^0$ invariant yield in 0--10\% and 60--80\% collisions and $R_{\rm CP}$(0--10\%/60--80\%). In the last column we also comment on the correlation in $p_{T}$ for each individual source. Later when reporting $p_{T}$--integrated yields or $R_{\rm CP}$, systematic uncertainties are calculated under the following considerations: a) for $p_{T}$ uncorrelated sources, we take the quadratic sum of various $p_{T}$ bins; b) for sources that are largely correlated in $p_{T}$, we take the arithmetic sum as a conservative estimate.


\section{Results and Discussion}
\label{result}

\subsection{$p_{T}$ Spectra and Integrated Yields}
\label{result:pt}

\begin{figure}
\centering
\includegraphics[width=0.43\textwidth]{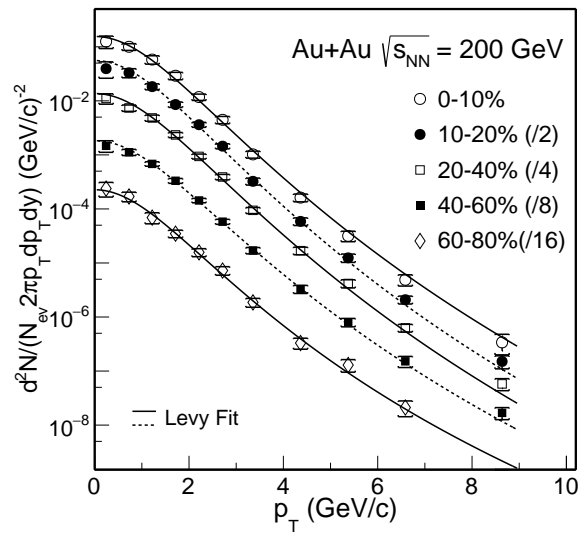}
\caption{$D^{0}$ invariant yield at mid-rapidity ($|y|<1$) vs. transverse momentum for different centrality classes. Error bars (not visible for many data points) indicate statistical uncertainties and brackets depict systematic uncertainties. Global systematic uncertainties in $B.R.$ are not plotted. Solid and dashed lines depict Levy function fits.}
\label{fig:D0_spectra} 
\end{figure}

Figure~\ref{fig:D0_spectra} shows the efficiency--corrected $D^0$ invariant yield at mid-rapidity ($|y|<1$) vs. $p_{T}$ in 0--10\%, 10--20\%, 20--40\%, 40--60\% and 60--80\% Au+Au collisions. $D^0$ spectra in some centrality bins are scaled with arbitrary factors indicated on the figure for clarity. Dashed and solid lines depict fits to the spectra with the Levy function:

\begin{equation}
  \begin{aligned}
    \frac{d^2N}{2\pi p_{T}dp_{T}dy} = 
   & \frac{1}{2\pi}\frac{dN}{dy}\frac{(n-1)(n-2)}{nT(nT+m_0(n-2))} \\
  & \times \bigg(1+\frac{\sqrt{p_{T}^2+m_0^2}-m_0}{nT}\bigg)^{-n},
  \end{aligned}
\label{equ:equation4}
\end{equation}
where $m_0$ is the $D^0$ mass (1.864 GeV/$c^2$) and $dN/dy$, $T$ and $n$ are free parameters. The Levy function fit describes the $D^0$ spectra nicely in all centrality bins in our measured $p_T$ region.

\begin{figure}
\centering
\includegraphics[width=0.42\textwidth]{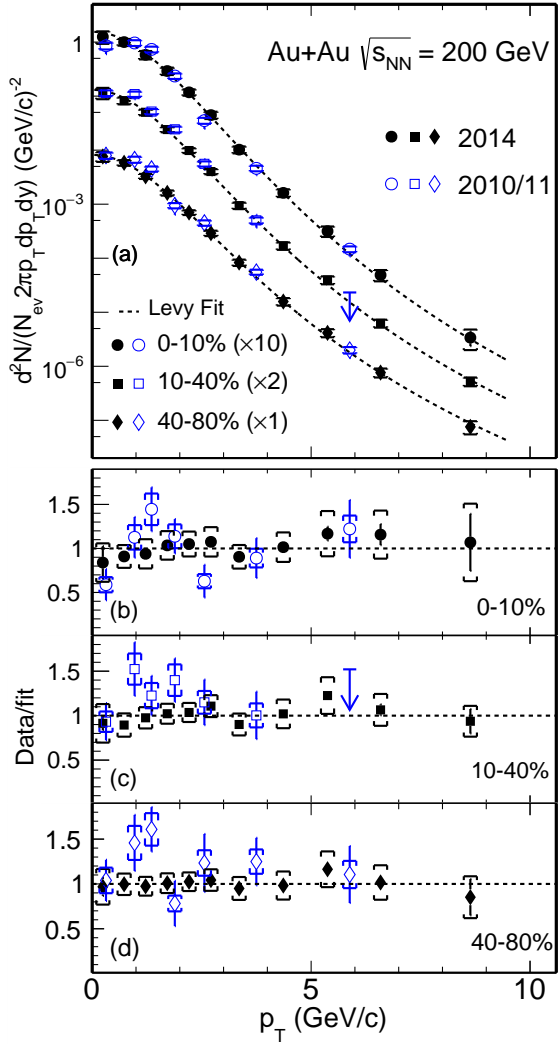}
  \caption{(a) Measured $D^{0}$ spectra from this analysis compared with the previous 2010/11 measurements for different centrality classes. Dashed lines depict Levy function fits to 2014 data. (b) - (d), Ratio of measured spectra to the fitted Levy functions in 0--10\%, 10--40\% and 40--80\% centrality bins, respectively.}
\label{fig:D0_compareSpectra_run10} 
\end{figure}

We compare our new measurements with previous measurements using the STAR TPC only. The previous measurements are recently corrected after fixing errors in the TOF PID efficiency calculation~\cite{Star_D_RAA}. Figure~\ref{fig:D0_compareSpectra_run10} shows the $p_{T}$ spectra comparison in 0--10\%, 10-40\% and 40--80\% centrality bins in panel (a) and the ratios to the Levy fit functions in panels (b), (c), and (d), respectively. The new measurement with the HFT detector shows a nice agreement with the measurement without the HFT, but with significantly improved precision.

\begin{figure}
\centering
\includegraphics[width=0.41\textwidth]{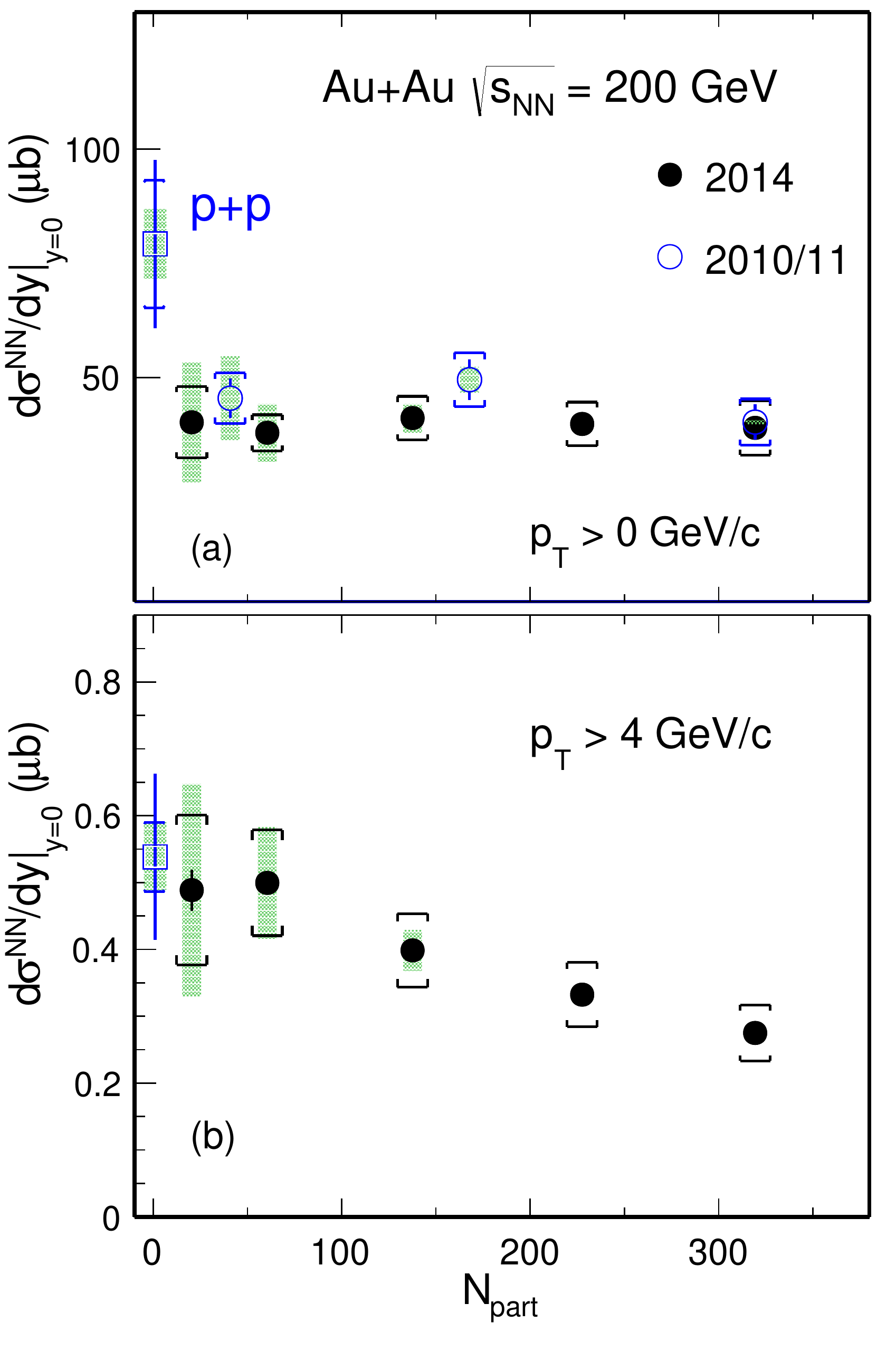}
  \caption{Integrated $D^{0}$ cross section per nucleon-nucleon collision at mid-rapidity for $p_{T}$\,$>$\,0 (a) and $p_{T}$\,$>$\,4\,GeV/$c$ (b) as a function of centrality $N_{\rm part}$. The statistical and systematic uncertainties are shown as error bars and brackets on the data points. The green boxes on the data points depict the overall normalization uncertainties in $p$+$p$ and Au+Au data respectively.}
\label{fig:Xsection_D0} 
\end{figure}

The measured $D^0$ spectra cover a wide $p_{T}$ region which allows us to extract the $p_{T}$--integrated total $D^0$ yield at mid-rapidity with good precision.\,\,\,Figure\,\,\ref{fig:Xsection_D0} shows the $p_{T}$--integrated cross section for $D^0$ production per nucleon-nucleon collision $d\sigma^{NN}/dy|_{y=0}$ from different centrality bins for the full $p_{T}$ range shown in the top panel and for $p_{T}$\,$>$\,4\,GeV/$c$ shown in the bottom panel. The result from the previous $p$+$p$ measurement is also shown in both panels~\cite{Star_D_pp}.

While $d\sigma^{NN}/dy|_{y=0}$ for $p_{T}$\,$>$\,4\,GeV/$c$ shows a clear decreasing trend from peripheral to mid-central and central collisions, that for the full $p_T$ range shows approximately a flat distribution as a function of $N_{\rm part}$, though the systematic uncertainty in the 60--80\% centrality bin is a bit large. The values for the full $p_T$ range in mid-central to central Au+Au collisions are smaller than that in $p$+$p$ collisions with $\sim1.5\sigma$ effect considering the large uncertainties from the $p$+$p$ measurements. The total charm quark yield in heavy-ion collisions is expected to follow the number-of-binary-collision scaling since charm quarks are believed to be predominately created at the initial hard scattering before the formation of the QGP at RHIC energies. However, the cold nuclear matter (CNM) effect including shadowing could also play an important role. In addition, hadronization through coalescence has been suggested to potentially modify the charm quark distribution in various charm hadron states which may lead to the reduction in the observed $D^0$ yields in Au+Au collisions~\cite{GRECO2004202} (as seen in Fig.~\ref{fig:Xsection_D0}). For instance, hadronization through coalescence can lead to an enhancement of the charmed baryon $\Lambda_{c}^+$ yield relative to $D^0$ yield~\cite{Oh2009,Zhao:2018jlw,Plumari:2017ntm}, and together with the strangeness enhancement in the hot QCD medium, can also lead to an enhancement in the charmed strange meson $D_{s}^+$ yield relative to $D^0$~\cite{He2013,Zhao:2018jlw,Plumari:2017ntm}. Therefore, determination of the total charm quark yield in heavy-ion collisions will require measurements of other charm hadron states over a broad momentum range.

\subsection{Collectivity}
\label{result:collectivity}

\subsubsection{$m_{T}$ Spectra}
\label{result:collectivity:mT}

\begin{figure}
\centering
\includegraphics[width=0.43\textwidth]{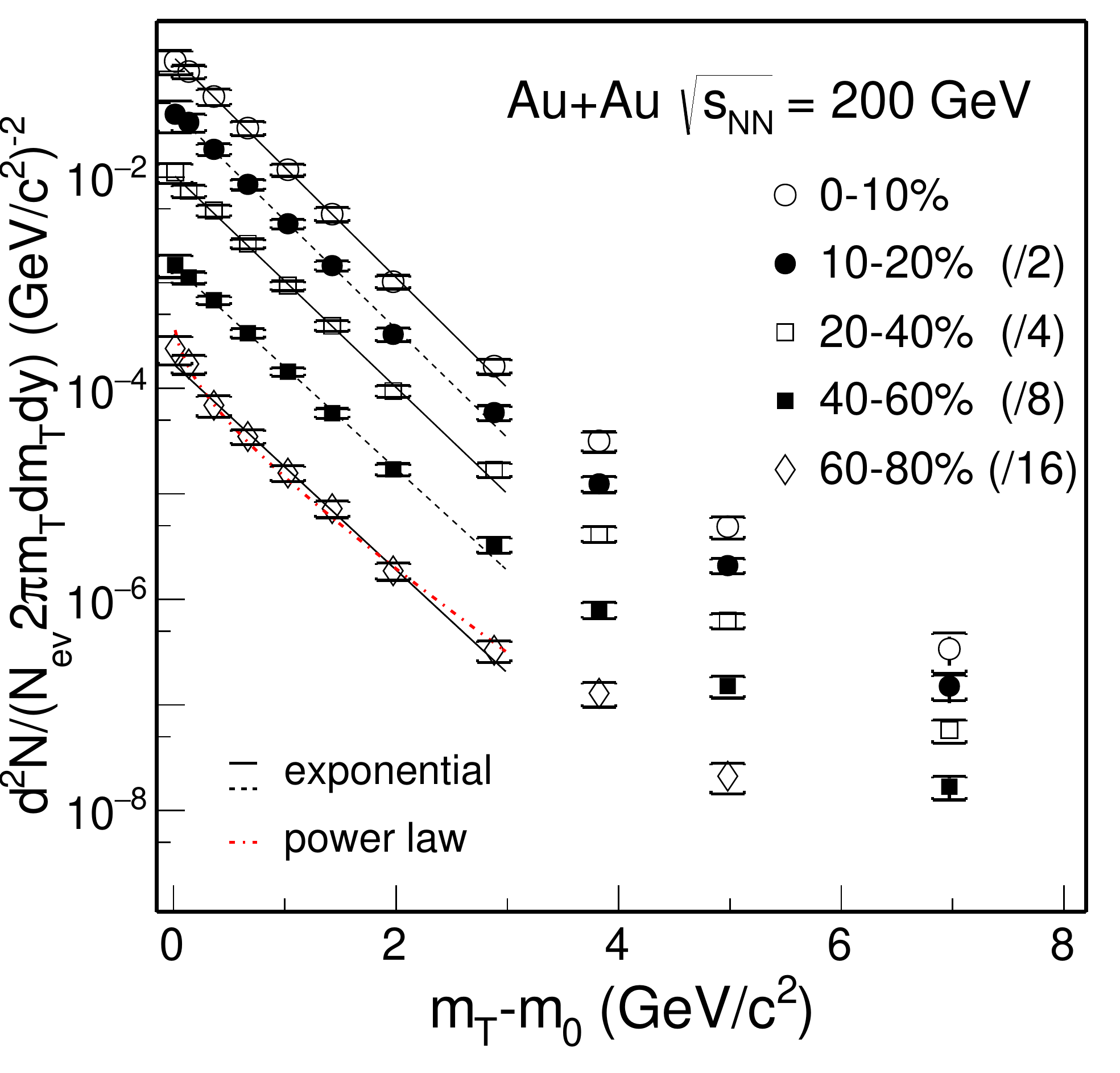}
\caption{$D^{0}$ invariant yield at mid-rapidity ($|y|$\,$<$\,1) vs. transverse kinetic energy ($m_{T}$-$m_{0}$) for different centrality classes. Error bars (not visible for many data points) indicate statistical uncertainties and brackets depict systematic uncertainties. Global systematic uncertainties in $B.R.$ are not plotted. Solid and dashed black lines depict exponential function fits and the dot-dashed line depicts a power-law function fit to the spectrum in the 60--80\% centrality bin.}
\label{fig:mTFit_D0} 
\end{figure}

Transverse mass spectra can be used to study the collectivity of produced hadrons in heavy-ion collisions. Figure~\ref{fig:mTFit_D0} shows the $D^{0}$ invariant yield at mid-rapidity ($|y|$\,$<$\,1) vs. transverse kinetic energy ($m_{T}$-$m_{0}$) for different centrality classes, where $m_{T} = \sqrt{p_{T}^2+m_0^2}$ and $m_0$ is the $D^0$ meson mass at rest. Solid and dashed black lines depict exponential function fits inspired by thermal models to data in various centrality bins up to $m_{T}-m_{0}$\,$<$\,3\,GeV/$c^2$ using the fit function shown below:
\begin{equation}
  \begin{aligned}
\frac{d^2N}{2\pi m_{T}dm_{T}dy} = \frac{dN/dy}{2\pi T_{\rm eff}(m_0+T_{\rm eff})}e^{-(m_{T}-m_0)/T_{\rm eff}}.
  \end{aligned}
\label{equ:equation5}
\end{equation}
Such a method has been often used to analyze the particle spectra and to understand kinetic freeze-out properties from the data in heavy--ion collisions~\cite{Kaneta:1999lnf,StarWhitePaper}.

A power-law function (shown below) is also used to fit the spectrum in the 60--80\% centrality bin:
\begin{equation}
  \begin{aligned}
\frac{d^2N}{2\pi p_{T}dp_{T}dy} = \frac{dN}{dy}\frac{4(n-1)(n-2)}{2\pi (n-3)^2\langle p_{T} \rangle ^2}\bigg(1+\frac{2p_{T}}{\langle p_{T} \rangle (n-3)}\bigg)^{-n},
  \end{aligned}
\label{equ:equation6}
\end{equation}
where $dN/dy$, $\langle p_{T}\rangle$, and $n$ are three free parameters.

The power-law function fit shows a good description of the 60--80\% centrality data indicating that the $D^0$ meson production in this peripheral bin is close to the expected feature of perturbative QCD. The $D^0$ meson spectra in more central collisions can be well described by the exponential function fit at $m_{T}-m_{0}$\,$<$\,3\,GeV/$c^2$ suggesting the $D^0$ mesons have gained collective motion in the medium evolution in these collisions.

\begin{figure}
\centering
\includegraphics[width=0.43\textwidth]{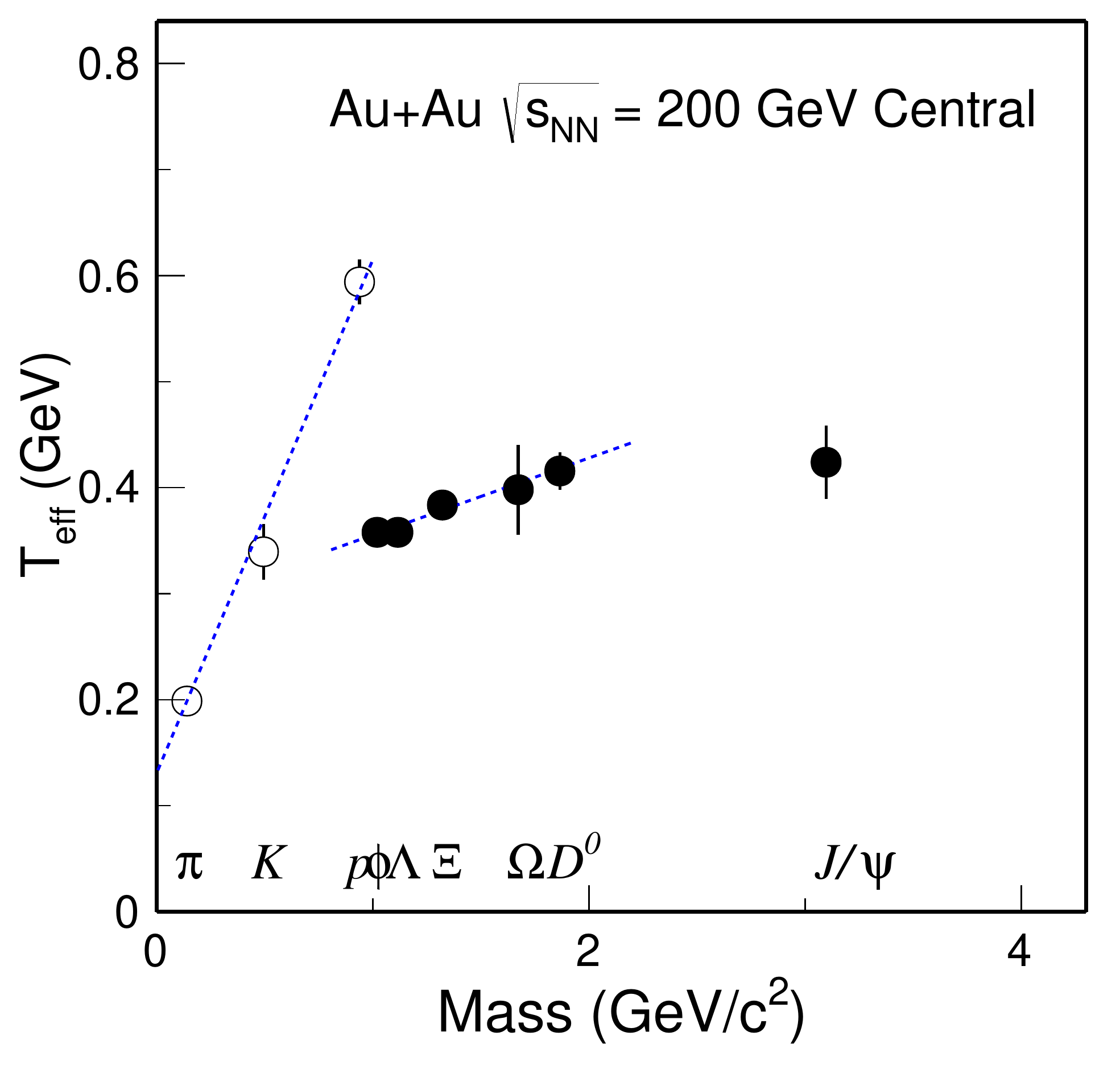}
\caption{Slope parameter $T_{\rm eff}$ for different particles in central Au+Au collisions~\cite{Adams:2003xp,Abelev:2007rw,Adams:2006ke,Adamczyk:2013tvk}. The dashed lines depict linear function fits to $\pi,K,p$ and $\phi,\Lambda,\Xi^{-},\Omega^{-},D^0$ respectively.}
\label{fig:Teff_ALL} 
\end{figure}

The obtained slope parameter $T_{\rm eff}$ for $D^0$ mesons is compared to other light and strange hadrons measured at RHIC. 
Figure~\ref{fig:Teff_ALL} summarizes the slope parameter $T_{\rm eff}$ for various identified hadrons ($\pi^{\pm}$, $K^{\pm}$, $p$/$\bar{p}$, $\phi$, $\Lambda$, $\Xi^-$, $\Omega$, $D^0$ and $J/\psi$) in central Au+Au collisions at $\sqrt{s_{_{\rm NN}}} {\rm = 200\,GeV}$~\cite{Adams:2003xp,Abelev:2007rw,Adams:2006ke,Adamczyk:2013tvk}. Point-by-point statistical and systematic uncertainties are added as a quadratic sum when performing these fits. All fits are performed up to $m_{T} - m_{0} <1\,\rm{GeV}/c^2$ for $\pi,\ K,\ p$, $<2$\,GeV/$c^2$ for $\phi,\ \Lambda,\ \Xi$, and $<3$\,GeV/$c^2$ for $\Omega,\ D^{0},\ J/\psi$, respectively. 

The slope parameter $T_{\rm eff}$ in a thermalized medium can be characterized by the random (generally interpreted as a kinetic freeze-out temperature $T_{\rm fo}$) and collective (radial flow velocity $\langle\beta_{T}\rangle$) components with a simple relation~\cite{StarWhitePaper,Csorgo:1995bi,Kolb:2003dz}:
\begin{equation}
  \begin{aligned}
T_{\rm eff} = T_{\rm fo} + m_0 \langle\beta_{T}\rangle^2.
  \end{aligned}
\label{equ:equation7}
\end{equation}
Therefore, $T_{\rm eff}$ will show a linear dependence as a function of particle mass $m_0$ with a slope that can be used to characterize the radial flow collective velocity.

The data points clearly show two different systematic trends. $\pi,\ K,\ p$ data points follow one linear dependence while $\phi,\ \Lambda,\ \Xi^{-},\ \Omega^{-},\ D^0$ data points follow another linear dependence, as represented by the dashed lines shown in Fig.~\ref{fig:Teff_ALL}. Particles, such as, $\pi,\ K,\ p$ gain radial collectivity through the whole system evolution, therefore the linear dependence exhibits a larger slope. On the other hand the linear dependence of $\phi,\ \Lambda,\ \Xi^{-},\ \Omega^{-},\ D^0$ data points shows a smaller slope indicating these particles may freeze out from the system earlier, and therefore receive less radial collectivity.

\subsubsection{Blast-wave fit}
\label{result:collectivity:BW}

\begin{figure}
\centering
\includegraphics[width=0.43\textwidth]{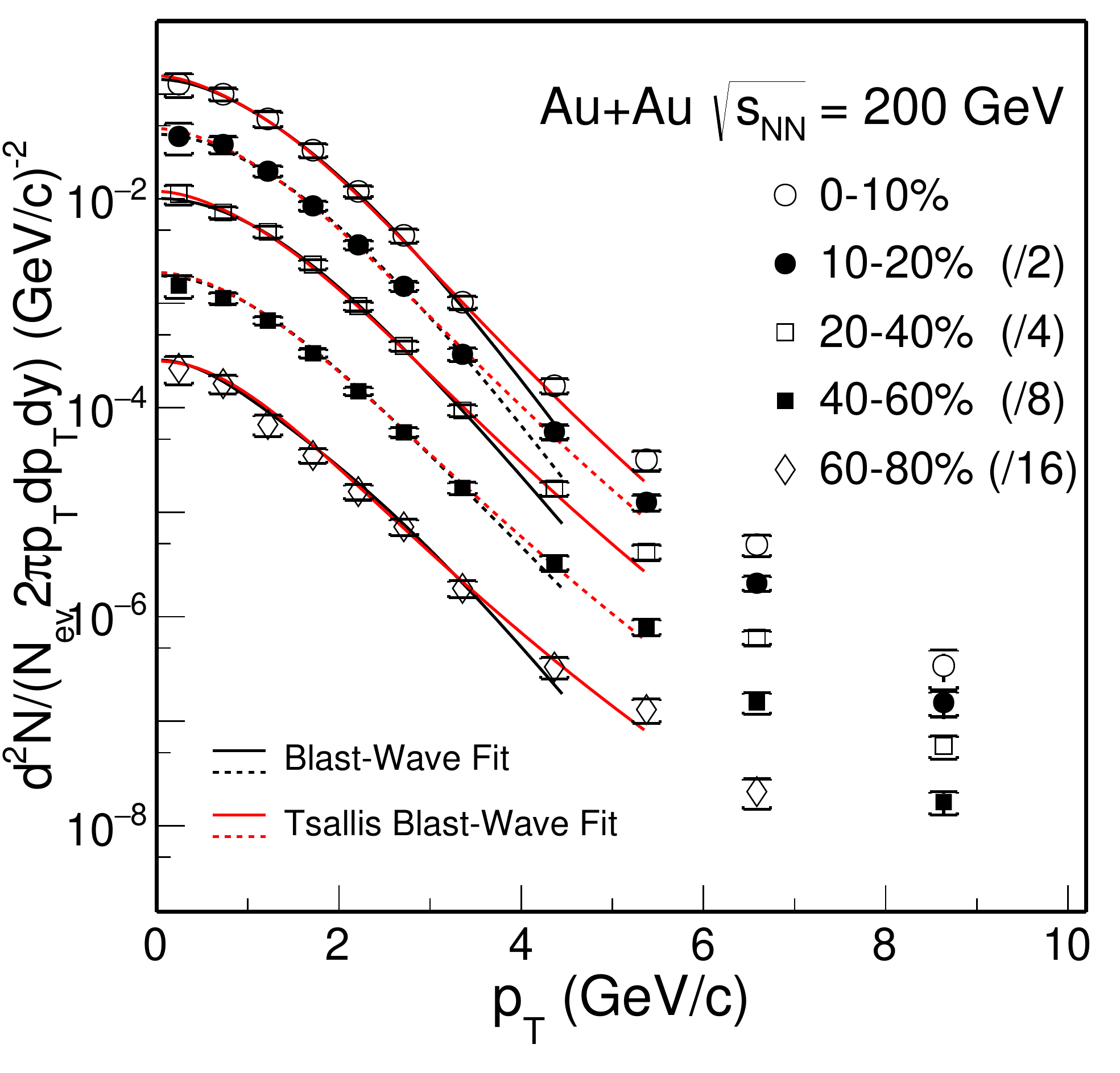}
\caption{$D^{0}$ invariant yield at mid-rapidity ($|y|$\,$<$\,1) vs. transverse momentum for different centrality classes. Black and red lines depict Blast-Wave (BW) and Tsallis Blast-Wave (TBW) fits for each centrality bin respectively.}
\label{fig:BWFit} 
\end{figure}

The Blast-Wave (BW) model is extensively used to study the particle kinetic freeze-out properties~\cite{Adams:2003xp,Adamczyk:2017iwn}. Assuming a hard-sphere uniform particle source with a kinetic freeze-out temperature $T_{\rm kin}$ and a transverse radial flow velocity $\beta$, the particle transverse momentum spectral shape is given by~\cite{Schnedermann:1993ws}:

\begin{equation}
  \begin{aligned}
  & \frac{dN}{p_{T}dp_{T}} = \frac{dN}{m_{T}dm_{T}} \propto \\
  & \int_0^R rdr m_{T} I_0\bigg(\frac{p_{T}\sinh\rho}{T_{\rm kin}}\bigg) K_1\bigg(\frac{m_{T}\cosh\rho}{T_{\rm kin}}\bigg),
  \end{aligned}
\label{equ:equation8}
\end{equation}
where $\rho = \tanh^{-1}\beta$, and $I_0$ and $K_1$ are the modified Bessel functions. The flow velocity profile is taken as:

\begin{equation}
  \begin{aligned}
    \beta = \beta_{s}\left(\frac{r}{R}\right)^{n},
  \end{aligned}
\label{equ:equation9}
\end{equation}
where $\beta_{s}$ is the maximum velocity at the surface and $r/R$ is the relative radial position in the thermal source. The choice of $R$ only affects the overall spectrum magnitude while the spectrum shape constrains the three free parameters $T_{\rm kin}$, $\langle\beta\rangle=2/(2+n)\beta_{s}$, and $n$.

In the modified Tsallis Blast-Wave (TBW) model, an additional parameter $q$ is introduced to account for the non-equilibrium feature in a system~\cite{Tang:2008ud}. The particle transverse momentum spectral shape is then described by: 

\begin{equation}
  \begin{aligned}
    & \frac{dN}{m_{T}dm_{T}} \propto m_{T}\int_{-Y}^{+Y}\cosh(y)dy \int_{-\pi}^{+\pi} d\phi \int_0^R rdr \\
    & \bigg(1+\frac{q-1}{T_{\rm kin}}\big(m_{T}\cosh(y)\cosh(\rho)-p_{T}\sinh(\rho)\cos(\phi)\big)\bigg)^{\textstyle -\frac{1}{q-1}}
  \end{aligned}
\label{equ:equation10}
\end{equation}
When $q-1$ approaches zero, the TBW function returns to the regular BW function shown in Eq.~\ref{equ:equation8}.

Figure~\ref{fig:BWFit} shows the Blast-Wave and Tsallis Blast-Wave fits to the data in different centrality bins. The parameter $n$ in these fits is fixed to be 1 due to the limited number of data points and is inspired by the fit result for light-flavor hadrons ($\pi,K$ and $p$)~\cite{Tang:2008ud}. The $p_{T}$ range in the BW fits is restricted to be less than 3$m_{0}$ where $m_{0}$ is the rest mass of $D^0$ mesons.

\begin{figure}
\centering
\includegraphics[width=0.43\textwidth]{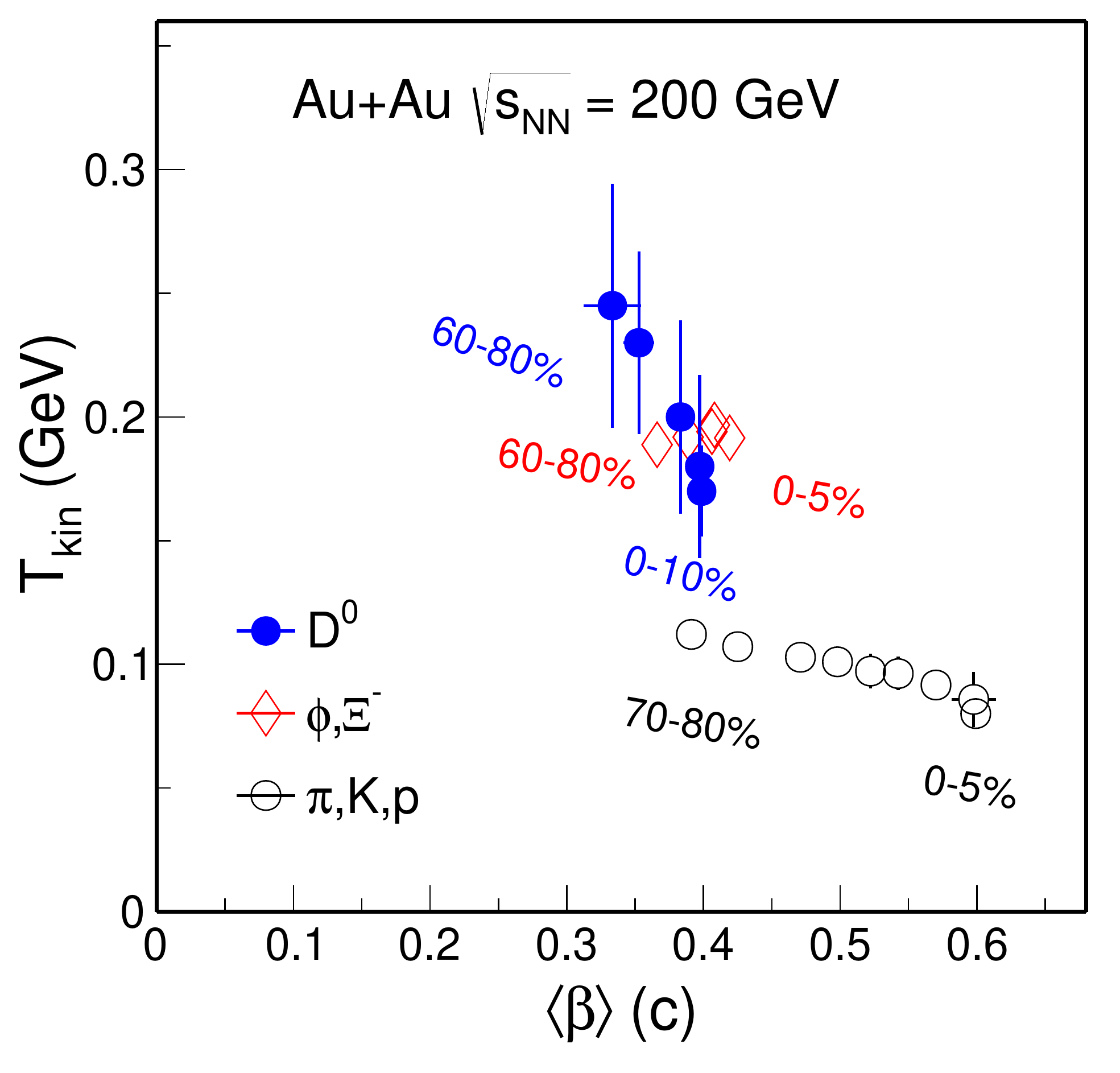}
\caption{Results of $T_{\rm kin}$ vs. $\langle\beta\rangle$ from the Blast-Wave model fits to different groups of particles. The data points for each group of particles present the results from different centrality bins with the most central data point at the largest $\langle\beta\rangle$.}
\label{fig:BWFitSummary} 
\end{figure}

Figure~\ref{fig:BWFitSummary} summarizes the fit parameters $T_{\rm kin}$ vs. $\langle\beta\rangle$ from the Blast-Wave model fits to different groups of particles: black markers for the simultaneous fit to $\pi,\ K,\ p$; red markers for the simultaneous fit to $\phi,\ \Xi^-$ and blue markers for the fit to $D^0$.\,\,\,The data points for each group of particles represent the fit results from different centrality bins with the most central data point at the largest $\langle\beta\rangle$ value.\,\,\,Similar as in the fit to the $m_{T}$ spectra, point-by-point statistical and systematic uncertainties are added in quadrature when performing the fit. The fit results for $\pi,\ K,\ p$ are consistent with previously published results~\cite{Tang:2008ud}. The fit results for multi-strangeness particles $\phi,\ \Xi^{-}$, and for $D^0$ show much smaller mean transverse velocity $\langle\beta\rangle$ and larger kinetic freeze-out temperature, suggesting these particles decouple from the system earlier and gain less radial collectivity compared to light hadrons. The resulting $T_{\rm kin}$ parameters for $\phi,\ \Xi^-$ and for $D^0$ are close to the pseudocritical temperature $T_{c}$ calculated from a lattice QCD calculation at zero baryon chemical potential~\cite{Bazavov:2011nk}, indicating negligible contribution from the hadronic stage to the observed radial flow of these particles.\,\,Therefore, the collectivity that $D^0$ mesons obtain is mostly through the partonic stage re-scatterings in the QGP phase. 

Table~\ref{table:TBW_fit} lists the fitting parameters, $\langle\beta\rangle$ and $(q-1)$ for the $D^0$ data in different centralities. Results show a similar trend as the regular BW fit, i.e. the most central data point is located at the largest $\langle\beta\rangle$ value. The $(q-1)$ parameter in TBW, which characterizes the degree of non-equilibrium in a system, is found to be close to zero, indicating that the system is approaching thermalization in these collisions.

\begin{table}[t]
\centering{
  \caption{ $\langle\beta\rangle$ and $(q-1)$ from the Tsallis Blast-Wave fits to the $D^0$ data in different centralities.}
\begin{tabular}{rcccccccc} \hline \hline
  \hspace{0.5cm}Centrality\hspace{1cm} & \multicolumn{3}{c}{$\langle\beta\rangle$($c$)} & \hspace{1cm} & \multicolumn{3}{c}{$q-1$} & \\ \hline
  0--10 \%\hspace{1cm}     & 0.263 & $\pm$ & 0.018 & \hspace{0.5cm} & 0.066 & $\pm$ & 0.008 \\
  10--20 \%\hspace{1cm}    & 0.255 & $\pm$ & 0.022 & \hspace{0.5cm} & 0.068 & $\pm$ & 0.010 \\
  20--40 \%\hspace{1cm}    & 0.264 & $\pm$ & 0.015 & \hspace{0.5cm} & 0.070 & $\pm$ & 0.007 \\
  40--60 \%\hspace{1cm}    & 0.251 & $\pm$ & 0.023 & \hspace{0.5cm} & 0.074 & $\pm$ & 0.011 \\
  60--80 \%\hspace{1cm}    & 0.217 & $\pm$ & 0.037 & \hspace{0.5cm} & 0.075 & $\pm$ & 0.010  \\ \hline \hline
\end{tabular}
\label{table:TBW_fit}
}
\end{table}

\subsection{Nuclear Modification Factors - $R_{\rm CP}$ and $R_{\rm AA}$}
\label{result:RCP}

\begin{figure}
\centering
\includegraphics[width=0.45\textwidth]{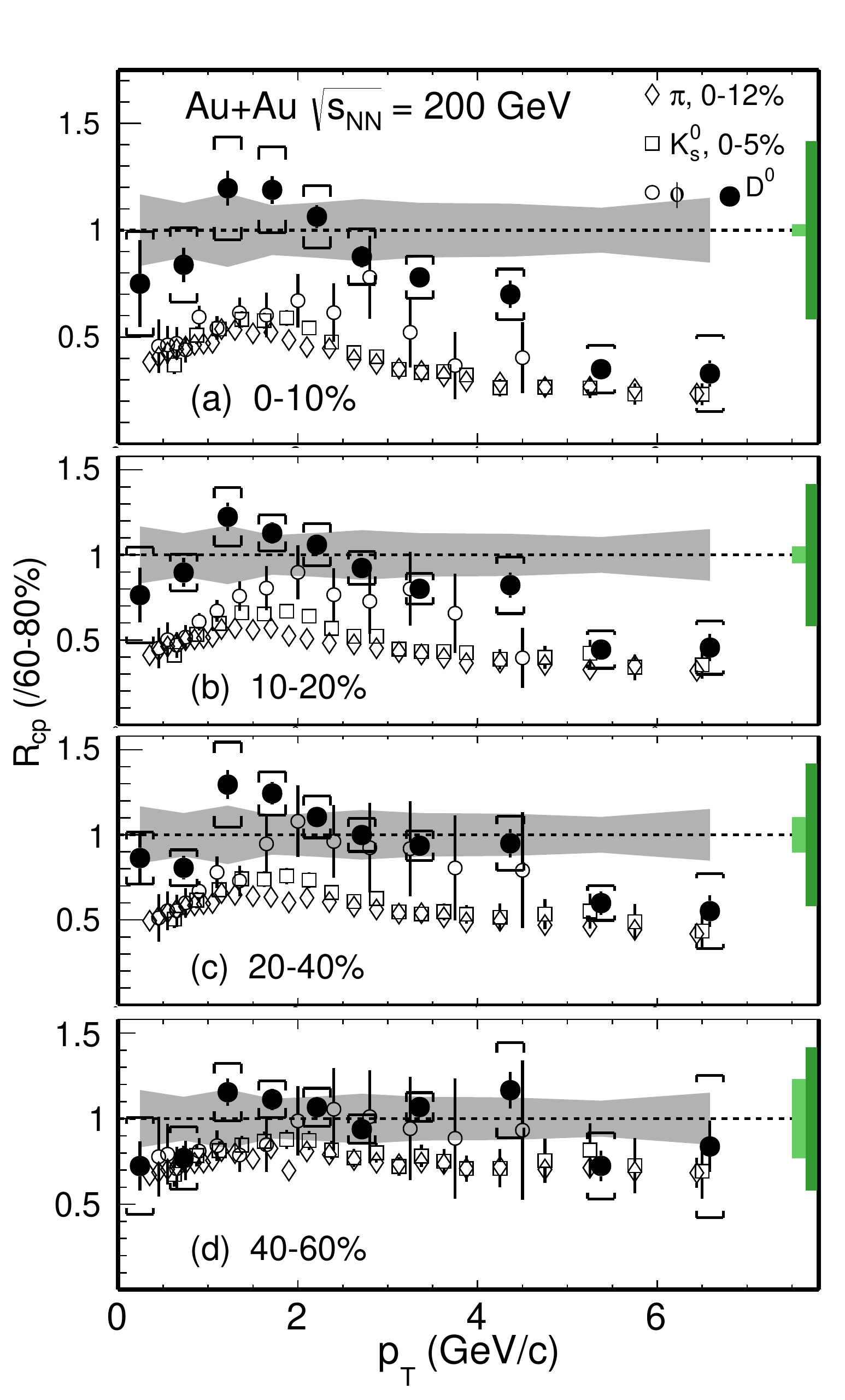}
\caption{$D^{0}$ $R_{\rm CP}$ with the 60--80\% spectrum as the reference for different centrality classes in Au+Au collisions compared to that of other light and strange mesons ($\pi^{\pm}$, $K^0_{S}$ and $\phi$)~\cite{Adams2006_Identified,Abelev2009,Agakishiev2012}. The statistical and systematic uncertainties are shown as error bars and brackets on the data points. The grey bands around unity depict the systematic uncertainty due to vertex resolution correction, mostly from the 60--80\% reference spectrum. The light and dark green boxes on the right depict the normalization uncertainties in determining the $N_{\rm bin}$ for each centrality (light green) and the 60--80\% centrality bin (dark green), respectively.}
\label{fig:D0_Rcp} 
\end{figure}

Figure~\ref{fig:D0_Rcp} shows the calculated $D^0$ $R_{\rm CP}$ (see Eq.~\ref{equ:equation3}) with the 60--80\% peripheral bin as the reference for different centrality bins 0--10\%, 10--20\%, 20--40\% and 40--60\%; the results are compared to other light and strange flavor mesons. The grey bands around unity depict the vertex resolution correction uncertainty on the measured $D^0$ data points, mostly originating from the 60--80\% reference spectrum. The dark and light green boxes around unity on the right side indicate the global $N_{\rm bin}$ systematic uncertainties for the 60--80\% centrality bin and for the corresponding centrality bin in each panel. The global $N_{\rm bin}$ systematic uncertainties should be applied to the data points of all particles in each panel.

The measured $D^0$ $R_{\rm CP}$ in central 0--10\% collisions shows a significant suppression at $p_{T}>$ 5\,GeV/$c$. The suppression level is similar to that of light and strange flavor mesons and the $R_{\rm CP}$ suppression gradually decreases when moving from central collisions to mid-central and peripheral collisions. The $D^0$ $R_{\rm CP}$ for $p_{T}$\,$<$\,4\,GeV/$c$ is consistent with no suppression, in contrast to light-flavor hadrons. Comparisons to dynamic model calculations for the $D^0$ $R_{\rm CP}$ will be discussed in Sec.~\ref{result:theory}.

\begin{figure}
\centering
\includegraphics[width=0.45\textwidth]{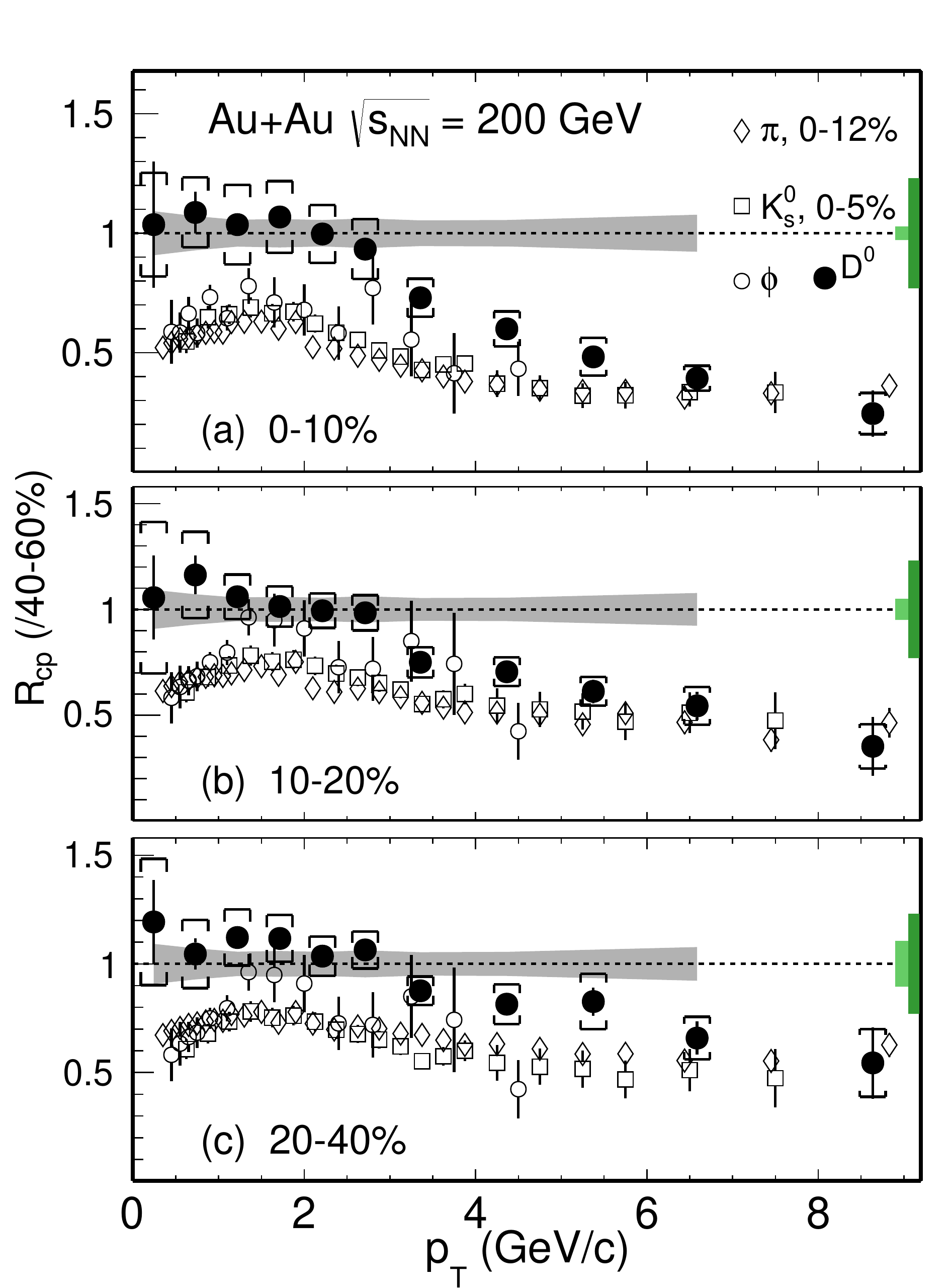}
\caption{$D^{0}$ $R_{\rm CP}$ with the 40--60\% spectrum as the reference for different centrality classes in Au+Au collisions compared to that of other light and strange mesons ($\pi^{\pm}$, $K^0_{S}$ and $\phi$)~\cite{Adams2006_Identified,Abelev2009,Agakishiev2012}. The statistical and systematic uncertainties are shown as error bars and brackets on the data points. The grey bands around unity depict the systematic uncertainty due to vertex resolution correction, mostly from the 40--60\% reference spectrum. The light and dark green boxes on the right depict the normalization uncertainties in determining the $N_{\rm bin}$ for each centrality (light green) and the 40--60\% centrality bin (dark green), respectively.}
\label{fig:D0_Rcp2} 
\end{figure}

The precision of the 60--80\% centrality spectrum is limited due to the large systematic uncertainty in determining the $N_{\rm bin}$ based on the MC Glauber model. Figure~\ref{fig:D0_Rcp2} shows the $D^0$ $R_{\rm CP}$ for different centralities as a function of $p_{T}$ with the 40--60\% centrality spectrum as the reference. The grey bands around unity in each panel represent the systematic uncertainties due to the vertex resolution contribution from the 40--60\% centrality. The green boxes around unity depict the global $N_{\rm bin}$ systematic uncertainties for the 40--60\% centrality bin and for each corresponding centrality bin. As a comparison, $R_{\rm CP}$ of charged pions, $K_{s}^{0}$ and $\phi$ in the corresponding centralities are also plotted in each panel. With much smaller systematic uncertainties, the observations seen before using the 60--80\% centrality spectrum as the reference still hold.

\begin{figure}
\centering
\includegraphics[width=0.45\textwidth]{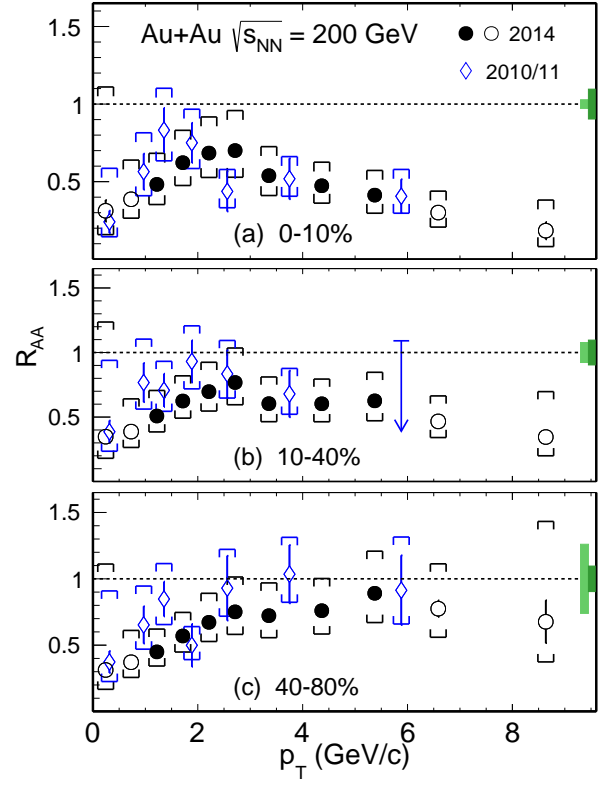}
\caption{$D^{0}$ $R_{\rm AA}$ in Au+Au collisions at $\sqrt{s_{_{\rm NN}}}$\,=\,200\,GeV for 0--10\% (a), 10--40\% (b) and 40--80\% (c) centrality bins, respectively. The first two and last two data points are presented as empty circles, indicating that the $p$+$p$ reference is extrapolated into these $p_T$ ranges. The statistical and systematic uncertainties are shown as error bars and brackets on the data points. The light and dark green boxes on the right depict the normalization uncertainties in determining the $N_{\rm bin}$ in Au+Au collisions and the total inelastic cross section in $p$+$p$ collisions, respectively.}
\label{fig:D0_RAA} 
\end{figure}

\begin{figure}
\centering
\includegraphics[width=0.45\textwidth]{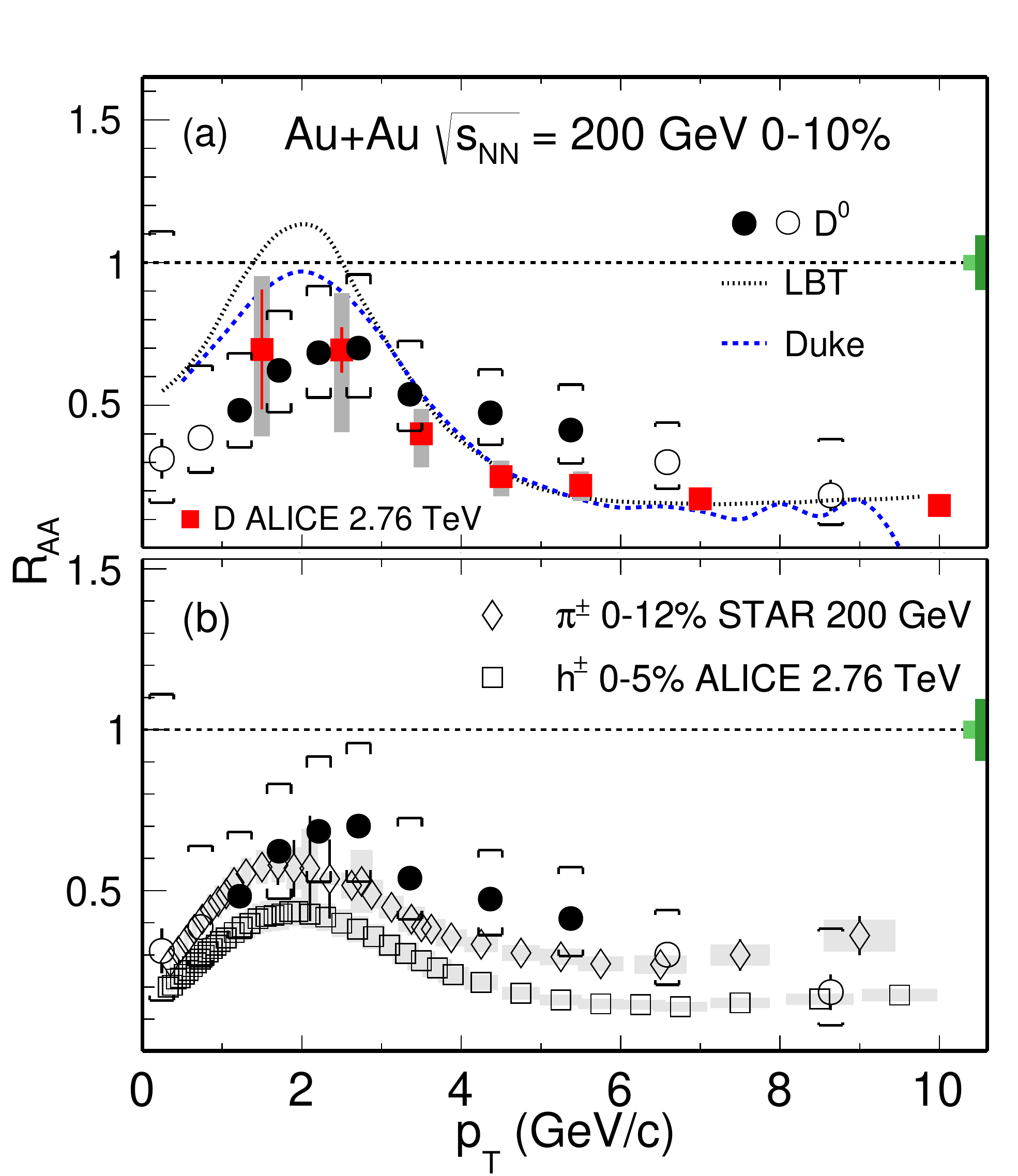}
  \caption{$D^{0}$ $R_{\rm AA}$ in 0--10\% Au+Au collisions at ${\sqrt{s_{\rm NN}} = \rm{200\,GeV}}$ compared to the ALICE $D$-meson result in 0--10\% Pb + Pb collisions at $\sqrt{s_{_{\rm NN}}}$ = 2.76\,TeV (a) and charged hadrons from ALICE and $\pi^{\pm}$ from STAR (b). Also shown in panel (a) are the model calculations from the LBT and Duke groups~\cite{Cao:2016gvr,LBT:private,Xu:2017obm}. Notations for statistical and systematic uncertainties are the same as in previous figures.}
\label{fig:D0_RAA_LHC} 
\end{figure}

Figure~\ref{fig:D0_RAA} shows the calculated $D^0$ $R_{\rm AA}$ (see Eq.~\ref{equ:equation0}) with the $p$+$p$ measurement~\cite{Star_D_pp} as the reference for different centrality bins 0--10\% (a), 10--40\% (b) and 40--80\% (c), respectively. The new $R_{\rm AA}$ measurements are also compared to the previous Au+Au measurements using the STAR TPC after the recent correction~\cite{Star_D_RAA}. The $p$+$p$ $D^0$ reference spectrum is updated using the latest global analysis of charm fragmentation ratios from~\cite{charm_frag} and also by taking into account the $p_T$ dependence of the fragmentation ratio between $D^0$ and $D^{*\pm}$ from PYTHIA. The new measurement with the HFT detector shows a nice agreement with the measurement without the HFT. The brackets on the data points depict the total systematic uncertainty dominated by the uncertainty in the $p$+$p$ reference spectrum. The first two and last two data points are empty circles indicating those are calculated with an extrapolated $p$+$p$ reference. The light and dark green boxes around unity on the right side indicate the global $N_{\rm bin}$ systematic uncertainties for the corresponding centrality bin in each panel and the total inelastic cross section uncertainty in $p$+$p$ collisions.

The measured $D^0$ $R_{\rm AA}$ in central (0--10\%) and mid-central (10--40\%) collisions show a significant suppression at the high $p_{T}$ range which reaffirms the strong interactions between charm quarks and the medium, while the new Au+Au data points from this analysis contain much improved precision. Figure~\ref{fig:D0_RAA_LHC} shows the $D^0$ $R_{\rm AA}$ in the 0--10\% most central collisions compared to that of (a) average $D$-meson from ALICE and (b) charged hadrons from ALICE and $\pi^{\pm}$ from STAR~\cite{Alice_D_RAA_2,Alice_hadron_RAA,StarPi0}. The $D^0$ $R_{\rm AA}$ from this measurement is comparable to that from the LHC measurements in Pb+Pb collisions at $\sqrt{s_{_{\rm NN}}}$ = 2.76\,TeV despite the large energy difference between these measurements.
The comparison to that of light hadrons shows a similar suppression at high $p_{T}$, while in the intermediate range, $D^0$ mesons seem to be less suppressed. From low to intermediate $p_{T}$ region, the $D^0$ $R_{\rm AA}$ in the central 0-10\% collisions shows a characteristic bump structure that is consistent with the expectation from model predictions that charm quarks gain sizable collective motion during the medium evolution. The large uncertainty in the $p$+$p$ baseline need to be further reduced before making more quantitative conclusions.

\subsection{$\overline{D}^{0}$ and $D^{0}$ spectra and double ratio}
\label{result:D0barD0ratio} 

\begin{figure}
\centering
\includegraphics[width=0.45\textwidth]{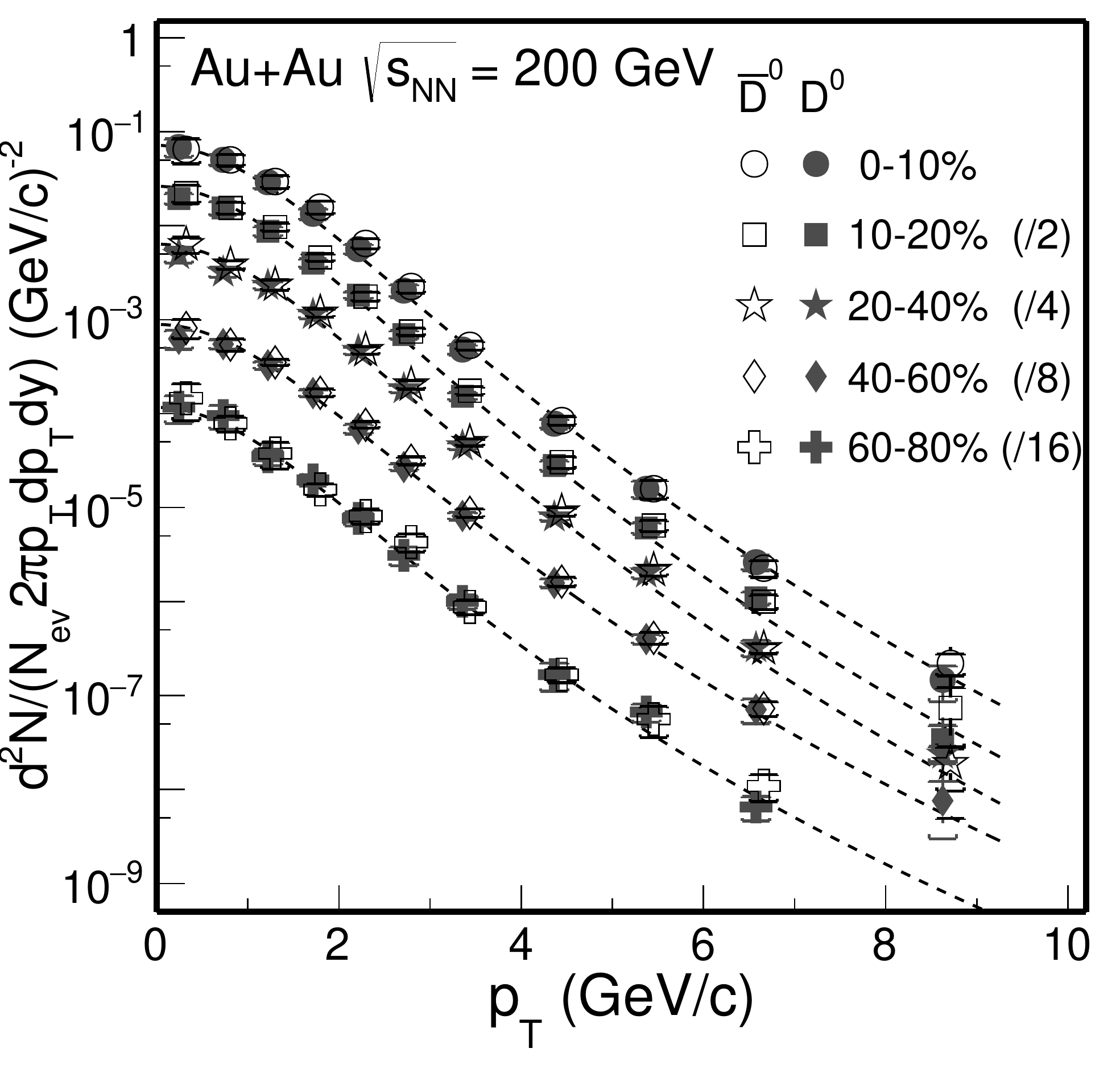}
\caption{$D^{0}$ and $\overline{D}^{0}$ invariant yields at mid-rapidity ($|y|$\,$<$\,1) vs. transverse momentum for different centrality classes. Error bars (not visible for many data points) indicate statistical uncertainties and brackets depict systematic uncertainties. Global systematic uncertainties in $B.R.$ and $N_{\rm bin}$ are not plotted. Solid lines depict Levy function fits.}
\label{fig:D0_spectra_bothposneg} 
\end{figure}

\begin{figure}
\centering
\includegraphics[width=0.45\textwidth]{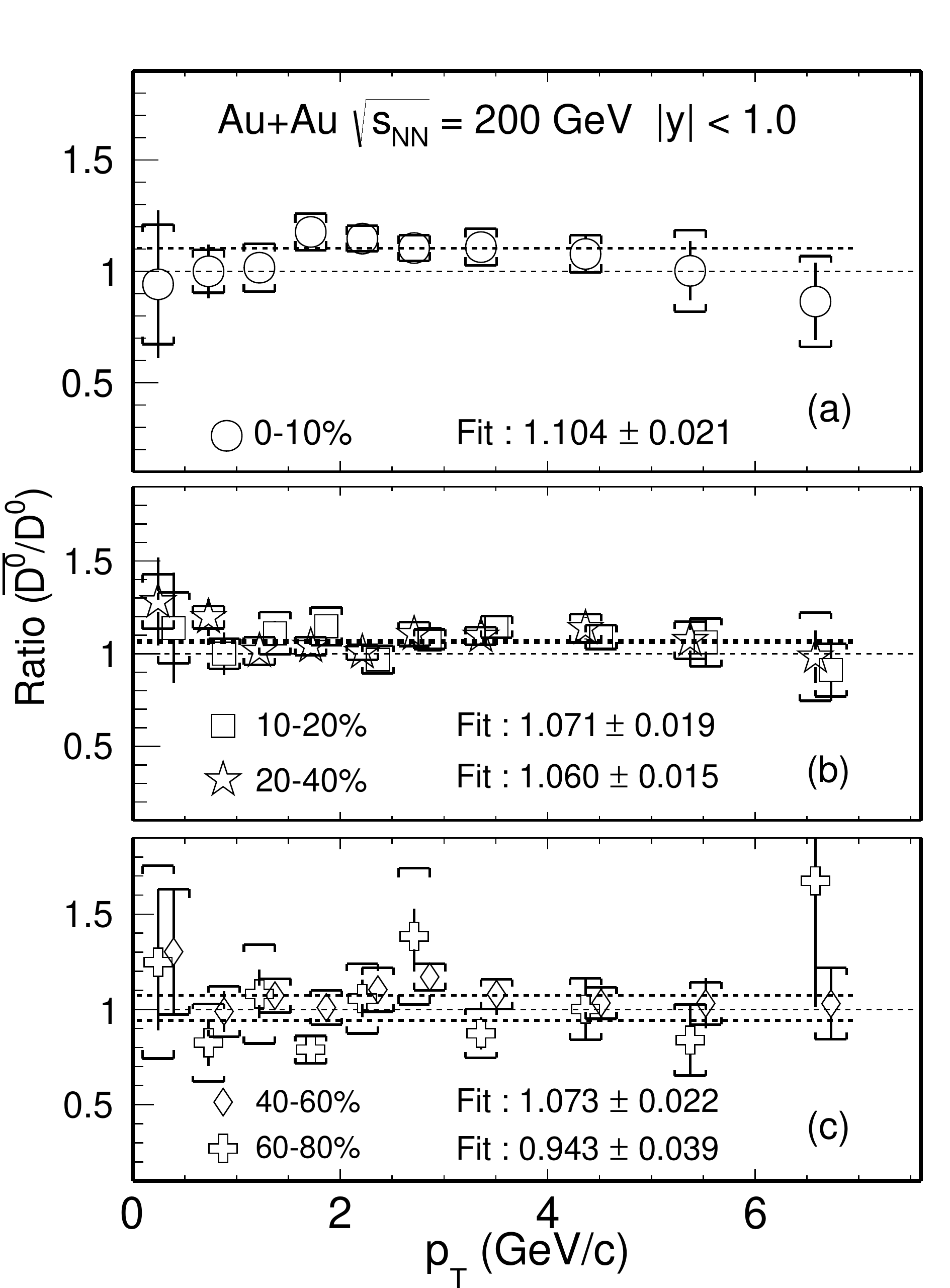}
\caption{$\overline{D}^{0}$/$D^{0}$ invariant yield ratio at mid-rapidity ($|y|$\,$<$\,1) vs. transverse momentum for different centrality classes. Error bars indicate statistical uncertainties and brackets depict systematic uncertainties. Dashed lines depict constant function fits to the $\overline{D}^{0}$/$D^{0}$ ratios.}
\label{fig:D0_spectra_ratioposneg} 
\end{figure}

Figure~\ref{fig:D0_spectra_bothposneg} shows the $p_{T}$ spectra of $\overline{D}^{0}$ and $D^0$ mesons separately in 0--10\%, 10--20\%, 20--40\%, 40--60\% and 60--80\% centrality bins. Figure~\ref{fig:D0_spectra_ratioposneg} shows the $\overline{D}^{0}$/$D^{0}$ ratio in various centrality bins. Dashed lines represent constant function fits to the $\overline{D}^{0}$/$D^{0}$ ratio in each centrality bin by combining the point-by-point statistical and systematic uncertainties. The $\overline{D}^0/D^0$ ratio has a small but significant deviation from unity in central and mid-central collisions. Table~\ref{table:d0bard0ratio} lists the fitted results for the $\overline{D}^{0}$/$D^0$ ratio from various centralities. In the most central collisions, $\overline{D}^{0}$ yield is higher than the $D^0$ yield by $\sim$4.9$\sigma$. The total charm quark and anti-charm quark should be conserved since they are created in pairs. A thermal model calculation predicts that the $\Lambda_{c}^-$/$\Lambda_{c}^+$ ratio will be smaller than unity at RHIC due to the finite baryon density~\cite{ANDRONIC200336}. This will then yield more 
$\overline{D}^{0}$ mesons formed than $D^0$ mesons in Au+Au collisions at RHIC. To verify the total charm quark conservation, one would need precise measurements of $D^{+}$/$D^{-}$, $D_{s}^{+}$/$D_{s}^{-}$ as well as $\Lambda_{c}^+$/$\Lambda_{c}^-$ ratios in the future.

\begin{table}[t]
\centering{
  \caption{ $\overline{D}^{0}$/$D^0$ ratio for various centrality bins obtained from the fit to data in Fig.~\ref{fig:D0_spectra_ratioposneg}.}
\begin{tabular}{rcccccccc} \hline \hline
  \hspace{1cm}Centrality\hspace{1cm} & \multicolumn{3}{c}{$\overline{D}^0$/$D^0$} & \hspace{1cm} \\ \hline
0--10 \%\hspace{1cm}      & 1.104 & $\pm$ & 0.021 \\
10--20 \%\hspace{1cm}     & 1.071 & $\pm$ & 0.019 \\
20--40 \%\hspace{1cm}     & 1.060 & $\pm$ & 0.015 \\
40--60 \%\hspace{1cm}     & 1.073 & $\pm$ & 0.022 \\
60--80 \%\hspace{1cm}     & 0.943 & $\pm$ & 0.039  \\ \hline \hline
\end{tabular}
\label{table:d0bard0ratio}
}
\end{table}

\subsection{Comparison to Models}
\label{result:theory}

\begin{figure}
\centering
\includegraphics[width=0.45\textwidth]{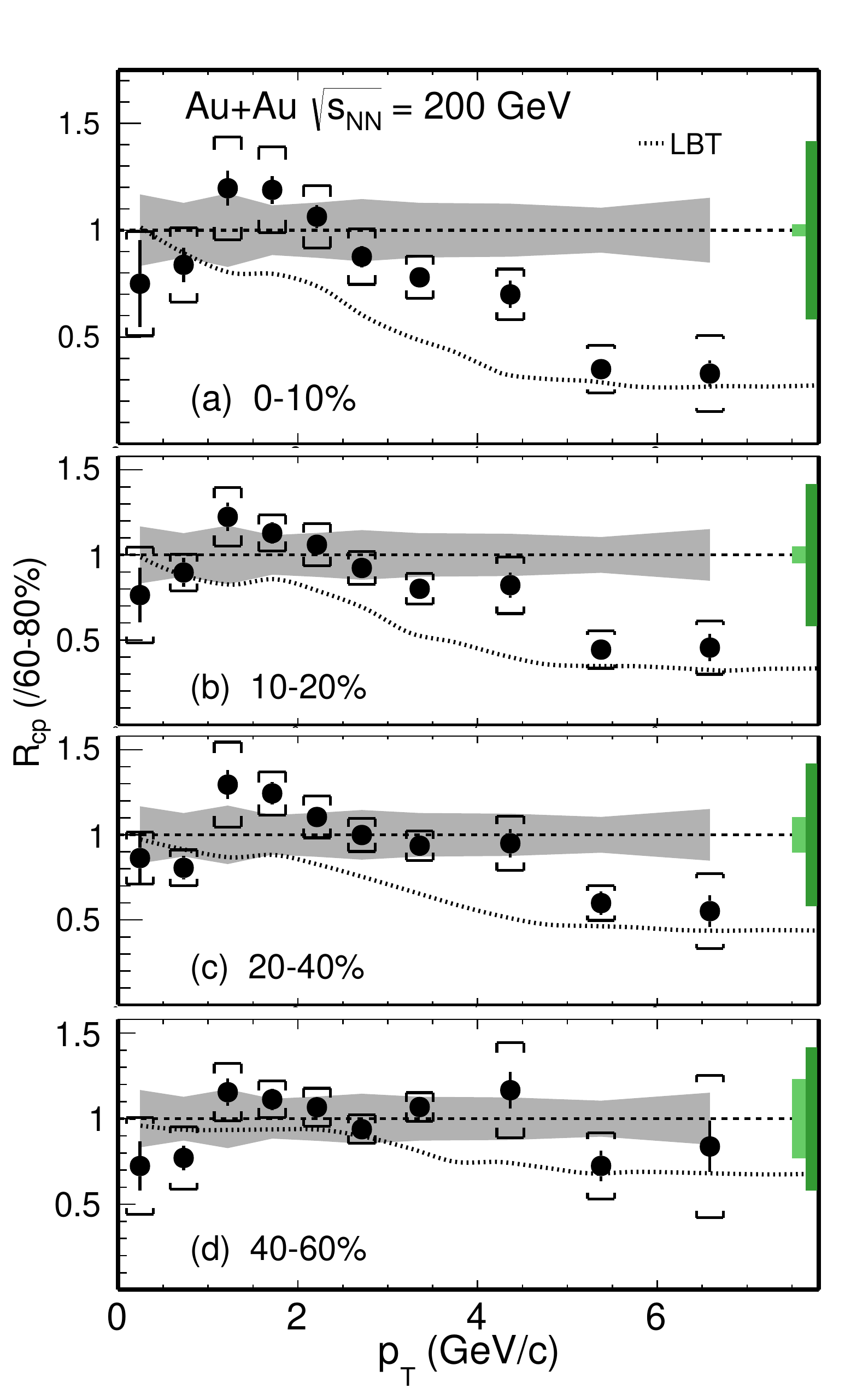}
\caption{$D^{0}$ $R_{\rm CP}$ with the 60--80\% spectrum as the reference for different centrality classes compared to the LBT model calculations shown by dashed lines~\cite{Cao:2016gvr,LBT:private}. Data points shown here are the same as in Fig.~\ref{fig:D0_Rcp}.}
\label{fig:D0_Rcp11} 
\end{figure}

\begin{figure}
\centering
\includegraphics[width=0.45\textwidth]{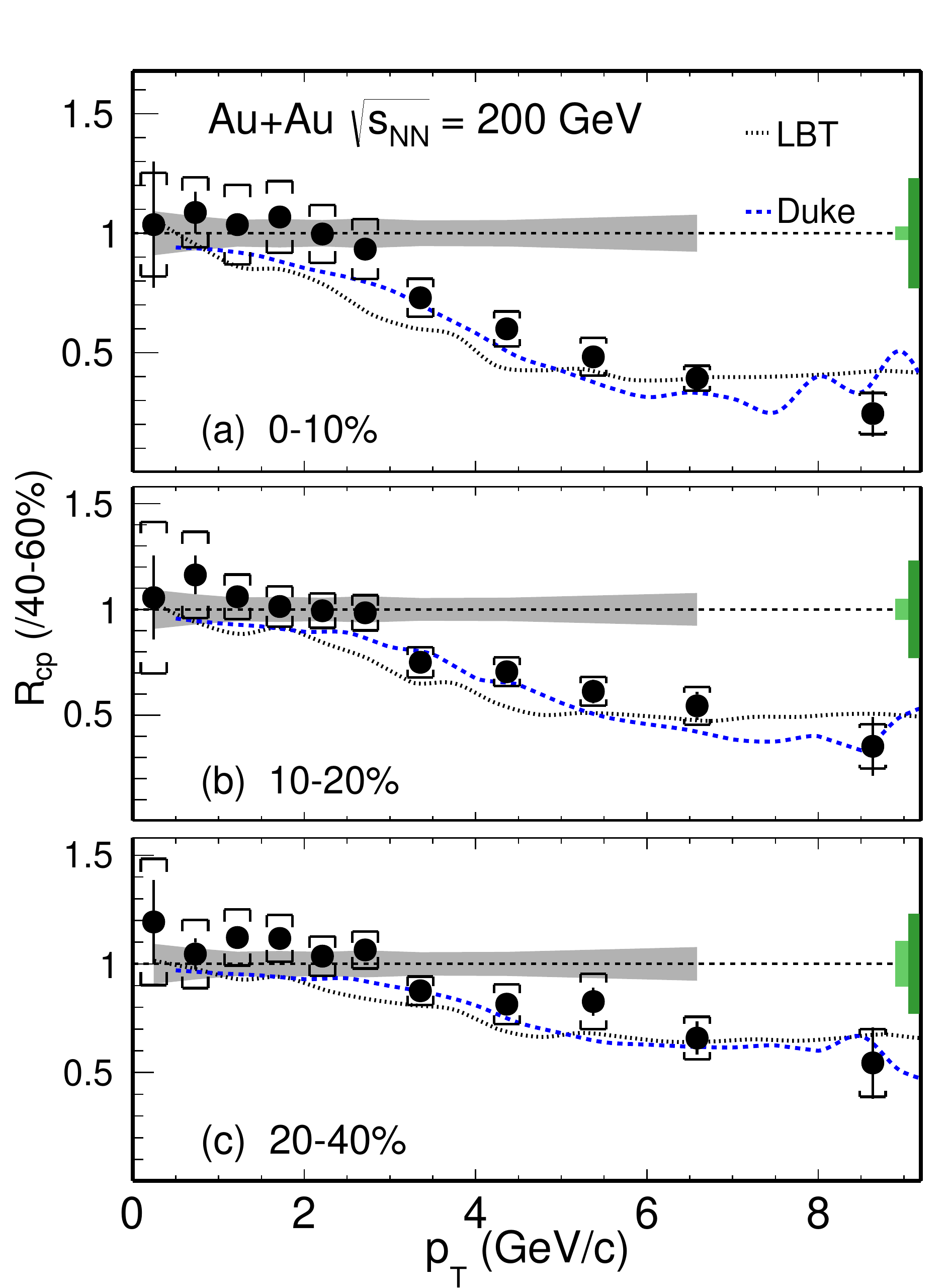}
  \caption{$D^{0}$ $R_{\rm CP}$ with the 40--60\% spectrum as the reference for different centrality classes compared to model calculations from LBT (black dotted lines) and the Duke (blue dashed lines) groups~\cite{Cao:2016gvr,LBT:private,Xu:2017obm}. Data points shown here are the same as in Fig.~\ref{fig:D0_Rcp2}.}
\label{fig:D0_Rcp22} 
\end{figure}

Over the past several years, there have been rapid developments in the theoretical calculations on the charm hadron production~\cite{Rapp:2018qla,Cao:2018ews}. Here we compare our measurements to several recent calculations based on the Duke model and the Linearized Boltzmann Transport (LBT) model~\cite{Cao:2016gvr,LBT:private,Xu:2017obm}.

The Duke model~\cite{Duke,Xu:2017obm} uses a Langevin stochastic simulation to trace the charm quark propagation inside the QGP medium. Both collisional and radiative energy losses are included in the calculation and charm quarks are hadronized via a hybrid approach combining both coalescence and fragmentation mechanisms. The bulk medium is simulated using a viscous hydrodynamic evolution followed by a hadronic cascade evolution using the UrQMD model~\cite{urQMD}. The charm quark interaction with the medium is characterized using a temperature and momentum-dependent diffusion coefficient. The medium parameters have been constrained via a statistical Bayesian analysis by fitting the previous experimental data of $R_{\rm AA}$ and $v_{2}$ of light, strange and charm hadrons~\cite{Xu:2017obm}. The extracted charm quark spatial diffusion coefficient at zero momentum $2\pi TD_s|_{p=0}$ is about 1--3 near $T_{c}$ and exhibits a positive slope for its temperature dependence above $T_{c}$.

The Linearized Boltzmann Transport (LBT) calculation~\cite{Cao:2016gvr} extends the LBT approach developed before to include both light and heavy flavor parton evolution in the QGP medium. The transport calculation includes all $2\rightarrow 2$ elastic scattering processes for collisional energy loss and the higher-twist formalism for medium induced radiative energy loss. It uses the same hybrid approach as in the Duke model for charm quark hadronization. The heavy quark transport is coupled with a 3D viscous hydrodynamic evolution which is tuned for light flavor hadron data. The charm quark spatial diffusion coefficient is estimated via the equation $2\pi TD_s =8\pi/\hat{q}$ ($\hat{q}$ is the quark transport coefficient due to elastic scatterings) at parton momentum $p$\,$=$\,10\,GeV/$c$. The $2\pi TD_s$ is $\sim$3 at $T_{c}$ and increases to $\sim$6 at $T = 500$\,MeV~\cite{LBT:private}.

Figures~\ref{fig:D0_Rcp11} and~\ref{fig:D0_Rcp22} show the measured $D^0$ $R_{\rm CP}$ compared to the Duke and LBT model calculations with the 60--80\% and 40--60\% reference spectra respectively. The $R_{\rm CP}$ curves from these models are calculated based on the $D^0$ spectra provided by each group~\cite{Cao:2016gvr,LBT:private,Xu:2017obm}. The Duke model did not calculate the spectra in the 60--80\% centrality bin due to a concern about the viscous hydrodynamic implementation. In Fig.~\ref{fig:D0_RAA_LHC} for the most central collisions, there are also calculations for the $D^0$ $R_{\rm AA}$ from the Duke and LBT groups, respectively. These two models also have the predictions for the $D^0$ $v_2$ measurements for Au+Au collisions at ${\sqrt{s_{\rm NN}} = \rm{200\,GeV}}$~\cite{Star_D_v2}. 
Both model calculations match our new measured $R_{\rm CP}$ data well. The much improved precision of these new measurements are expected to further constrain the theoretical model uncertainties in these calculations.

\section{Summary}
\label{summary}

In summary, we report the improved measurement of $D^0$ production invariant yield at mid-rapidity ($|y|$\,$<$\,1) in Au+Au collisions at ${\sqrt{s_{\rm NN}} = \rm{200\,GeV}}$ with the STAR HFT detector.\,\,$D^0$ invariant yields are presented as a function of $p_{T}$ in various centrality classes.\,\,The $p_{T}$--integrated $D^0$ production cross section per nucleon-nucleon collisions in mid-central and central Au+Au collisions seem to be smaller than that measured in $p$+$p$ collisions by 1.5$\sigma$, indicating that CNM effects and/or hadronization through quark coalescence may play an important role in Au+Au collisions. This calls for precise measurements of $D^0$ production in $p/d$+A collisions to understand the CNM effects as well as other charm hadron states in heavy-ion collisions to better constrain the total charm quark yield. 

The $\overline{D}^{0}$ yield is observed to be higher than the $D^0$ yield in the most central collisions, by $\sim$4.9$\sigma$ on average. This is potentially consistent with the expectation, due to the finite baryon density of the system at RHIC, that the $\Lambda_{c}^-$/$\Lambda_{c}^+$ ratio should be smaller than unity, which would result in larger $\overline{D}^{0}$ yield than $D^0$.

The $D^0$ spectra at low $p_{T}$ and $m_{T}$ regions are fit by the exponential function and the (Tsallis) Blast-Wave model to study the $D^0$ meson radial collectivity. The slope parameter extracted from the exponential function fit for $D^0$ mesons follows the same linearly increasing trend vs. particle mass as $\phi$, $\Lambda$, $\Xi^-$, $\Omega^-$ particles, but is different from the trend of $\pi,K$ and $p$ particles. The extracted kinetic freeze-out temperature and transverse velocity from the Blast-Wave model fit are comparable to the fit results of $\phi,\Xi^-$ multi-strange-quark hadrons, but different from those of $\pi,K$ and $p$.\,\,These observations suggest that $D^0$ hadrons show a radial collective behavior with the medium, but freeze out from the system earlier and gain less radial collectivity compared to $\pi,K$ and $p$ particles. This observation is consistent with collective behavior observed in $v_2$ measurements. The fit results also suggest that $D^0$ mesons have similar kinetic freeze-out properties as multi-strange-quark hadrons $\phi,\Xi^-$.

The nuclear modification factors $R_{\rm CP}$ of $D^0$ mesons are presented with both 60--80\% and 40--60\% centrality spectra as the reference, respectively. The $D^0$ $R_{\rm CP}$ is significantly suppressed at high $p_{T}$ and the suppression level is comparable to that of light hadrons at $p_{T}> 5\,\rm{GeV}/c$, re-affirming our previous observation~\cite{Star_D_RAA}. This indicates that charm quarks lose significant energy when traversing through the hot QCD medium. The $D^0$ $R_{\rm CP}$ is above the light hadron $R_{\rm CP}$ at low $p_{T}$. We compare our $D^0$ $R_{\rm CP}$ measurements to two recent theoretical model calculations from LBT and the Duke groups. These two models have the $2\pi TD_s$ value around 1-3 near $T_{c}$ and agree with our new $R_{\rm CP}$ measurements. The nuclear modification factor $R_{\rm AA}$ of $D^0$ mesons in 0-10\% central Au+Au collisions at ${\sqrt{s_{\rm NN}} = \rm{200\,GeV}}$ is comparable to that from the ALICE measurement in Pb+Pb at ${\sqrt{s_{\rm NN}} = \rm{2.76\,TeV}}$. At $p_{T}$\,$<$\,5\,GeV/$c$, the $D^0$ $R_{\rm AA}$ shows a characteristic bump structure. Model calculations that predict sizable collective motion for charm quarks during the medium evolution can qualitatively describe our data. We expect the new data points with much improved precision can be used in the future to further constrain our understanding of the charm-medium interactions as well as to better determine the medium transport parameter.

\section{Acknowledgement}
\label{acknowledgement}

We thank the RHIC Operations Group and RCF at BNL, the NERSC Center at LBNL, and the Open Science Grid consortium for providing resources and support. This work is supported in part by the Office of Nuclear Physics within the U.S. DOE Office of Science, the U.S. National Science Foundation, the Ministry of Education and Science of the Russian Federation, National Natural Science Foundation of China, Chinese Academy of Science, the Ministry of Science and Technology of China and the Chinese Ministry of Education, the National Research Foundation of Korea, GA and MSMT of the Czech Republic, Department of Atomic Energy and Department of Science and Technology of the Government of India; the National Science Centre of Poland, National Research Foundation, the Ministry of Science, Education and Sports of the Republic of Croatia, RosAtom of Russia and German Bundesministerium fur Bildung, Wissenschaft, Forschung and Technologie (BMBF) and the Helmholtz Association.

\bibliography{D0spectra}

\begin{thebibliography}{58}%
\makeatletter
\providecommand \@ifxundefined [1]{%
 \@ifx{#1\undefined}
}%
\providecommand \@ifnum [1]{%
 \ifnum #1\expandafter \@firstoftwo
 \else \expandafter \@secondoftwo
 \fi
}%
\providecommand \@ifx [1]{%
 \ifx #1\expandafter \@firstoftwo
 \else \expandafter \@secondoftwo
 \fi
}%
\providecommand \natexlab [1]{#1}%
\providecommand \enquote  [1]{``#1''}%
\providecommand \bibnamefont  [1]{#1}%
\providecommand \bibfnamefont [1]{#1}%
\providecommand \citenamefont [1]{#1}%
\providecommand \href@noop [0]{\@secondoftwo}%
\providecommand \href [0]{\begingroup \@sanitize@url \@href}%
\providecommand \@href[1]{\@@startlink{#1}\@@href}%
\providecommand \@@href[1]{\endgroup#1\@@endlink}%
\providecommand \@sanitize@url [0]{\catcode `\\12\catcode `\$12\catcode
  `\&12\catcode `\#12\catcode `\^12\catcode `\_12\catcode `\%12\relax}%
\providecommand \@@startlink[1]{}%
\providecommand \@@endlink[0]{}%
\providecommand \url  [0]{\begingroup\@sanitize@url \@url }%
\providecommand \@url [1]{\endgroup\@href {#1}{\urlprefix }}%
\providecommand \urlprefix  [0]{URL }%
\providecommand \Eprint [0]{\href }%
\providecommand \doibase [0]{http://dx.doi.org/}%
\providecommand \selectlanguage [0]{\@gobble}%
\providecommand \bibinfo  [0]{\@secondoftwo}%
\providecommand \bibfield  [0]{\@secondoftwo}%
\providecommand \translation [1]{[#1]}%
\providecommand \BibitemOpen [0]{}%
\providecommand \bibitemStop [0]{}%
\providecommand \bibitemNoStop [0]{.\EOS\space}%
\providecommand \EOS [0]{\spacefactor3000\relax}%
\providecommand \BibitemShut  [1]{\csname bibitem#1\endcsname}%
\let\auto@bib@innerbib\@empty
\bibitem [{\citenamefont {Adams}\ \emph {et~al.}(2005)\citenamefont {Adams}
  \emph {et~al.}}]{StarWhitePaper}%
  \BibitemOpen
  \bibfield  {author} {\bibinfo {author} {\bibfnamefont {J.}~\bibnamefont
  {Adams}} \emph {et~al.} (\bibinfo {collaboration} {STAR}),\ }\href {\doibase
  10.1016/j.nuclphysa.2005.03.085} {\bibfield  {journal} {\bibinfo  {journal}
  {Nucl. Phys.}\ }\textbf {\bibinfo {volume} {A757}},\ \bibinfo {pages} {102}
  (\bibinfo {year} {2005})}\BibitemShut {NoStop}%
\bibitem [{\citenamefont {Adcox}\ \emph {et~al.}(2005)\citenamefont {Adcox}
  \emph {et~al.}}]{PhenixWhitePaper}%
  \BibitemOpen
  \bibfield  {author} {\bibinfo {author} {\bibfnamefont {K.}~\bibnamefont
  {Adcox}} \emph {et~al.} (\bibinfo {collaboration} {PHENIX}),\ }\href
  {\doibase 10.1016/j.nuclphysa.2005.03.086} {\bibfield  {journal} {\bibinfo
  {journal} {Nucl. Phys.}\ }\textbf {\bibinfo {volume} {A757}},\ \bibinfo
  {pages} {184} (\bibinfo {year} {2005})}\BibitemShut {NoStop}%
\bibitem [{\citenamefont {Muller}\ \emph {et~al.}(2012)\citenamefont {Muller},
  \citenamefont {Schukraft},\ and\ \citenamefont {Wyslouch}}]{LhcSummary}%
  \BibitemOpen
  \bibfield  {author} {\bibinfo {author} {\bibfnamefont {B.}~\bibnamefont
  {Muller}}, \bibinfo {author} {\bibfnamefont {J.}~\bibnamefont {Schukraft}}, \
  and\ \bibinfo {author} {\bibfnamefont {B.}~\bibnamefont {Wyslouch}},\ }\href
  {\doibase 10.1146/annurev-nucl-102711-094910} {\bibfield  {journal} {\bibinfo
   {journal} {Ann. Rev. Nucl. Part. Sci.}\ }\textbf {\bibinfo {volume} {62}},\
  \bibinfo {pages} {361} (\bibinfo {year} {2012})}\BibitemShut {NoStop}%
\bibitem [{\citenamefont {Adamczyk}\ \emph {et~al.}(2016)\citenamefont
  {Adamczyk} \emph {et~al.}}]{Adamczyk:2015ukd}%
  \BibitemOpen
  \bibfield  {author} {\bibinfo {author} {\bibfnamefont {L.}~\bibnamefont
  {Adamczyk}} \emph {et~al.} (\bibinfo {collaboration} {STAR}),\ }\href
  {\doibase 10.1103/PhysRevLett.116.062301} {\bibfield  {journal} {\bibinfo
  {journal} {Phys. Rev. Lett.}\ }\textbf {\bibinfo {volume} {116}},\ \bibinfo
  {pages} {062301} (\bibinfo {year} {2016})}\BibitemShut {NoStop}%
\bibitem [{\citenamefont {Abelev}\ \emph {et~al.}(2015)\citenamefont {Abelev}
  \emph {et~al.}}]{Abelev:2014pua}%
  \BibitemOpen
  \bibfield  {author} {\bibinfo {author} {\bibfnamefont {B.~B.}\ \bibnamefont
  {Abelev}} \emph {et~al.} (\bibinfo {collaboration} {ALICE}),\ }\href
  {\doibase 10.1007/JHEP06(2015)190} {\bibfield  {journal} {\bibinfo  {journal}
  {JHEP}\ }\textbf {\bibinfo {volume} {06}},\ \bibinfo {pages} {190} (\bibinfo
  {year} {2015})}\BibitemShut {NoStop}%
\bibitem [{\citenamefont {Lin}\ and\ \citenamefont
  {Gyulassy}(1995)}]{Ziwei_Lin}%
  \BibitemOpen
  \bibfield  {author} {\bibinfo {author} {\bibfnamefont {Z.}~\bibnamefont
  {Lin}}\ and\ \bibinfo {author} {\bibfnamefont {M.}~\bibnamefont {Gyulassy}},\
  }\href {\doibase 10.1103/PhysRevC.51.2177} {\bibfield  {journal} {\bibinfo
  {journal} {Phys. Rev. C}\ }\textbf {\bibinfo {volume} {51}},\ \bibinfo
  {pages} {2177} (\bibinfo {year} {1995})}\BibitemShut {NoStop}%
\bibitem [{\citenamefont {Cacciari}\ \emph {et~al.}(2005)\citenamefont
  {Cacciari}, \citenamefont {Nason},\ and\ \citenamefont {Vogt}}]{Cacciari}%
  \BibitemOpen
  \bibfield  {author} {\bibinfo {author} {\bibfnamefont {M.}~\bibnamefont
  {Cacciari}}, \bibinfo {author} {\bibfnamefont {P.}~\bibnamefont {Nason}}, \
  and\ \bibinfo {author} {\bibfnamefont {R.}~\bibnamefont {Vogt}},\ }\href
  {\doibase 10.1103/PhysRevLett.95.122001} {\bibfield  {journal} {\bibinfo
  {journal} {Phys. Rev. Lett.}\ }\textbf {\bibinfo {volume} {95}},\ \bibinfo
  {pages} {122001} (\bibinfo {year} {2005})}\BibitemShut {NoStop}%
\bibitem [{\citenamefont {Moore}\ and\ \citenamefont {Teaney}(2005)}]{Moore}%
  \BibitemOpen
  \bibfield  {author} {\bibinfo {author} {\bibfnamefont {G.~D.}\ \bibnamefont
  {Moore}}\ and\ \bibinfo {author} {\bibfnamefont {D.}~\bibnamefont {Teaney}},\
  }\href {\doibase 10.1103/PhysRevC.71.064904} {\bibfield  {journal} {\bibinfo
  {journal} {Phys. Rev. C}\ }\textbf {\bibinfo {volume} {71}},\ \bibinfo
  {pages} {064904} (\bibinfo {year} {2005})}\BibitemShut {NoStop}%
\bibitem [{\citenamefont {Abelev}\ \emph {et~al.}(2012)\citenamefont {Abelev}
  \emph {et~al.}}]{Alice_D_RAA_1}%
  \BibitemOpen
  \bibfield  {author} {\bibinfo {author} {\bibfnamefont {B.}~\bibnamefont
  {Abelev}} \emph {et~al.} (\bibinfo {collaboration} {ALICE}),\ }\href
  {\doibase 10.1007/JHEP09(2012)112} {\bibfield  {journal} {\bibinfo  {journal}
  {JHEP}\ }\textbf {\bibinfo {volume} {09}},\ \bibinfo {pages} {112} (\bibinfo
  {year} {2012})}\BibitemShut {NoStop}%
\bibitem [{\citenamefont {Adam}\ \emph {et~al.}(2016)\citenamefont {Adam} \emph
  {et~al.}}]{Alice_D_RAA_2}%
  \BibitemOpen
  \bibfield  {author} {\bibinfo {author} {\bibfnamefont {J.}~\bibnamefont
  {Adam}} \emph {et~al.} (\bibinfo {collaboration} {ALICE}),\ }\href {\doibase
  10.1007/JHEP03(2016)081} {\bibfield  {journal} {\bibinfo  {journal} {JHEP}\
  }\textbf {\bibinfo {volume} {03}},\ \bibinfo {pages} {081} (\bibinfo {year}
  {2016})}\BibitemShut {NoStop}%
\bibitem [{\citenamefont {Sirunyan}\ \emph
  {et~al.}(2018{\natexlab{a}})\citenamefont {Sirunyan} \emph
  {et~al.}}]{CMS_D_RAA_5TeV}%
  \BibitemOpen
  \bibfield  {author} {\bibinfo {author} {\bibfnamefont {A.~M.}\ \bibnamefont
  {Sirunyan}} \emph {et~al.} (\bibinfo {collaboration} {CMS}),\ }\href
  {\doibase https://doi.org/10.1016/j.physletb.2018.05.074} {\bibfield
  {journal} {\bibinfo  {journal} {Phys. Lett. B}\ }\textbf {\bibinfo {volume}
  {782}},\ \bibinfo {pages} {474 } (\bibinfo {year}
  {2018}{\natexlab{a}})}\BibitemShut {NoStop}%
\bibitem [{\citenamefont {Adamczyk}\ \emph
  {et~al.}(2014{\natexlab{a}})\citenamefont {Adamczyk} \emph
  {et~al.}}]{Star_D_RAA}%
  \BibitemOpen
  \bibfield  {author} {\bibinfo {author} {\bibfnamefont {L.}~\bibnamefont
  {Adamczyk}} \emph {et~al.} (\bibinfo {collaboration} {STAR}),\ }\href
  {\doibase 10.1103/PhysRevLett.113.142301} {\bibfield  {journal} {\bibinfo
  {journal} {Phys. Rev. Lett.}\ }\textbf {\bibinfo {volume} {113}},\ \bibinfo
  {pages} {142301} (\bibinfo {year} {2014}{\natexlab{a}})},\ \bibinfo {note}
  {erratum: Phys. Rev. Lett. $\bold{121}$, 229901 (2018)}\BibitemShut {NoStop}%
\bibitem [{\citenamefont {Abelev}\ \emph
  {et~al.}(2013{\natexlab{a}})\citenamefont {Abelev} \emph
  {et~al.}}]{Alice_D_v2_276TeV_PRL}%
  \BibitemOpen
  \bibfield  {author} {\bibinfo {author} {\bibfnamefont {B.}~\bibnamefont
  {Abelev}} \emph {et~al.} (\bibinfo {collaboration} {ALICE}),\ }\href
  {\doibase 10.1103/PhysRevLett.111.102301} {\bibfield  {journal} {\bibinfo
  {journal} {Phys. Rev. Lett.}\ }\textbf {\bibinfo {volume} {111}},\ \bibinfo
  {pages} {102301} (\bibinfo {year} {2013}{\natexlab{a}})}\BibitemShut
  {NoStop}%
\bibitem [{\citenamefont {Abelev}\ \emph {et~al.}(2014)\citenamefont {Abelev}
  \emph {et~al.}}]{Alice_D_v2_276TeV_PRC}%
  \BibitemOpen
  \bibfield  {author} {\bibinfo {author} {\bibfnamefont {B.}~\bibnamefont
  {Abelev}} \emph {et~al.} (\bibinfo {collaboration} {ALICE}),\ }\href
  {\doibase 10.1103/PhysRevC.90.034904} {\bibfield  {journal} {\bibinfo
  {journal} {Phys. Rev. C}\ }\textbf {\bibinfo {volume} {90}},\ \bibinfo
  {pages} {034904} (\bibinfo {year} {2014})}\BibitemShut {NoStop}%
\bibitem [{\citenamefont {Sirunyan}\ \emph
  {et~al.}(2018{\natexlab{b}})\citenamefont {Sirunyan} \emph
  {et~al.}}]{CMS_D_vn_5TeV}%
  \BibitemOpen
  \bibfield  {author} {\bibinfo {author} {\bibfnamefont {A.~M.}\ \bibnamefont
  {Sirunyan}} \emph {et~al.} (\bibinfo {collaboration} {CMS}),\ }\href
  {\doibase 10.1103/PhysRevLett.120.202301} {\bibfield  {journal} {\bibinfo
  {journal} {Phys. Rev. Lett.}\ }\textbf {\bibinfo {volume} {120}},\ \bibinfo
  {pages} {202301} (\bibinfo {year} {2018}{\natexlab{b}})}\BibitemShut
  {NoStop}%
\bibitem [{\citenamefont {Adamczyk}\ \emph
  {et~al.}(2017{\natexlab{a}})\citenamefont {Adamczyk} \emph
  {et~al.}}]{Star_D_v2}%
  \BibitemOpen
  \bibfield  {author} {\bibinfo {author} {\bibfnamefont {L.}~\bibnamefont
  {Adamczyk}} \emph {et~al.} (\bibinfo {collaboration} {STAR}),\ }\href
  {\doibase 10.1103/PhysRevLett.118.212301} {\bibfield  {journal} {\bibinfo
  {journal} {Phys. Rev. Lett.}\ }\textbf {\bibinfo {volume} {118}},\ \bibinfo
  {pages} {212301} (\bibinfo {year} {2017}{\natexlab{a}})}\BibitemShut
  {NoStop}%
\bibitem [{\citenamefont {Contin}\ \emph {et~al.}(2018)\citenamefont {Contin}
  \emph {et~al.}}]{Contin:2017mck}%
  \BibitemOpen
  \bibfield  {author} {\bibinfo {author} {\bibfnamefont {G.}~\bibnamefont
  {Contin}} \emph {et~al.},\ }\href {\doibase 10.1016/j.nima.2018.03.003}
  {\bibfield  {journal} {\bibinfo  {journal} {Nucl. Instrum. Meth.}\ }\textbf
  {\bibinfo {volume} {A907}},\ \bibinfo {pages} {60} (\bibinfo {year}
  {2018})}\BibitemShut {NoStop}%
\bibitem [{\citenamefont {Anderson}\ \emph {et~al.}(2003)\citenamefont
  {Anderson} \emph {et~al.}}]{TPC}%
  \BibitemOpen
  \bibfield  {author} {\bibinfo {author} {\bibfnamefont {M.}~\bibnamefont
  {Anderson}} \emph {et~al.},\ }\href {\doibase 10.1016/S0168-9002(02)01964-2}
  {\bibfield  {journal} {\bibinfo  {journal} {Nucl. Instrum. Meth.}\ }\textbf
  {\bibinfo {volume} {A499}},\ \bibinfo {pages} {659} (\bibinfo {year}
  {2003})}\BibitemShut {NoStop}%
\bibitem [{\citenamefont {Llope}(2012)}]{TOF}%
  \BibitemOpen
  \bibfield  {author} {\bibinfo {author} {\bibfnamefont {W.~J.}\ \bibnamefont
  {Llope}} (\bibinfo {collaboration} {STAR}),\ }\href {\doibase
  10.1016/j.nima.2010.07.086} {\bibfield  {journal} {\bibinfo  {journal} {Nucl.
  Instrum. Meth.}\ }\textbf {\bibinfo {volume} {A661}},\ \bibinfo {pages}
  {S110} (\bibinfo {year} {2012})}\BibitemShut {NoStop}%
\bibitem [{\citenamefont {Adamczyk}\ \emph {et~al.}(2012)\citenamefont
  {Adamczyk} \emph {et~al.}}]{Star_D_pp}%
  \BibitemOpen
  \bibfield  {author} {\bibinfo {author} {\bibfnamefont {L.}~\bibnamefont
  {Adamczyk}} \emph {et~al.} (\bibinfo {collaboration} {STAR}),\ }\href
  {\doibase 10.1103/PhysRevD.86.072013} {\bibfield  {journal} {\bibinfo
  {journal} {Phys. Rev. D}\ }\textbf {\bibinfo {volume} {86}},\ \bibinfo
  {pages} {072013} (\bibinfo {year} {2012})}\BibitemShut {NoStop}%
\bibitem [{\citenamefont {Llope}\ \emph {et~al.}(2004)\citenamefont {Llope}
  \emph {et~al.}}]{VPD}%
  \BibitemOpen
  \bibfield  {author} {\bibinfo {author} {\bibfnamefont {W.~J.}\ \bibnamefont
  {Llope}} \emph {et~al.},\ }\href {\doibase 10.1016/j.nima.2003.11.414}
  {\bibfield  {journal} {\bibinfo  {journal} {Nucl. Instrum. Meth.}\ }\textbf
  {\bibinfo {volume} {A522}},\ \bibinfo {pages} {252} (\bibinfo {year}
  {2004})}\BibitemShut {NoStop}%
\bibitem [{\citenamefont {Kalman}(1960)}]{Kalman}%
  \BibitemOpen
  \bibfield  {author} {\bibinfo {author} {\bibfnamefont {R.~E.}\ \bibnamefont
  {Kalman}},\ }\href@noop {} {\bibfield  {journal} {\bibinfo  {journal}
  {Journal of Basic Engineering}\ }\textbf {\bibinfo {volume} {82}},\ \bibinfo
  {pages} {35} (\bibinfo {year} {1960})}\BibitemShut {NoStop}%
\bibitem [{\citenamefont {Voss}\ \emph {et~al.}(2007)\citenamefont {Voss} \emph
  {et~al.}}]{TMVA}%
  \BibitemOpen
  \bibfield  {author} {\bibinfo {author} {\bibfnamefont {H.}~\bibnamefont
  {Voss}} \emph {et~al.},\ }\href@noop {} {\bibfield  {journal} {\bibinfo
  {journal} {PoS}\ }\textbf {\bibinfo {volume} {ACAT}},\ \bibinfo {pages} {040}
  (\bibinfo {year} {2007})}\BibitemShut {NoStop}%
\bibitem [{\citenamefont {Tanabashi}\ \emph {et~al.}(2018)\citenamefont
  {Tanabashi} \emph {et~al.}}]{pdg}%
  \BibitemOpen
  \bibfield  {author} {\bibinfo {author} {\bibfnamefont {M.}~\bibnamefont
  {Tanabashi}} \emph {et~al.} (\bibinfo {collaboration} {Particle Data
  Group}),\ }\href {\doibase 10.1103/PhysRevD.98.030001} {\bibfield  {journal}
  {\bibinfo  {journal} {Phys. Rev. D}\ }\textbf {\bibinfo {volume} {98}},\
  \bibinfo {pages} {030001} (\bibinfo {year} {2018})}\BibitemShut {NoStop}%
\bibitem [{\citenamefont {Brun}\ \emph {et~al.}(1994)\citenamefont {Brun} \emph
  {et~al.}}]{GEANT3}%
  \BibitemOpen
  \bibfield  {author} {\bibinfo {author} {\bibfnamefont {R.}~\bibnamefont
  {Brun}} \emph {et~al.},\ }\href {\doibase 10.17181/CERN.MUHF.DMJ1} {\
  (\bibinfo {year} {1994}),\ 10.17181/CERN.MUHF.DMJ1}\BibitemShut {NoStop}%
\bibitem [{\citenamefont {Gyulassy}\ and\ \citenamefont {Wang}(1994)}]{HIJING}%
  \BibitemOpen
  \bibfield  {author} {\bibinfo {author} {\bibfnamefont {M.}~\bibnamefont
  {Gyulassy}}\ and\ \bibinfo {author} {\bibfnamefont {X.-N.}\ \bibnamefont
  {Wang}},\ }\href {\doibase https://doi.org/10.1016/0010-4655(94)90057-4}
  {\bibfield  {journal} {\bibinfo  {journal} {Comp. Phys. Comm.}\ }\textbf
  {\bibinfo {volume} {83}},\ \bibinfo {pages} {307 } (\bibinfo {year}
  {1994})}\BibitemShut {NoStop}%
\bibitem [{\citenamefont {Adams}\ \emph {et~al.}(2004)\citenamefont {Adams}
  \emph {et~al.}}]{Adams:2003xp}%
  \BibitemOpen
  \bibfield  {author} {\bibinfo {author} {\bibfnamefont {J.}~\bibnamefont
  {Adams}} \emph {et~al.} (\bibinfo {collaboration} {STAR}),\ }\href {\doibase
  10.1103/PhysRevLett.92.112301} {\bibfield  {journal} {\bibinfo  {journal}
  {Phys. Rev. Lett.}\ }\textbf {\bibinfo {volume} {92}},\ \bibinfo {pages}
  {112301} (\bibinfo {year} {2004})}\BibitemShut {NoStop}%
\bibitem [{\citenamefont {Shao}\ \emph {et~al.}(2006)\citenamefont {Shao} \emph
  {et~al.}}]{Shao:2005iu}%
  \BibitemOpen
  \bibfield  {author} {\bibinfo {author} {\bibfnamefont {M.}~\bibnamefont
  {Shao}} \emph {et~al.},\ }\href {\doibase 10.1016/j.nima.2005.11.251}
  {\bibfield  {journal} {\bibinfo  {journal} {Nucl. Instrum. Meth.}\ }\textbf
  {\bibinfo {volume} {A558}},\ \bibinfo {pages} {419} (\bibinfo {year}
  {2006})}\BibitemShut {NoStop}%
\bibitem [{\citenamefont {Xu}\ \emph {et~al.}(2010)\citenamefont {Xu} \emph
  {et~al.}}]{Xu:2008th}%
  \BibitemOpen
  \bibfield  {author} {\bibinfo {author} {\bibfnamefont {Y.}~\bibnamefont {Xu}}
  \emph {et~al.},\ }\href {\doibase 10.1016/j.nima.2009.12.011} {\bibfield
  {journal} {\bibinfo  {journal} {Nucl. Instrum. Meth.}\ }\textbf {\bibinfo
  {volume} {A614}},\ \bibinfo {pages} {28} (\bibinfo {year}
  {2010})}\BibitemShut {NoStop}%
\bibitem [{\citenamefont {Greco}\ \emph {et~al.}(2004)\citenamefont {Greco},
  \citenamefont {Ko},\ and\ \citenamefont {Rapp}}]{GRECO2004202}%
  \BibitemOpen
  \bibfield  {author} {\bibinfo {author} {\bibfnamefont {V.}~\bibnamefont
  {Greco}}, \bibinfo {author} {\bibfnamefont {C.}~\bibnamefont {Ko}}, \ and\
  \bibinfo {author} {\bibfnamefont {R.}~\bibnamefont {Rapp}},\ }\href {\doibase
  https://doi.org/10.1016/j.physletb.2004.06.064} {\bibfield  {journal}
  {\bibinfo  {journal} {Phys. Lett. B}\ }\textbf {\bibinfo {volume} {595}},\
  \bibinfo {pages} {202 } (\bibinfo {year} {2004})}\BibitemShut {NoStop}%
\bibitem [{\citenamefont {Oh}\ \emph {et~al.}(2009)\citenamefont {Oh} \emph
  {et~al.}}]{Oh2009}%
  \BibitemOpen
  \bibfield  {author} {\bibinfo {author} {\bibfnamefont {Y.}~\bibnamefont {Oh}}
  \emph {et~al.},\ }\href {\doibase 10.1103/PhysRevC.79.044905} {\bibfield
  {journal} {\bibinfo  {journal} {Phys. Rev. C}\ }\textbf {\bibinfo {volume}
  {79}},\ \bibinfo {pages} {1} (\bibinfo {year} {2009})}\BibitemShut {NoStop}%
\bibitem [{\citenamefont {Zhao}\ \emph {et~al.}(2018)\citenamefont {Zhao} \emph
  {et~al.}}]{Zhao:2018jlw}%
  \BibitemOpen
  \bibfield  {author} {\bibinfo {author} {\bibfnamefont {J.}~\bibnamefont
  {Zhao}} \emph {et~al.},\ }\href@noop {} {\  (\bibinfo {year} {2018})},\
  \Eprint {http://arxiv.org/abs/1805.10858} {arXiv:1805.10858 [hep-ph]}
  \BibitemShut {NoStop}%
\bibitem [{\citenamefont {Plumari}\ \emph {et~al.}(2018)\citenamefont {Plumari}
  \emph {et~al.}}]{Plumari:2017ntm}%
  \BibitemOpen
  \bibfield  {author} {\bibinfo {author} {\bibfnamefont {S.}~\bibnamefont
  {Plumari}} \emph {et~al.},\ }\href {\doibase 10.1140/epjc/s10052-018-5828-7}
  {\bibfield  {journal} {\bibinfo  {journal} {Eur. Phys. J.}\ }\textbf
  {\bibinfo {volume} {C78}},\ \bibinfo {pages} {348} (\bibinfo {year}
  {2018})}\BibitemShut {NoStop}%
\bibitem [{\citenamefont {He}\ \emph {et~al.}(2013)\citenamefont {He},
  \citenamefont {Fries},\ and\ \citenamefont {Rapp}}]{He2013}%
  \BibitemOpen
  \bibfield  {author} {\bibinfo {author} {\bibfnamefont {M.}~\bibnamefont
  {He}}, \bibinfo {author} {\bibfnamefont {R.~J.}\ \bibnamefont {Fries}}, \
  and\ \bibinfo {author} {\bibfnamefont {R.}~\bibnamefont {Rapp}},\ }\href
  {\doibase 10.1103/PhysRevLett.110.112301} {\bibfield  {journal} {\bibinfo
  {journal} {Phys. Rev. Lett.}\ }\textbf {\bibinfo {volume} {110}},\ \bibinfo
  {pages} {112301} (\bibinfo {year} {2013})}\BibitemShut {NoStop}%
\bibitem [{\citenamefont {Kaneta}(1999)}]{Kaneta:1999lnf}%
  \BibitemOpen
  \bibfield  {author} {\bibinfo {author} {\bibfnamefont {M.}~\bibnamefont
  {Kaneta}},\ }\emph {\bibinfo {title} {{Thermal and Chemical Freeze-out in
  Heavy Ion Collisions}}},\ \href
  {https://www.phenix.bnl.gov/~kaneta/phd_thesis/phd.pdf} {Ph.D. thesis},\
  \bibinfo  {school} {Hiroshima U.} (\bibinfo {year} {1999})\BibitemShut
  {NoStop}%
\bibitem [{\citenamefont {Abelev}\ \emph
  {et~al.}(2007{\natexlab{a}})\citenamefont {Abelev} \emph
  {et~al.}}]{Abelev:2007rw}%
  \BibitemOpen
  \bibfield  {author} {\bibinfo {author} {\bibfnamefont {B.~I.}\ \bibnamefont
  {Abelev}} \emph {et~al.} (\bibinfo {collaboration} {STAR}),\ }\href {\doibase
  10.1103/PhysRevLett.99.112301} {\bibfield  {journal} {\bibinfo  {journal}
  {Phys. Rev. Lett.}\ }\textbf {\bibinfo {volume} {99}},\ \bibinfo {pages}
  {112301} (\bibinfo {year} {2007}{\natexlab{a}})}\BibitemShut {NoStop}%
\bibitem [{\citenamefont {Adams}\ \emph {et~al.}(2007)\citenamefont {Adams}
  \emph {et~al.}}]{Adams:2006ke}%
  \BibitemOpen
  \bibfield  {author} {\bibinfo {author} {\bibfnamefont {J.}~\bibnamefont
  {Adams}} \emph {et~al.} (\bibinfo {collaboration} {STAR}),\ }\href {\doibase
  10.1103/PhysRevLett.98.062301} {\bibfield  {journal} {\bibinfo  {journal}
  {Phys. Rev. Lett.}\ }\textbf {\bibinfo {volume} {98}},\ \bibinfo {pages}
  {062301} (\bibinfo {year} {2007})}\BibitemShut {NoStop}%
\bibitem [{\citenamefont {Adamczyk}\ \emph
  {et~al.}(2014{\natexlab{b}})\citenamefont {Adamczyk} \emph
  {et~al.}}]{Adamczyk:2013tvk}%
  \BibitemOpen
  \bibfield  {author} {\bibinfo {author} {\bibfnamefont {L.}~\bibnamefont
  {Adamczyk}} \emph {et~al.} (\bibinfo {collaboration} {STAR}),\ }\href
  {\doibase 10.1103/PhysRevC.90.024906} {\bibfield  {journal} {\bibinfo
  {journal} {Phys. Rev. C}\ }\textbf {\bibinfo {volume} {90}},\ \bibinfo
  {pages} {024906} (\bibinfo {year} {2014}{\natexlab{b}})}\BibitemShut
  {NoStop}%
\bibitem [{\citenamefont {Csorgo}\ and\ \citenamefont
  {Lorstad}(1996)}]{Csorgo:1995bi}%
  \BibitemOpen
  \bibfield  {author} {\bibinfo {author} {\bibfnamefont {T.}~\bibnamefont
  {Csorgo}}\ and\ \bibinfo {author} {\bibfnamefont {B.}~\bibnamefont
  {Lorstad}},\ }\href {\doibase 10.1103/PhysRevC.54.1390} {\bibfield  {journal}
  {\bibinfo  {journal} {Phys. Rev. C}\ }\textbf {\bibinfo {volume} {54}},\
  \bibinfo {pages} {1390} (\bibinfo {year} {1996})}\BibitemShut {NoStop}%
\bibitem [{\citenamefont {Kolb}\ and\ \citenamefont
  {Heinz}(2003)}]{Kolb:2003dz}%
  \BibitemOpen
  \bibfield  {author} {\bibinfo {author} {\bibfnamefont {P.~F.}\ \bibnamefont
  {Kolb}}\ and\ \bibinfo {author} {\bibfnamefont {U.~W.}\ \bibnamefont
  {Heinz}},\ }\href@noop {} {\bibfield  {journal} {\bibinfo  {journal} {Quark
  Gluon Plasma 3}\ ,\ \bibinfo {pages} {634}} (\bibinfo {year}
  {2003})}\BibitemShut {NoStop}%
\bibitem [{\citenamefont {Adamczyk}\ \emph
  {et~al.}(2017{\natexlab{b}})\citenamefont {Adamczyk} \emph
  {et~al.}}]{Adamczyk:2017iwn}%
  \BibitemOpen
  \bibfield  {author} {\bibinfo {author} {\bibfnamefont {L.}~\bibnamefont
  {Adamczyk}} \emph {et~al.} (\bibinfo {collaboration} {STAR}),\ }\href
  {\doibase 10.1103/PhysRevC.96.044904} {\bibfield  {journal} {\bibinfo
  {journal} {Phys. Rev. C}\ }\textbf {\bibinfo {volume} {96}},\ \bibinfo
  {pages} {044904} (\bibinfo {year} {2017}{\natexlab{b}})}\BibitemShut
  {NoStop}%
\bibitem [{\citenamefont {Schnedermann}\ \emph {et~al.}(1993)\citenamefont
  {Schnedermann} \emph {et~al.}}]{Schnedermann:1993ws}%
  \BibitemOpen
  \bibfield  {author} {\bibinfo {author} {\bibfnamefont {E.}~\bibnamefont
  {Schnedermann}} \emph {et~al.},\ }\href {\doibase 10.1103/PhysRevC.48.2462}
  {\bibfield  {journal} {\bibinfo  {journal} {Phys. Rev. C}\ }\textbf {\bibinfo
  {volume} {48}},\ \bibinfo {pages} {2462} (\bibinfo {year}
  {1993})}\BibitemShut {NoStop}%
\bibitem [{\citenamefont {Tang}\ \emph {et~al.}(2009)\citenamefont {Tang} \emph
  {et~al.}}]{Tang:2008ud}%
  \BibitemOpen
  \bibfield  {author} {\bibinfo {author} {\bibfnamefont {Z.}~\bibnamefont
  {Tang}} \emph {et~al.},\ }\href {\doibase 10.1103/PhysRevC.79.051901}
  {\bibfield  {journal} {\bibinfo  {journal} {Phys. Rev. C}\ }\textbf {\bibinfo
  {volume} {79}},\ \bibinfo {pages} {051901} (\bibinfo {year}
  {2009})}\BibitemShut {NoStop}%
\bibitem [{\citenamefont {Bazavov}\ \emph {et~al.}(2012)\citenamefont {Bazavov}
  \emph {et~al.}}]{Bazavov:2011nk}%
  \BibitemOpen
  \bibfield  {author} {\bibinfo {author} {\bibfnamefont {A.}~\bibnamefont
  {Bazavov}} \emph {et~al.},\ }\href {\doibase 10.1103/PhysRevD.85.054503}
  {\bibfield  {journal} {\bibinfo  {journal} {Phys. Rev. D}\ }\textbf {\bibinfo
  {volume} {85}},\ \bibinfo {pages} {054503} (\bibinfo {year}
  {2012})}\BibitemShut {NoStop}%
\bibitem [{\citenamefont {Abelev}\ \emph {et~al.}(2006)\citenamefont {Abelev}
  \emph {et~al.}}]{Adams2006_Identified}%
  \BibitemOpen
  \bibfield  {author} {\bibinfo {author} {\bibfnamefont {B.~I.}\ \bibnamefont
  {Abelev}} \emph {et~al.} (\bibinfo {collaboration} {STAR}),\ }\href {\doibase
  10.1103/PhysRevLett.97.152301} {\bibfield  {journal} {\bibinfo  {journal}
  {Phys. Rev. Lett.}\ }\textbf {\bibinfo {volume} {97}},\ \bibinfo {pages}
  {152301} (\bibinfo {year} {2006})}\BibitemShut {NoStop}%
\bibitem [{\citenamefont {Abelev}\ \emph {et~al.}(2009)\citenamefont {Abelev}
  \emph {et~al.}}]{Abelev2009}%
  \BibitemOpen
  \bibfield  {author} {\bibinfo {author} {\bibfnamefont {B.~I.}\ \bibnamefont
  {Abelev}} \emph {et~al.} (\bibinfo {collaboration} {STAR}),\ }\href {\doibase
  10.1103/PhysRevC.79.064903} {\bibfield  {journal} {\bibinfo  {journal} {Phys.
  Rev. C}\ }\textbf {\bibinfo {volume} {79}},\ \bibinfo {pages} {064903}
  (\bibinfo {year} {2009})}\BibitemShut {NoStop}%
\bibitem [{\citenamefont {Agakishiev}\ \emph {et~al.}(2012)\citenamefont
  {Agakishiev} \emph {et~al.}}]{Agakishiev2012}%
  \BibitemOpen
  \bibfield  {author} {\bibinfo {author} {\bibfnamefont {G.}~\bibnamefont
  {Agakishiev}} \emph {et~al.} (\bibinfo {collaboration} {STAR}),\ }\href
  {\doibase 10.1103/PhysRevLett.108.072301} {\bibfield  {journal} {\bibinfo
  {journal} {Phys. Rev. Lett.}\ }\textbf {\bibinfo {volume} {108}},\ \bibinfo
  {pages} {072301} (\bibinfo {year} {2012})}\BibitemShut {NoStop}%
\bibitem [{\citenamefont {Cao}\ \emph {et~al.}(2016)\citenamefont {Cao} \emph
  {et~al.}}]{Cao:2016gvr}%
  \BibitemOpen
  \bibfield  {author} {\bibinfo {author} {\bibfnamefont {S.}~\bibnamefont
  {Cao}} \emph {et~al.},\ }\href {\doibase 10.1103/PhysRevC.94.014909}
  {\bibfield  {journal} {\bibinfo  {journal} {Phys. Rev. C}\ }\textbf {\bibinfo
  {volume} {94}},\ \bibinfo {pages} {014909} (\bibinfo {year}
  {2016})}\BibitemShut {NoStop}%
\bibitem [{\citenamefont {Cao}()}]{LBT:private}%
  \BibitemOpen
  \bibfield  {author} {\bibinfo {author} {\bibfnamefont {S.}~\bibnamefont
  {Cao}},\ }\href@noop {} {}\bibinfo {howpublished} {private
  communication}\BibitemShut {NoStop}%
\bibitem [{\citenamefont {Xu}\ \emph {et~al.}(2018)\citenamefont {Xu} \emph
  {et~al.}}]{Xu:2017obm}%
  \BibitemOpen
  \bibfield  {author} {\bibinfo {author} {\bibfnamefont {Y.}~\bibnamefont {Xu}}
  \emph {et~al.},\ }\href {\doibase 10.1103/PhysRevC.97.014907} {\bibfield
  {journal} {\bibinfo  {journal} {Phys. Rev. C}\ }\textbf {\bibinfo {volume}
  {97}},\ \bibinfo {pages} {014907} (\bibinfo {year} {2018})}\BibitemShut
  {NoStop}%
\bibitem [{\citenamefont {Lisovyi}\ \emph {et~al.}(2016)\citenamefont
  {Lisovyi}, \citenamefont {Verbytskyi},\ and\ \citenamefont
  {Zenaiev}}]{charm_frag}%
  \BibitemOpen
  \bibfield  {author} {\bibinfo {author} {\bibfnamefont {M.}~\bibnamefont
  {Lisovyi}}, \bibinfo {author} {\bibfnamefont {A.}~\bibnamefont {Verbytskyi}},
  \ and\ \bibinfo {author} {\bibfnamefont {O.}~\bibnamefont {Zenaiev}},\
  }\href@noop {} {\bibfield  {journal} {\bibinfo  {journal} {Eur. Phys. J. C}\
  }\textbf {\bibinfo {volume} {76}},\ \bibinfo {pages} {397} (\bibinfo {year}
  {2016})}\BibitemShut {NoStop}%
\bibitem [{\citenamefont {Abelev}\ \emph
  {et~al.}(2013{\natexlab{b}})\citenamefont {Abelev} \emph
  {et~al.}}]{Alice_hadron_RAA}%
  \BibitemOpen
  \bibfield  {author} {\bibinfo {author} {\bibfnamefont {B.}~\bibnamefont
  {Abelev}} \emph {et~al.} (\bibinfo {collaboration} {ALICE}),\ }\href
  {\doibase https://doi.org/10.1016/j.physletb.2013.01.051} {\bibfield
  {journal} {\bibinfo  {journal} {Phys. Lett. B}\ }\textbf {\bibinfo {volume}
  {720}},\ \bibinfo {pages} {52 } (\bibinfo {year}
  {2013}{\natexlab{b}})}\BibitemShut {NoStop}%
\bibitem [{\citenamefont {Abelev}\ \emph
  {et~al.}(2007{\natexlab{b}})\citenamefont {Abelev} \emph {et~al.}}]{StarPi0}%
  \BibitemOpen
  \bibfield  {author} {\bibinfo {author} {\bibfnamefont {B.}~\bibnamefont
  {Abelev}} \emph {et~al.} (\bibinfo {collaboration} {STAR}),\ }\href {\doibase
  https://doi.org/10.1016/j.physletb.2007.06.035} {\bibfield  {journal}
  {\bibinfo  {journal} {Phys. Lett. B}\ }\textbf {\bibinfo {volume} {655}},\
  \bibinfo {pages} {104 } (\bibinfo {year} {2007}{\natexlab{b}})}\BibitemShut
  {NoStop}%
\bibitem [{\citenamefont {Andronic}\ \emph {et~al.}(2003)\citenamefont
  {Andronic} \emph {et~al.}}]{ANDRONIC200336}%
  \BibitemOpen
  \bibfield  {author} {\bibinfo {author} {\bibfnamefont {A.}~\bibnamefont
  {Andronic}} \emph {et~al.},\ }\href {\doibase
  https://doi.org/10.1016/j.physletb.2003.07.066} {\bibfield  {journal}
  {\bibinfo  {journal} {Phys. Lett. B}\ }\textbf {\bibinfo {volume} {571}},\
  \bibinfo {pages} {36 } (\bibinfo {year} {2003})}\BibitemShut {NoStop}%
\bibitem [{\citenamefont {Rapp}\ \emph {et~al.}(2018)\citenamefont {Rapp} \emph
  {et~al.}}]{Rapp:2018qla}%
  \BibitemOpen
  \bibfield  {author} {\bibinfo {author} {\bibfnamefont {R.}~\bibnamefont
  {Rapp}} \emph {et~al.},\ }\href {\doibase 10.1016/j.nuclphysa.2018.09.002}
  {\bibfield  {journal} {\bibinfo  {journal} {Nucl. Phys.}\ }\textbf {\bibinfo
  {volume} {A979}},\ \bibinfo {pages} {21} (\bibinfo {year}
  {2018})}\BibitemShut {NoStop}%
\bibitem [{\citenamefont {Cao}\ \emph {et~al.}(2018)\citenamefont {Cao} \emph
  {et~al.}}]{Cao:2018ews}%
  \BibitemOpen
  \bibfield  {author} {\bibinfo {author} {\bibfnamefont {S.}~\bibnamefont
  {Cao}} \emph {et~al.},\ }\href@noop {} {\  (\bibinfo {year} {2018})},\
  \Eprint {http://arxiv.org/abs/1809.07894} {arXiv:1809.07894 [nucl-th]}
  \BibitemShut {NoStop}%
\bibitem [{\citenamefont {Cao}\ \emph {et~al.}(2015)\citenamefont {Cao},
  \citenamefont {Qin},\ and\ \citenamefont {Bass}}]{Duke}%
  \BibitemOpen
  \bibfield  {author} {\bibinfo {author} {\bibfnamefont {S.}~\bibnamefont
  {Cao}}, \bibinfo {author} {\bibfnamefont {G.}~\bibnamefont {Qin}}, \ and\
  \bibinfo {author} {\bibfnamefont {S.}~\bibnamefont {Bass}},\ }\href {\doibase
  10.1103/PhysRevC.92.024907} {\bibfield  {journal} {\bibinfo  {journal} {Phys.
  Rev. C}\ }\textbf {\bibinfo {volume} {92}},\ \bibinfo {pages} {024907}
  (\bibinfo {year} {2015})}\BibitemShut {NoStop}%
\bibitem [{\citenamefont {Bleicher}\ \emph {et~al.}(1999)\citenamefont
  {Bleicher} \emph {et~al.}}]{urQMD}%
  \BibitemOpen
  \bibfield  {author} {\bibinfo {author} {\bibfnamefont {M.}~\bibnamefont
  {Bleicher}} \emph {et~al.},\ }\href
  {http://stacks.iop.org/0954-3899/25/i=9/a=308} {\bibfield  {journal}
  {\bibinfo  {journal} {J. Phys. G}\ }\textbf {\bibinfo {volume} {25}},\
  \bibinfo {pages} {1859} (\bibinfo {year} {1999})}\BibitemShut {NoStop}%
\end{thebibliography}%

\end{document}